\PassOptionsToPackage{colorlinks=true, linkcolor=magenta, urlcolor=blue, citecolor=blue, unicode}{hyperref}
\documentclass[twocolumn,superscriptaddress,amsmath,amssymb,aps,prx,nofootinbib, 10pt,longbibliography]{revtex4-2}

\usepackage{color}
\usepackage{CJKutf8}
\usepackage[T1]{fontenc}
\usepackage{amsmath}
\usepackage{amssymb}
\usepackage{graphicx}
\graphicspath{ {./images/} }
\usepackage{romannum}
\usepackage{physics}
\usepackage[caption=false]{subfig}
\usepackage{comment}
\usepackage{mathtools}
\usepackage{import}
\usepackage{natbib}
\usepackage{algorithm}
\usepackage[noend]{algpseudocode}
\usepackage[normalem]{ulem}
\usepackage{orcidlink}
\usepackage{suffix}
\usepackage{bbm}
\usepackage{diagbox}

\definecolor{MAGENTA}{cmyk}{0,1,0,0}
\definecolor{LightPink}{rgb}{0.858, 0.188, 0.478}
\definecolor{RED4}{rgb}{0.55,0,0}
\def\qq{\hbox{\foreignlanguage{greek}{\coppa}}}

\newcommand{\renyi}{R\'enyi}

\usepackage{amsfonts}
\usepackage{tabularx}
\usepackage{dcolumn}
\usepackage{bm}
\usepackage{graphicx}
\usepackage{epstopdf}
\usepackage[greek,english]{babel}
\usepackage{teubner}

\usepackage[dvipsnames]{xcolor}

\usepackage{tensors}
\usepackage{braket}

\usetikzlibrary{positioning,shapes.misc}

\newcommand{\ranglee}{\rangle\!\rangle}
\newcommand{\langlee}{\langle\!\langle}
\newcommand{\la}{\langle}

\usepackage[capitalise,compress]{cleveref}

\begin{document}

\hypersetup{
  colorlinks=true,
  linkcolor=magenta,
  urlcolor=blue,
  citecolor=blue,
  unicode=true
}

\begin{CJK}{UTF8}{gbsn}
\pagenumbering{arabic}

\title{Mixed-state phases induced by power-law quantum channels}

\author{Christopher Fechisin\,\orcidlink{0000-0002-2590-8221}}
    \email[Corresponding author: ]{fechisin@umd.edu}
  \affiliation{Joint Quantum Institute, NIST/University of Maryland, College Park, MD, 20742, USA}\affiliation{Joint Center for Quantum Information and Computer Science, NIST/University of Maryland, College Park, MD, 20742, USA}

  \author{Tsung-Cheng Lu\,\orcidlink{0000-0001-9438-1067}}
  \affiliation{Joint Center for Quantum Information and Computer Science, NIST/University of Maryland, College Park, MD, 20742, USA}
  \affiliation{Department of Physics and Astronomy, Center for Quantum Research and Technology,
University of Oklahoma, Norman, OK 73069, USA}
  
  \author{Zhi-Yuan Wei (魏志远)\,\orcidlink{0000-0003-4465-2361}}
  \email[Corresponding author: ]{zywei@umd.edu}
    \affiliation{Joint Quantum Institute, NIST/University of Maryland, College Park, MD, 20742, USA}\affiliation{Joint Center for Quantum Information and Computer Science, NIST/University of Maryland, College Park, MD, 20742, USA}

  \author{Jeet Shah\,\orcidlink{0000-0001-5873-8129}}
  \affiliation{Joint Quantum Institute, NIST/University of Maryland, College Park, MD, 20742, USA}\affiliation{Joint Center for Quantum Information and Computer Science, NIST/University of Maryland, College Park, MD, 20742, USA}

\author{Yu-Xin Wang (王语馨)\,\orcidlink{0000-0003-2848-1216}}
  \affiliation{Joint Center for Quantum Information and Computer Science, NIST/University of Maryland, College Park, MD, 20742, USA}

\author{Alexey~V.~Gorshkov\,\orcidlink{0000-0003-0509-3421}}
  \affiliation{Joint Quantum Institute, NIST/University of Maryland, College Park, MD, 20742, USA}\affiliation{Joint Center for Quantum Information and Computer Science, NIST/University of Maryland, College Park, MD, 20742, USA}

  \author{Cheng-Ju Lin\,\orcidlink{0000-0001-7898-0211}}
  \affiliation{Joint Quantum Institute, NIST/University of Maryland, College Park, MD, 20742, USA}\affiliation{Joint Center for Quantum Information and Computer Science, NIST/University of Maryland, College Park, MD, 20742, USA}
  \affiliation{BlueQubit Inc., San Francisco, CA, 94105, USA}
  \affiliation{Department of Physics, University of Miami, Coral Gables, FL, 33146, USA}
 
\date{\today}
\begin{abstract}
Open quantum systems can exhibit inherently mixed quantum orders, such as strong-to-weak spontaneous symmetry breaking (SW-SSB), which have no analogue in pure states. It is generally expected that a finite-depth local quantum channel cannot induce a phase transition to SW-SSB in one spatial dimension. We circumvent this obstruction by introducing a long-range quantum channel, which implements two-body dephasing on pairs of sites with probability falling off according to a power-law in the site separation. Applying this channel to a one-parameter family of matrix product states interpolating between trivial and symmetry-protected topological (SPT) phases, we construct a phase diagram that hosts trivial, mixed SPT, and SW-SSB phases with respect to \renyi-2 diagnostics. Notably, the mixed SPT phase is distinct from previously studied decohered SPT phases: it hosts two logical qubits and exhibits nonvanishing \renyi-2 string order, despite vanishing conventional string order. Through a series of mappings to a classical statistical mechanics model, we establish a direct connection between the loss of logical information and the onset of SW-SSB in this model. Finally, we consider the feasibility of experimental realization of the phenomena we have studied. We emphasize that our family of mixed states can be tuned to regimes of high purity, substantially reducing measurement overhead and enabling their experimental realization and detection on near-term quantum hardware. 
\end{abstract}

{\begingroup
\maketitle
		\endgroup}
\end{CJK}

\makeatletter
\def\l@subsubsection#1#2{}
\makeatother

\tableofcontents

\section{Introduction}
\label{sec:intro}

The exploration of possible phases of matter is one central goal in quantum many-body physics. In the context of ground states of local Hamiltonians, there has been significant progress in classifying distinct phases and their transitions. What happens if the system of interest is coupled to an external environment and undergoes a non-equilibrium process? How should we define a phase for such a non-equilibrium many-body mixed state, where the notion of adiabatic continuation in gapped Hamiltonians is no longer applicable? More broadly, are there new phenomena that can occur in many-body mixed states without any counterpart in isolated systems described by pure states? These questions motivate the exploration of ``mixed-state phases’’ and their transitions. This research is particularly timely due to the rapid development of controllable quantum devices, which can be subject to both natural and engineered noisy dynamics ~\cite{ma2019dissipative,gertler2021protecting,harringtonEngineered2022,brown2022trade,cole2022resource,malinowski2022generation,vanmourik2024experimental}.

Symmetry provides a simple reasoning for why the mixed state phases can exhibit new phenomena absent in pure states. Unlike pure states, mixed states can be symmetric in two distinct ways, known as strong (or exact) and weak (or average) symmetry~\cite{buca2012note,albert2014symmetries,albert2018lindbladians,lieu2020symmetry}. Given a set of unitary transformations $\{U_h\mid h\in G\}$, a mixed state $\rho$ is said to have a strong $G$ symmetry if $U_h \rho=e^{i\theta_h}\rho$, for some phases $\theta_h$, for all $h\in G$. It is said to have a weak $G$ symmetry if $U_h \rho U_h^\dagger=\rho$ for all $h\in G$. Based on the definitions, it is easy to see that strong symmetry implies weak symmetry, while the converse is not necessarily true. In particular, when restricting to pure states, the strong symmetry and weak symmetry become equivalent. As such, many-body mixed states naturally have a richer symmetry structure, thereby suggesting a diverse possibility of new phenomena associated with distinct symmetry-breaking patterns that do not occur in pure states. 

Since strong symmetry necessarily implies weak symmetry, a strong symmetry constitutes a large symmetry group that contains the weak symmetry as a subgroup. This algebraic structure suggests an interesting possibility: strong-to-weak spontaneous symmetry breaking (SW-SSB)~\cite{lee2023quantum,lessa_strong--weak_2024,sala_spontaneous_2024} (see Ref.~\cite{wang2026strongtoweakspontaneoussymmetrybreaking} for a review), which occurs when a strong symmetry is spontaneously broken down to a weak one. SW-SSB is an intrinsically mixed-state phenomenon; it cannot occur in pure states due to the equivalence of strong and weak symmetry in that setting. Consequently, there has been significant interest in exploring SW-SSB from a variety of perspectives~\cite{gu2024spontaneous,ma2024symmetry,sala_spontaneous_2024,kuno2024strong,lessa2025higher,huang2025hydrodynamics,Sun2025scheme,ziereis2025strong,guo2025quantum,lu2025holographicdualitybulktopological}. Notably, this phenomenon has emerged as a unified framework with implications for many different areas of study, including stability of topological quantum memories and error correction~\cite{bao2023mixed-state,fan_2024_breakdown,Grover_separability_2024,Lu_boundary_noise_2024,vijay2025informationcriticalphasesdecoherence,Zhang2025strong,Sa2025exactly}, thermalization~\cite{lessa_strong--weak_2024,kim2025persistent}, and hydrodynamics~\cite{huang2025hydrodynamics,hauser2026strongtoweaksymmetrybreakingopen}.

Given these developments, an immediate question with direct experimental relevance is the following: When can evolution under a quantum channel induce SW-SSB? In two and higher dimensions, there are known examples where a symmetric initial state is mapped by a symmetric finite-depth local channel to an SW-SSB state over a finite range of parameters, such as the channel strength \cite{lessa_strong--weak_2024,sala_spontaneous_2024,kuno2024strong,ding2026strong}. On the other hand, it is generally believed that an extended SW-SSB phase cannot occur in one dimension (1D). This is based on the fact that the SW-SSB phase in certain classes of noisy mixed states admits a dual description in terms of the ordered phase in a finite-temperature classical statistical mechanics model with short-range interactions, where tuning a temperature corresponds to tuning the noise strength $p$ of the channel \cite{sala_spontaneous_2024}. The absence of a finite-temperature ordered phase in 1D short-range statistical mechanics models \cite{thouless1969longrange} then suggests the absence of SW-SSB in 1D over an extended range of parameters. 

However, there is much precedent for overcoming limitations of short-ranged models by introducing long-range interactions whose strength decays as a power law with distance (see Ref.~\cite{defenu2023longrange} for a review). The classical 1D Ising model is an illustrative example. Although it does not host a finite-temperature phase transition with nearest-neighbor interactions~\cite{thouless1969longrange}, such a phase transition emerges when long-ranged interactions are introduced~\cite{dyson1969existence,fisher1972critical,cardy1981onedimensional}. This raises a natural question: Can a one-dimensional channel with \textit{power-law} correlations induce a transition to an extended SW-SSB phase?  We answer this question affirmatively by constructing such a channel, which implements two-body dephasing on pairs of sites with probability falling off according to a power law in the site separation, and demonstrating that it exhibits a finite-$p$ transition to SW-SSB in the \renyi-2 sense. 

A second ingredient in our construction is the choice of the initial pure state to which the power-law channel is applied. We consider the family of matrix-product states $\ket{\psi_g}$ parametrized by $g$ introduced in Ref.~\cite{wolf2006quantum}, to be defined precisely in \cref{eq:psi_g_MPS}. This family:
\begin{enumerate}
    \item respects an antiunitary $\mathbb{Z}_2^T$ symmetry,
    \item interpolates between the cluster-state fixed point (in the SPT phase) at $g=-1$ and the product-state fixed point (in the trivial phase) at $g=1$,
    \item undergoes a gap closing at $g=0$ where $\ket{\psi_{g}}$ is the GHZ state, 
 which is invariant under the channel we consider.
\end{enumerate}
In choosing this family of initial states, we build on a sizable body of existing work focused on local decoherence applied to the $|g|=1$ fixed points~\cite{ma2024symmetry,lee2025robust,chirame_stable_2024,kuno2024strong,kuno2025renyimarkovlengthonedimensional,zhang_quantum_2024,lee2024symmetry,chen_symmetry-enforced_2024,weinstein2025efficient,sala_spontaneous_2024}. We extend these results in two directions, introducing both long-range noise and non-fixed-point initial states. The payoff of adding long-range noise is clear: we obtain an extended SW-SSB phase in 1D. In venturing beyond the fixed points, we uncover mixed states with large purity which can be more efficiently probed on quantum devices, as we will soon discuss.

\begin{figure*}[t]
    \centering
    \begin{picture}(0,0)
    \put(5,135){\large (a)}
    \end{picture}
    \raisebox{30mm}{
    \begin{tikzpicture}[
    font=\normalsize,
    Zsub/.style={inner sep=0.05cm, shape=circle, fill=BrickRed,
                 label={[label distance=-0.1cm]0:$Z_{#1}$}},
  ]
  %%%%%%%%%%%%%%%%% frame
  \draw[rounded corners=6pt] (-0.55,-1.75) rectangle (6.55,1.15);

  %%%%%%%%%%%%%%%%% action of the channel (above the frame)
  \node[NavyBlue] at (1.5,1.85) {$\ket{\psi_g}$};
  \node at (4.0,1.85) {$\rho(\alpha;g,p)$};
  \draw[-stealth] (2.05,1.85) -- (3.10,1.85);
  \node[BrickRed] at (2.575,2.25) {$\mathcal{E}^{(\alpha)}_p$};
  \draw (2.575,1.75) -- (2.575,1.15);

  %%%%%%%%%%%%%%%%% chain of tensors with physical legs
  \draw (-0.35,0) -- (6.35,0);
  \foreach \i in {1,...,9}{
    \pgfmathsetmacro{\x}{0.75*(\i-1)}
    \coordinate (s\i) at (\x,0);
    \draw (\x,0) -- (\x,0.75);
  }

  %%%%%%%%%%%%%%%%% power-law-correlated pairs
      \def\arcang{48}          % common tangent angle at the shared node
  \begin{scope}
    \clip[rounded corners=6pt] (-0.35,-1.75) rectangle (6.55,1.15);
    \foreach \k/\o in {1/0.75, 2/0.5, 3/0.35, 4/0.25, 5/0.21, 6/0.18, 7/0.12}{
      \pgfmathsetmacro{\xl}{0.75-0.75*\k}
      \pgfmathsetmacro{\xr}{0.75+0.75*\k}
      \pgfmathsetmacro{\lo}{pow(\k,-0.22)}
      \draw[opacity=\o] (s2) to[out=180+\arcang, in=-\arcang, looseness=\lo] (\xl,0);
      \ifnum\k=5\relax\else   % k=5 on the right is the highlighted pair
        \draw[opacity=\o] (s2) to[out=-\arcang, in=180+\arcang, looseness=\lo] (\xr,0);
      \fi
    }
    % the highlighted pair (i,j)
    \pgfmathsetmacro{\lo}{pow(5,-0.22)}
    \draw[BrickRed] (s2) to[out=-\arcang, in=180+\arcang, looseness=\lo] (s7);
  \end{scope}

  %%%%%%%%%%%%%%%%% dephasing operators on the physical legs
  \node[Zsub=i] at (0.75,0.4) {};
  \node[Zsub=j] at (4.50,0.4) {};

  %%%%%%%%%%%%%%%%% pair probability
  \node[BrickRed] at (3.0,-1.25)
    {$\mathrm{Pr}(Z_iZ_j)\propto\dfrac{p}{|i-j|^{\alpha}}$};
    \foreach \i in {1,...,9}{
      \pgfmathsetmacro{\x}{0.75*(\i-1)};
        \node[blankg] at (\x,0) {};
      } 
\end{tikzpicture}}
    \hfill
\includegraphics[width=0.53\textwidth]{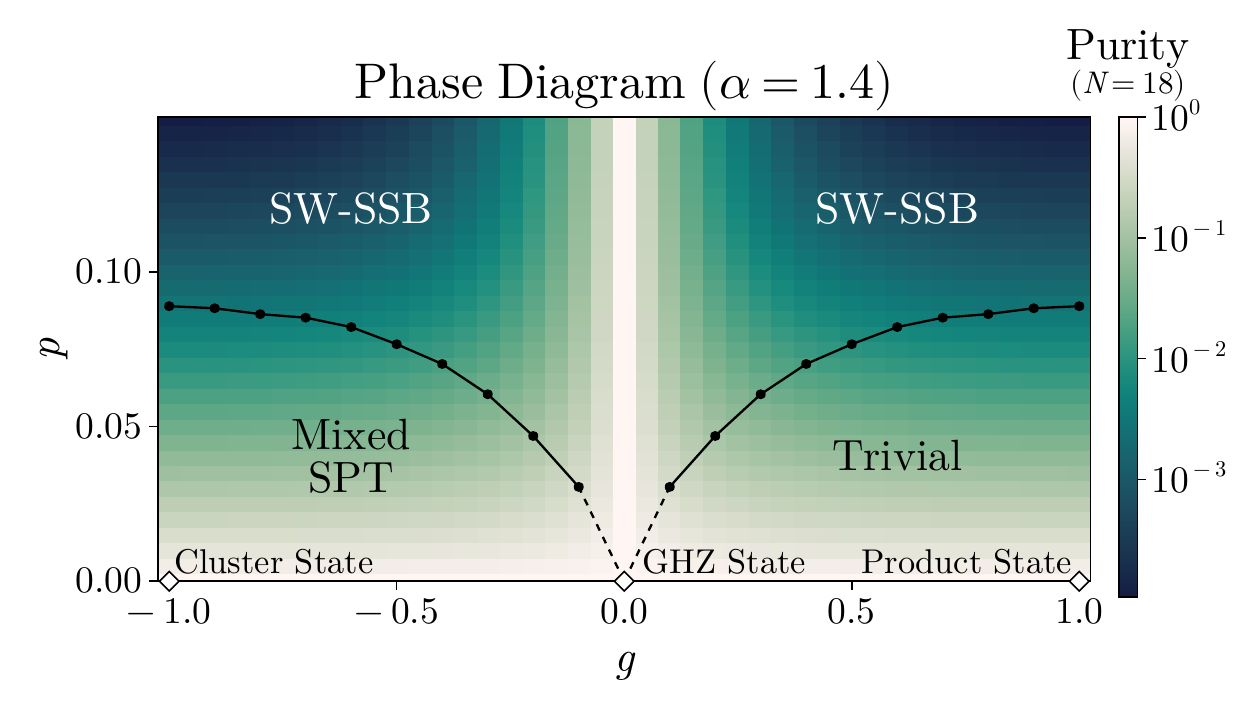}
    \begin{picture}(0,0)
    \put(-270,135){\large (b)}
    \end{picture}
    \caption{(a) Summary of the construction considered in this work. We produce a family of mixed states $\rho(\alpha;g,p)$ by decohering the MPS $\ket{\psi_g}$ via the power-law-correlated quantum channel $\mathcal{E}^{(\alpha)}_p$ introduced in \cref{eq:main_channel}. (b) Phase diagram classifying $\rho(\alpha;g,p)$ for one choice of fixed $\alpha$, namely $\alpha = 1.4$. The phase diagram features two regions in which mixed analogues of the pure-state orders persist, giving way at the phase boundary $p=p_c(g)$ to SW-SSB in the \renyi-2 sense. Each point denotes a measured phase boundary, with solid lines interpolating linearly between the measured values. The dashed lines are extrapolations based on the argument that $p_c(g)\to 0$ as $|g|\to 0$ within the statistical mechanics model we derive. We emphasize that we find mixed-state phase transitions at relatively large purity by exploring pure states beyond the RG fixed points, which are denoted by white diamonds along the $p=0$ axis.}
    \label{fig:summary}
\end{figure*}

Constructing the phase diagram as a function of $g$ and dephasing strength $p$ for representative values of the power-law exponent is a central goal of this work. Our task is partially simplified by the existence of a finite-depth local unitary---namely, the cluster state entangler---relating $\ket{\psi_g}$ to $\ket{\psi_{-g}}$. This mapping guarantees that both states undergo SW-SSB at the same critical $p=p_c(|g|)$. While the $0<p<p_c$ phase is featureless for $g>0$, we show that it is a non-trivial mixed SPT phase protected by an antiunitary $\mathbb{Z}_2^T$ symmetry for $g<0$. An overview of our construction and a representative phase diagram are presented in \cref{fig:summary}. 

Characterizing the mixed SPT order is especially subtle because the string order parameter which diagnoses SPT order in the pure SPT phase vanishes in the mixed SPT phase, as we will discuss in \cref{sec:string_order_params}. We circumvent this difficulty by examining the coherent information~\cite{schumacher1996quantum,lloyd1997capacity}, a measure of how much logical information encoded in the edge modes survives decoherence. This perspective reveals that the phase obtained by decohering the pure SPT phase is indeed a nontrivial mixed SPT phase, and we are able to construct a string order parameter nonlinear in the density matrix which can diagnose this order. 

From a practical standpoint, the family of mixed states which we construct is also well-suited to address an important experimental challenge: A mixed state with smaller purity will require more samples to accurately resolve observables of interest on a quantum device. Fundamentally, this bottleneck arises because diagnosing mixed-state order requires observables that are nonlinear in the density matrix.
The most efficiently measurable observables that can detect mixed-state order are quadratic in the density matrix and we refer to them as \renyi-2 quantities. Given a density matrix $\rho$, the purity $\Tr[\rho^2]$ sets the scale of all \renyi-2 observables, and in turn strongly influences the resources required to measure signatures of mixed-state order with sufficient accuracy. The purity is typically exponentially small in system size, which presents a significant obstacle to any experimental realization of mixed-state order. 

Because the GHZ state $(g=0)$ is invariant under the action of our dephasing channel, continuity implies that nearby initial states experience a slow decay in purity with system size and channel strength, which we confirm numerically. We thereby identify a family of mixed states which (i) exhibit a finite-$p$ transition from novel mixed-state SPT order to SW-SSB and (ii) have large purity. Furthermore, because our initial states are matrix product states (MPS) with small bond dimension, there are a variety of proposals for efficiently preparing them on near-term hardware~\cite{Schon2005,wei2025state,malz2023preparation,smith2024constant,zyadi}, with several approaches already implemented experimentally~\cite{smith2022crossing,scheer2025renormalization}. For these reasons, the family of mixed states we construct constitutes a prime candidate for observing mixed-state phases and their transitions on a quantum device. In \cref{sec:device}, we propose an explicit experimental protocol for detecting our SW-SSB transition on quantum hardware and quantify the sample advantage enabled by the higher purity discussed above.

The remainder of the text is organized as follows. In \cref{sec:model}, we introduce a family of $\mathbb{Z}_2^T$-symmetric pure states and the power-law quantum channels. We also review the relevant pure-state SPT physics in this section. In \cref{sec:SW-SSB}, we present observables which diagnose SW-SSB and use these observables to identify the critical $p_c(g)$ at which our channel induces SW-SSB. In \cref{sec:mixed_SPT}, we characterize the mixed SPT phase which persists below $p_c(g)$. Finally, in \cref{sec:device}, we present a protocol for realization of the SW-SSB transition on a quantum device. We demonstrate that certain initial states---namely, those with small $|g|$---maintain comparatively large purity upon application of our channel, which dramatically reduces shot overhead in a potential experimental realization.

\section{Model}\label{sec:model}
In this section, we introduce the family of initial pure states and the power-law channel used to decohere them.

\subsection{Family of matrix product states}
We begin with a one-parameter family of matrix product states on an open chain, which are given (up to normalization\footnote{Here and throughout, we will not be overly concerned with constant multiplicative factors in wavefunctions and amplitudes. We assume that physical wavefunctions are appropriately normalized.}) by ~\cite{wolf2006quantum}
\begin{equation}\label{eq:psi_g_MPS}
    \ket{\psi_g}=\sum_{\{i_j\}=0,1}\bra{L}M_g^{[i_1]}\cdots M_g^{[i_N]}\ket{R}\ket{i_1,\ldots,i_N},
\end{equation}
where the MPS tensor is
\begin{equation}\label{eq:M-tensors-def}
    M_g^{[0]}=\begin{pmatrix}
        0&0\\1&1
    \end{pmatrix},\quad M_g^{[1]}=\begin{pmatrix}
        1&g\\0&0
    \end{pmatrix},
\end{equation}
and the vectors $\bra{L}$ and $\ket{R}$ encode the boundary conditions, which we will discuss shortly. We take system size $N$ to be even for concreteness, though this is not essential to the physics. Diagrammatically, we introduce the bulk and boundary tensors
\begin{subequations}
\begin{equation}
    \begin{tikzpicture}
    \draw (-0.5,0)--(0.5,0);
    \draw (0,0.0) -- (0,0.5);
    \node[plusg] at (0,0) {};
    \node[] at (0,0.65) {$i$};
    \node[] at (-0.65,0) {$j$};
    \node[] at (0.65,0) {$k$};
  \end{tikzpicture}=\left(M_g^{[i]}\right)_{jk}, 
\end{equation}
\begin{equation}
  \begin{tikzpicture}[baseline=0]
    \draw (-0.5,0)--(0,0);
    \node[gedge] at (-0.5,0) {};
  \end{tikzpicture}=\bra{L},\quad \begin{tikzpicture}[baseline=0]
    \draw (-0.5,0)--(0,0);
    \node[gedge] at (0,0) {};
  \end{tikzpicture}=\ket{R},
\end{equation}
\end{subequations}

so that the state $\ket{\psi_g}$ can be expressed as
\begin{equation}
    \ket{\psi_g}=\begin{tikzpicture}
        \draw (0,0)--(5.2,0);
        \node[gedge] at (0,0) {};
        \node[gedge] at (5.2,0) {};
        \foreach \x in {0.5,1.2,1.9,3.3,4.0,4.7}{
        \draw (\x,0)--(\x,0.5);
        \node[plusg] at (\x,0) {};
        }
        \node[fill=white] at (2.6,0) {$\dots$};
    \end{tikzpicture}.
\end{equation}
When unambiguous, we will suppress the ``$g$'' label on the MPS tensors. The leg labeled by $i$ is the \textit{physical leg} of the tensor and corresponds to the local Hilbert space of the model at a particular site. The legs labeled by $j$ and $k$ are the \textit{virtual legs} of the tensor, which encode correlations across sites.

Next, we introduce a tensor identity which will simplify the coming analysis. Notice that 
\begin{equation}\label{eq:one-over-g}
    \begin{tikzpicture}
    \draw (-0.5,0)--(1.0,0);
    \draw (0,0.0) -- (0,0.5);
    \draw (0.5,0.0) -- (0.5,0.5);
    \node[plusg] at (0,0) {};
    \node[plusg] at (0.5,0) {};
  \end{tikzpicture} = g\cdot\left(
  \begin{tikzpicture}
    \draw (-0.8,0)--(1.6,0);
    \draw (0,0.0) -- (0,0.5);
    \draw (0.8,0.0) -- (0.8,0.5);
    \node[overg] at (0,0) {};
    \node[overg] at (0.8,0) {};
    \node[Xv] at (-0.4,0) {};
    \node[Xv] at (1.2,0) {};
    \node[X] at (0,0.3) {};
  \end{tikzpicture} \right),
\end{equation}
where ``$\cdot$'' indicates the usual multiplication of a tensor by a scalar, namely scalar multiplication of all entries. The $X$ operators appearing on the virtual legs cancel pairwise in the bulk, which allows us to conclude that $\ket{\psi_g}=\ket{\psi_{1/g}}$ up to a relabeling of the local physical basis and a rotation in each edge Hilbert space, once normalization is taken into account. Hence, nothing novel is gained in considering $g$ outside of the interval $[-1,1]$, so we restrict to $g\in[-1,1]$ from this point forward. 

One can show that $\ket{\psi_g}$ interpolates between the cluster state fixed point ($g=-1$) in the SPT phase and the product state fixed point ($g=1$) in the trivial phase via the GHZ state ($g=0$)~\cite{wolf2006quantum}. These phases are defined with respect to an anti-unitary $\mathbb{Z}_2^T$ symmetry which we now introduce. 

\subsubsection{Symmetries}

The antiunitary $\mathbb{Z}_2^T$ symmetry which protects the SPT phase is given by $\prod_i X_i\mathcal{K}$, where $\mathcal{K}$ denotes complex conjugation. We can verify that $\ket{\psi_g}$ respects this symmetry by examining the local tensors $M_g$. Notice that $M_g$ satisfies \cite{wolf2006quantum}
\begin{equation}\label{eq:pull_thru_V}
    \begin{tikzpicture}
        \draw (-0.5,0)--(0.5,0);
        \draw (0,0.0) -- (0,0.5);
        \node[blankg] at (0,0) {};
        \node[X] at (0,0.3) {};
    \end{tikzpicture}=\begin{tikzpicture}
        \draw (-0.8,0)--(0.8,0);
        \draw (0,0.0) -- (0,0.5);
        \node[blankg] at (0,0) {};
        \node[Vginv] at (-0.5,0.0) {};
        \node[Vg] at (0.5,0.0) {};
    \end{tikzpicture},\qquad V_g=\begin{pmatrix}
			    0&g\\1&0
			\end{pmatrix},
\end{equation}
so that an $X$ acting on the physical index pulls through to a $V_g^{-1}$ on the left virtual index and a $V_g$ on the right virtual index.  \cref{eq:pull_thru_V} implies that $\prod_i X_i$ is a symmetry of $\ket{\psi_g}$ due to the pairwise cancellation of $V_g$ and $V_g^{-1}$ in the bulk, which may act nontrivially at the edges. The tensor $M_g$ is also invariant under the action of complex conjugation $\mathcal{K}$ because it is real-valued. This implies that $\ket{\psi_g}$ is invariant in the bulk under complex conjugation, though the edges may transform nontrivially. 

Although $\ket{\psi_g}$ is symmetric under the action of $\prod_i X_i$ or $\mathcal{K}$ alone, the SPT phase is only protected by the combined symmetry $\mathcal{T}:=\prod_i X_i\mathcal{K}$.  This can be readily seen from the group cohomology classification of 1D SPTs~\cite{chen2012symmetry}, as $\mathcal{H}^2(\mathbb{Z}_2^T,U_T(1))$ contains a nontrivial element, while $\mathcal{H}^2(\mathbb{Z}_2,U(1))$ does not. More explicitly, we demonstrate in \cref{sec:MPS-SPT} that $\mathcal{T}$ acts non-trivially at the edges of $\ket{\psi_g}$, a characteristic feature of SPT order.

\subsubsection{Boundary conditions}

For most of the manuscript, we will work with a choice of boundary conditions such that $\ket{\psi_g}$ is a $+1$ eigenstate of $\prod_i X_i$. (However, we will need an entire logical basis to compute the coherent information, which we define in \cref{sec:coh_info}.) We refer to this choice as the \textit{symmetric boundary conditions}. In particular, we will use the symmetric boundary conditions when computing diagnostics of SW-SSB in \cref{sec:SW-SSB}. To configure these boundary conditions, we set $\bra{L}$ and $\ket{R}$ to appropriate eigenstates of $V_g^{-1}$ and $V_g$, respectively. One such choice is  

 \begin{equation}\label{eq:triv_BCs}
	     \bra{L}=\begin{pmatrix} \frac{1}{\sqrt{g}}&1\end{pmatrix},\quad \ket{R}=\begin{pmatrix}{\sqrt{g}}\\1\end{pmatrix}.
 \end{equation}
 We also note that this choice of boundary conditions ensures that $\ket{\psi_g}$ is symmetric under the full symmetry $\mathcal{T}$ when $g>0$. Hereafter, when we write $\ket{\psi_g}$, we mean the state with $\bra{L}$ and $\ket{R}$ defined as in \cref{eq:triv_BCs}.

We gain a better physical understanding of the intermediate states $\ket{\psi_g}$ by observing that the amplitude of the computational basis state $\ket{\boldsymbol{i}}\equiv \ket{i_1,i_2,\ldots,i_N}$ is given by
\begin{equation}
\label{eq:mps-coeff-g-domain}
    \bra{L}M_g^{[i_1]}M_g^{[i_2]}\cdots M_g^{[i_N]}\ket{R} \propto g^{\text{ DW(}\boldsymbol{i})/2},
\end{equation}
 where the function $\mathrm{DW}(\boldsymbol{i})$ counts the number of domain walls in the configuration $\boldsymbol{i}$. The proof of \cref{eq:mps-coeff-g-domain} is presented in \cref{app:DW_perspective}. From \cref{eq:mps-coeff-g-domain}, we see that the fixed-point states can be brought into the forms
 \begin{equation}\label{eq:fixed_points}
 \begin{split}
     &\ket{\psi_{g=1}}=\ket{+}^{\otimes N},\\
     &\ket{\psi_{g=-1}}=S_1S_N\left(\prod_{i=2}^{N-1} Z_i\right)\left(\prod_{j=1}^{N-1}CZ_{j,j+1}\ket{+}^{\otimes N}\right),
\end{split}
 \end{equation}
where $S=\sqrt{Z}$, $CZ_{i,j}$ is the controlled-Z gate, and all subscripts indicate site indices. We arrive at this form by recognizing that
\begin{equation}
    (-1)^{\mathrm{DW}(\boldsymbol{i})/2}=(-1)^{\sum_{i=1}^{N-1}\frac{1-Z_iZ_{i+1}}{4}}=\prod_{i=1}^{N-1}S_iS_{i+1}CZ_{i,i+1}.
\end{equation}
 Notice that $\prod_{j=1}^{N-1}CZ_{j,j+1}\ket{+}^{\otimes N}$ is the ordinary cluster state, so that $\ket{\psi_{g=-1}}$ is the ordinary cluster state up to a basis rotation on each site. 
 
 Using \cref{eq:mps-coeff-g-domain}, we are able to express the state $\ket{\psi_g}$ via imaginary-time evolution of the fixed-point states given in \cref{eq:fixed_points}:
\begin{equation}\label{eq:imag_time_symm}
     \ket{\psi_g}\propto e^{\beta\sum_{i=1}^{N-1}Z_i Z_{i+1}}\times \begin{cases}
     \ket{\psi_{g=1}},&g>0\\
     \ket{\psi_{g=-1}},&g<0
     \end{cases},
 \end{equation}
where $\beta=-\frac{1}{4}\ln|g|$ and we have once again used the fact that $\frac{1}{2}\sum_{i=1}^{N-1}(1- Z_i Z_{i+1})$ counts the domain walls in each computational basis state. Notice that as $g\to 0$ $(\beta\to\infty)$, the imaginary-time evolution operator becomes a projector onto the GHZ state. 

For the symmetric boundary conditions introduced in \cref{eq:triv_BCs}, there exists a finite-depth local unitary transformation $T$ relating $\ket{\psi_{g}}$ to $\ket{\psi_{-g}}$. Explicitly, for $g\neq 0$ we have
\begin{equation}\label{eq:T_def}
	T\ket{\psi_g}=\ket{\psi_{-g}},\quad T=S_1S_N\prod_{i=2}^{N-1}Z_i \prod_{j=1}^{N-1}CZ_{j,j+1},
\end{equation}
which is precisely the unitary appearing in \cref{eq:fixed_points}. In \cref{ap:CZ_trans}, we prove that $T$ maps $\ket{\psi_{g}}$ to $\ket{\psi_{-g}}$ and show that it commutes with $\prod_i X_i$. 

\subsubsection{Review of MPS with antiunitary symmetry}\label{sec:MPS-SPT}

 Given a symmetric MPS, it is natural to ask and straightforward to answer whether it belongs to the trivial or nontrivial SPT phase with respect to this symmetry. We can diagnose the SPT order of $\ket{\psi_g}$ as a function of $g$ by examining how the symmetry fractionalizes to the edges of the chain. Using complex conjugation and \cref{eq:pull_thru_V}, we have 
\begin{equation}
    \mathcal{T}\ket{\psi_g}=
\begin{tikzpicture}
        %%%%%%%%%%%%%%%%% ket layer
        \draw (-0.4,0)--(5.5,0);
        \node[gedge] at (-0.4,0) {};
        \node[gedge] at (5.5,0) {};
        \foreach \x in {0.8,1.5,2.2,3.6,4.3}{
        \draw (\x,0)--(\x,0.5);
        \node[blankg] at (\x,0) {};
        }
        \node[fill=white] at (2.9,0) {$\dots$};
        %%%%%%%%%%%%%%%%% virtuals
        \node[conj] at (-0.05,0.0) {};
        \node[Vginv] at (0.3,0.0) {};
        \node[conj] at (5.15,0.0) {};
        \node[Vg] at (4.8,0.0) {};
    \end{tikzpicture},
\end{equation}
which means that the anti-unitary symmetry $\mathcal{T}$ induces the following effective action on the edge Hilbert spaces:
\begin{equation}\label{eq:edge_alg}
\begin{split}
    \mathcal{T}:&\ket{R}\mapsto V_g\mathcal{K}\ket{R},\\&\bra{L}\mapsto \bra{L}\mathcal{K}V_g^{-1}.
\end{split}
\end{equation}

We now summarize the well-known classification result \cite{pollmann2010entanglement,chen2011classification,chen2011complete} as it applies to this model. The key is to recognize that 
\begin{equation}\label{eq:T_squared_action}
\begin{split}
    \mathcal{T}^2:&\ket{R}\mapsto VV^*\ket{R},\\&\bra{L}\mapsto \bra{L}\left(V^{-1}\right)^*V^{-1}.
\end{split}
\end{equation}
In order for \cref{eq:T_squared_action} to be compatible with the fact that $\mathcal{T}^2=+I$, we find that $VV^*=\pm I$. This sign is a discrete quantity which cannot change continuously as one moves throughout the phase, as expected for a topological invariant. The case where $VV^*=-I$ is identified with the non-trivial SPT phase. For the $V_g$ introduced in \cref{eq:pull_thru_V}, we arrive at $V_gV_g^*=gI$, which is equal to $\mathrm{sgn}(g)I$ up to normalization. This tells us that $g<0$ is the non-trivial SPT phase, and $g>0$ is the trivial phase.

\subsection{Power-law quantum channel}
We now introduce the quantum channel that will be considered in this work. We define the following quantum channel, parameterized by an overall strength $p$ and a power-law exponent $\alpha$:
\begin{equation}\label{eq:main_channel}
\begin{gathered}
    \mathcal{E}^{(\alpha)}_p(\rho)=\prod_{i<j}\mathcal{E}^{(\alpha)}_{ij;p}(\rho),\\ 
    \mathcal{E}^{(\alpha)}_{ij;p}(\rho)=p_{ij}^{(\alpha)} Z_iZ_j\rho Z_iZ_j+\left(1-p_{ij}^{(\alpha)}\right)\rho,
\end{gathered}
\end{equation}
where the $p^{(\alpha)}_{ij}$ are defined as
\begin{equation}\label{eq:pij-def}
p_{ij}^{(\alpha)} = 
\begin{cases}
  \frac{p/N^{1-\alpha}}{|i-j|^\alpha}   & \text{if } 0 \leq \alpha < 1\\
  \frac{p}{|i-j|^\alpha}   & \text{if }  \alpha > 1
\end{cases}.
\end{equation}
By tuning $\alpha$, we can realize a nearest-neighbor channel ($\alpha=\infty$), an all-to-all channel ($\alpha=0$), or any power-law channel in between. We introduce the normalization for $\alpha<1$ in anticipation of the mapping to a statistical mechanics model which we present in \cref{sec:sw-ssb-stat-mech}; this normalization ensures that the energy in the statistical mechanics model is extensive and corresponds to the asymptotic form of the Kac normalization for power-law Ising models~\cite{kac1963van,kastner2011diverging,defenu2024out}. From the perspective of our quantum channel, this normalization scales down the overall strength of the channel for very long-ranged interactions, so that the onset of SW-SSB remains at nonzero $p$ in these cases. 

Taking $\ketbra{\psi_g}{\psi_g}$ as our initial state, we can fix $\alpha$, construct the family of density matrices
\begin{equation}\label{eq:rho_agp}
    \rho(\alpha;g,p)=\mathcal{E}^{(\alpha)}_p(\ketbra{\psi_g}{\psi_g}),
\end{equation}
and ask to what mixed-state phase each $\rho(\alpha;g,p)$ belongs. This process yields a two-variable phase diagram for each choice of $\alpha$. Furthermore, the phases which may emerge are constrained by symmetry. Because $\mathcal{E}^{(\alpha)}_p$ preserves the strong symmetry of the initial state $\ketbra{\psi_g}{\psi_g}$, the final state $\rho(\alpha;g,p)$ must either respect this symmetry or break it spontaneously. As we will discuss in \cref{sec:sw-ssb_num}, we find adjacent regions of symmetric and SW-SSB order in our phase diagrams, furnishing a 1D decoherence-induced SW-SSB transition.

We also emphasize that $\mathcal{E}_p^{(\alpha)}$ leaves the GHZ state $\ket{\psi_{g=0}}$ invariant. This is significant because it implies that the phase diagram in the $g-p$ plane contains a cross section of unit purity. Consequently, we anticipate nearby regions of the phase diagram where the density matrices maintain large purity while exhibiting the onset of SW-SSB. This is promising for experimental realization of  novel mixed-state physics, as we discuss in \cref{sec:device}.

\section{Onset of strong-to-weak SSB}\label{sec:SW-SSB}

In this section, we will discuss the onset of SW-SSB in the states $\ket{\psi_g}$ induced by the channel $\mathcal{E}_p^{(\alpha)}$. We briefly review the physics of SW-SSB in \cref{sec:review_SW-SSB}, demonstrate that the ordinary two-point correlator vanishes in \cref{sec:linear-corr}, then identify how the \renyi-2 two-point correlator can be mapped to an observable in a classical statistical mechanics model in \cref{sec:sw-ssb-stat-mech}, and finally present the numerical results obtained via Monte Carlo numerics in the classical model in \cref{sec:sw-ssb_num}.

\subsection{Review of SW-SSB}\label{sec:review_SW-SSB}
A set of operators $\{U_h\mid h\in G\}$ is said to realize a strong $G$ symmetry of the mixed state $\rho$ if it leaves $\rho$ invariant (up to a phase) under the action of $U_g$ from a single side: $U_h\rho = e^{i\theta_h}\rho$ (or, equivalently, $\rho U_h^\dagger = e^{-i\theta_h} \rho$). Thinking of $\rho$ as an ensemble of pure states, a strong symmetry implies that every pure state in the mixture is symmetric with the same symmetry charge. In contrast, the same set of operators is said to be a weak symmetry of $\rho$ if it leaves $\rho$ invariant only when acting simultaneously from the left and the right: $U_h\rho U_h^\dagger = \rho$. From the ensemble perspective, weak symmetry is a property of the entire ensemble $\rho$ which cannot be attributed to the invariance of individual states.

Suppose a mixed state $\rho$ has a strong symmetry labeled by a group $G$ and realized by a set of operators $\{U_h\mid h\in G\}$. In this case, there is in fact a unique symmetry transformation for each $(h,k)\in G\times G$, as any action of the form $S_h\rho S_k^\dagger$ for $h$, $k\in G$ leaves $\rho$ invariant up to a phase. From this perspective, SW-SSB can be understood as the spontaneous breaking of $G\times G$ down to its diagonal subgroup, with only the transformations of the form $S_h \rho S_h^\dagger$ still leaving $\rho$ invariant. 

Spontaneous symmetry breaking is generally diagnosed through two-point correlators of charged operators, which probe long-range order. In the mixed-state context, one also has access to nontrivial observables which are nonlinear in the density matrix. These nonlinear observables can be helpful diagnostics for probing the structure of mixed states, and are in fact necessary 
to fully characterize the pattern of SSB in a mixed state. To diagnose SW-SSB, we will use two observables: the linear and \renyi-2 two-point correlators~\cite{lessa_strong--weak_2024,Shah2025instabilityofsteady}
\begin{equation}
\label{eq:correlators}
\begin{split}
    C_{\text{I}, mn} &= \frac{\Tr[ Z_m Z_n \rho]}{\Tr[\rho]},\\
    C_{\text{II}, mn} &= \frac{\Tr[ Z_m Z_n \rho Z_m Z_n \rho]}{\Tr[\rho^2]},
\end{split}
\end{equation}
where $\rho$ is a strongly symmetric state which potentially breaks the symmetry spontaneously. We remark that the fidelity correlator introduced in Ref.~\cite{lessa_strong--weak_2024} is a more robust diagnostic of SW-SSB than the \renyi-2 correlator $C_{\mathrm{II}}$, but we focus on the latter because it is significantly easier to measure and has been a meaningful diagnostic in many other examples~\cite{kuno2024strong,sala_spontaneous_2024,Zhang2025strong,orito2025strong,Sun2025scheme,kuno2026hierarchymixedsymmetryprotected,lu2026nonequilibriumtopologicalresponsecharge,ando2026gauge}. Furthermore, we expect that the $p_c$ diagnosed by the \renyi-2 correlator provides an upper bound on the $p_c$ diagnosed by the fidelity correlator. This bound has held in other relevant models~\cite{lessa_strong--weak_2024,fan_2024_breakdown} and an analogous bound has been proven for $p_c$ as witnessed by the coherent information (to be discussed in \cref{sec:coh_info}) in stabilizer codes subject to a special class of Pauli noise~\cite{vijay2026hierarchy}.

The linear correlator $C_{\mathrm{I}}$ is simply the expectation value of the two-point operator $Z_m Z_n$ in the state $\rho$. The operator $Z_i$ is charged under the weak symmetry (in addition to the strong symmetry from one side), so $C_\mathrm{I}$ acquires a nonzero value in the thermodynamic limit only when the weak symmetry is spontaneously broken, i.e. $\mathbb{Z}_2^\text{bra}\times\mathbb{Z}_2^\text{ket}\to \mathbb{Z}_1$. Consequently, $C_\mathrm{I}$ does not differentiate between a symmetry-unbroken state and SW-SSB; in either case, $C_\mathrm{I}\to0$ in the thermodynamic limit. To diagnose SW-SSB, we also measure the \renyi-2 correlator $C_{\mathrm{II}}$. In order to see most clearly what this correlator measures, we introduce the doubled Hilbert space, which is obtained via the mapping
\begin{equation}\label{eq:doubled-mapping}
    \rho = \sum_{i,j} \rho_{ij} \ketbra{i}{j} \mapsto | \rho \ranglee \equiv \sum_{i,j} \rho_{ij} \ket{i}\ket{j}.
\end{equation}
In particular, we find that
\begin{equation}
    C_{\mathrm{II},mn}\mapsto\frac{\langlee \rho|(Z_m\otimes Z_m)(Z_n\otimes Z_n)|\rho\ranglee}{\langlee \rho|\rho\ranglee}
\end{equation}
under the mapping introduced in \cref{eq:doubled-mapping}. Notice in particular that $C_{\mathrm{II},mn}$ maps to the two-point function of an operator charged under the strong symmetry, but not the weak. For this reason, a vanishing value of $C_\mathrm{I}$ along with a nonzero value of $C_\mathrm{II}$ at large separation is suggestive of SW-SSB.

In summary, we expect the following behavior in the thermodynamic limit: a symmetric state will have vanishing values for both correlators defined in \cref{eq:correlators}, a state exhibiting SW-SSB will have a nonzero value for $C_{\mathrm{II}}$ only, and a completely symmetry-broken state will have a nonzero value for both correlators. To construct the phase diagram, we must therefore measure both correlators for each pair $(g,p)$. 

\subsection{Linear correlator}\label{sec:linear-corr}
The linear correlator $C_\mathrm{I}$ introduced in \cref{eq:correlators} can be determined exactly throughout the phase diagram. In particular, we have
\begin{equation}
\label{eq:C_I_derivation}
\begin{split}
    C_{\text{I}, mn} = \frac{\sum_{\{s_i\}=\pm 1} s_m s_n e^{2\beta\sum_{i=1}^{N-1}s_i s_{i+1}}}{\sum_{\{s_i\}=\pm 1}e^{2\beta\sum_{i=1}^{N-1}s_i s_{i+1}}},
\end{split}
\end{equation}
where $\beta=-\frac{1}{4}\ln|g|$ is introduced by the imaginary-time evolution defined in \cref{eq:imag_time_symm} and the variables $\{s_i\}$ enter as eigenvalues of the $\{Z_i\}$ after expanding in the computational basis (the full derivation may be found in \cref{ap:nn-ising}). Notice that \cref{eq:C_I_derivation} is precisely the two-point correlator in a 1D nearest-neighbor Ising model with coupling $\beta_\text{eff}=-\frac{1}{2}\ln|g|$. Because the 1D nearest-neighbor Ising model is disordered for all finite $\beta$~\cite{thouless1969longrange}, we may conclude that $C_\text{I}$ vanishes for all values of $\alpha$ and $p$ so long as $g\neq 0$.
This tells us that there are no regions of the phase diagram (aside from the GHZ line at $g=0$) where the symmetry is completely broken. However, we must still measure the \renyi-2 correlator to determine whether each phase is symmetric or exhibits SW-SSB.

\subsection{Mapping to a statistical mechanics model}\label{sec:sw-ssb-stat-mech}

We can efficiently compute the \renyi-2 correlator introduced in \cref{eq:correlators} by mapping it to a correlation function in a classical 1D statistical mechanics model, which we will introduce in this section. We first make a brief technical aside. Examining the form of the correlators presented in \cref{eq:correlators}, it is clear that they are unchanged if $\rho$ is taken to $T\rho T^\dagger$, where $T$ is the SPT entangler defined in \cref{eq:T_def}. Furthermore, because $T\ket{\psi_g}=\ket{\psi_{-g}}$ and $T$ commutes with the decoherence channel $\mathcal{E}_p^{(\alpha)}$, the correlators are in fact unchanged if we take $\rho(\alpha;g,p)$ to $\rho(\alpha;-g,p)$. This implies that SW-SSB occurs at the same critical $p_c(|g|)$ whether in the trivial or SPT phase. We will therefore assume that $g>0$ in the remainder of this section for simplicity, as the same results immediately apply when $g<0$.

Under the standard mapping from matrices to vectors in the doubled Hilbert space introduced in \cref{eq:doubled-mapping}, the channel $\mathcal{E}_{ij;p}^{(\alpha)}$ becomes a linear operator $E_{ij;p}^{(\alpha)}$ acting on the doubled space
\begin{equation}
    \mathcal{E}_{ij;p}^{(\alpha)} \mapsto E_{ij;p}^{(\alpha)} \equiv (1-p_{ij}^{(\alpha)})   I + p_{ij}^{(\alpha)} Z_{i}^L Z_{j}^L Z_{i}^R Z_j^R,
\end{equation}
where $Z_{i}^L$ acts on the ket-part (left) of the vector $|\rho \ranglee$ while $Z_{i}^R$ acts on the bra-part (right).

To obtain the form of a partition function, we can write $E_{ij;p}^{(\alpha)}$ as an operator exponential
\begin{equation}\label{eq:}
    E_{ij;p}^{(\alpha)} = c_{ij;p}^{(\alpha)} e^{ J_{ij}^{(\alpha)} Z_i^L Z_j^L Z_i^R Z_j^R},
\end{equation}
where $c_{ij;p}^{(\alpha)}$ and $ J_{ij}^{(\alpha)}$ are defined via
\begin{equation}
\label{eq:beta-c-def}
\tanh{J_{ij}^{(\alpha)}} = \frac{p_{ij}^{(\alpha)}}{1-p_{ij}^{(\alpha)}}~,\;\;\; c_{ij;p}^{(\alpha)} = \sqrt{\left(1-p_{ij}^{(\alpha)}\right)^2 - \left(p_{ij}^{(\alpha)}\right)^2}~,
\end{equation}
and we have used the fact that $(Z_i^L Z_j^L Z_i^R Z_j^R)^2 = \I$. Here we assume that $p^{(\alpha)}_{ij} < 1/2$ for all pairs of $i$ and $j$, so that $J_{ij}^{(\alpha)}$ and $c_{ij;p}^{(\alpha)}$ are real. The formal assumption that $p^{(\alpha)}_{ij} < 1/2$ is unproblematic for the following reason: the \renyi-2 observables of interest exhibit a symmetry under $p^{(\alpha)}_{ij}\mapsto 1-p^{(\alpha)}_{ij}$, which allows us to replace any large $p^{(\alpha)}_{ij}$ with $1-p^{(\alpha)}_{ij}$ without altering the result. We will discuss this symmetry in greater detail momentarily.

The state obtained after applying the channel can be written as 
\begin{equation}
	|\rho(\alpha;g,p) \ranglee =  C_p^{(\alpha)} e^{\sum_{i<j}J_{ij}^{(\alpha)} Z_i^L Z_j^L Z_i^R Z_j^R } \ket{\psi_g}\ket{\psi_g}, 
\end{equation}
where $C_p^{(\alpha)} = \prod_{i<j} c_{ij;p}^{(\alpha)}$. Using the imaginary-time representation of $\ket{\psi_g}$ presented in \cref{eq:imag_time_symm}, we can finally write 
\begin{equation}
    |\rho(\alpha;g,p) \ranglee =  C_p^{(\alpha)} e^{-H } \ket{+}^{\otimes N}\ket{+}^{\otimes N}~,
\end{equation}
where 
\begin{multline}\label{eq:stat_mech_Ham}
    H = - \sum_{i<j} J_{ij}^{(\alpha)} Z_i^L Z_j^L Z_i^R Z_j^R \\
    - \frac{\ln g}{4} \sum_i \left( 2- Z_i^L Z_{i+1}^L - Z_i^R Z_{i+1}^R \right)~.
\end{multline}
 
The \renyi-2 correlator is given by
\begin{equation}
    C_{\text{II}, mn} = \langlee Z_m^L Z_n^L Z_m^R Z_n^R \ranglee,
\end{equation}
where the expectation value is evaluated in $| \rho(\alpha;g, p) \ranglee$.
We can further write $\langlee Z_m^L Z_n^L Z_m^R Z_n^R \ranglee$ and $\langlee Z_m^L  Z_m^R \ranglee $ as expectation values in a classical statistical-mechanical model with the partition function
\begin{equation}
\label{eq:partition-stat-mech-LR}
\begin{split}
    \mathcal{Z} = \sum_{\{s_i^L, s_i^R\}}  e^{-\beta\mathcal{H}[\mathbf{s}^L, \mathbf{s}^R]},
\end{split}
\end{equation}
where 
\begin{equation}
\label{eq:hamiltonian-stat-mech-LR}
\begin{split}
    \beta\mathcal{H}[\mathbf{s}^L, \mathbf{s}^R] = -2\sum_{i<j}&  J_{ij}^{(\alpha)} s_i^L s_i^R s_j^L s_j^R \\
    &+\frac{\ln |g|}{2} \sum_i  \left(s_i^L s_{i+1}^L + s_i^R s_{i+1}^R \right) .
\end{split}
\end{equation}
\cref{eq:hamiltonian-stat-mech-LR} describes a classical spin model consisting of a pair of spin chains, which arise from the bra and ket spaces of the quantum model, respectively. The Hamiltonian contains two types of ferromagnetic interactions: the first term encodes a long-range inter-chain four-body interaction, while the second term encodes a nearest-neighbor interaction within each chain. We note that the second term vanishes when $|g|=1$, and the Hamiltonian reduces to a long-ranged Ising model in the composite variable $s_i:=s_i^Ls_i^R$ with interaction strength $J_{ij}^{(\alpha)}$ (which decays not as a pure power law, but with the more complicated spatial profile defined via \cref{eq:beta-c-def,eq:pij-def}.) We discuss this limit more explicitly in the following section.

Within this construction, we may identify the \renyi-2 correlator with a two-point correlator in the classical model: 
\begin{equation}\label{eq:weak_corr}
    C_{\text{II}, mn} = \langle s_m^L s_m^R s_n^L s_n^R  \rangle_{\mathcal{Z}},
\end{equation}
where the expectation values are with respect to weights $e^{-\beta \mathcal{H}}$:
\begin{equation}
\langle \mathcal{O} \rangle_{\mathcal{Z}}\equiv \frac{1}{\mathcal{Z}} \sum_{\{ s_i^L, s_i^R \}} e^{-\beta \mathcal{H}} \mathcal{O}.
\end{equation}

Recall that we assumed in \cref{eq:beta-c-def} that $p_{ij}^{(\alpha)} < 1/2$  for all pairs $(i,j)$, so that $J_{ij}^{(\alpha)}$ would be real. If for any pair $(i,j)$ we have $p_{ij}^{(\alpha)} > 1/2$, we can rewrite $E_{ij;p}^{(\alpha)}$ as
\begin{equation}
    E_{ij;p}^{(\alpha)} = Z_{i}^L Z_{j}^L Z_{i}^R Z_j^R \left[ p_{ij}^{(\alpha)}\I + \left(1-p_{ij}^{(\alpha)}\right) Z_{i}^L Z_{j}^L Z_{i}^R Z_j^R \right].
\end{equation}
The operator in the square brackets can be written as an exponential with $\tanh\left[J_{ij}^{(\alpha)}\right]  = (1-p_{ij}^{(\alpha)})/p_{ij}^{(\alpha)}$, and the extra $Z_{i}^L Z_{j}^L Z_{i}^R Z_j^R$ squares to the identity when evaluating $C_{\text{II}, mn}$ or the purity.
Thus the mapping to the statistical mechanics model described by the partition function \cref{eq:partition-stat-mech-LR} again works if the definition of $J_{ij}^{(\alpha)}$ is modified to
\begin{equation}
    \tanh \left[J_{ij}^{(\alpha)} \right] = \min\left( \frac{p_{ij}^{(\alpha)}}{1-p_{ij}^{(\alpha)}}, \frac{1-p_{ij}^{(\alpha)}}{p_{ij}^{(\alpha)}}\right).
\end{equation}
In summary, we have shown that the \renyi-2 two-point correlator maps to a four-point correlator in a 1D classical model with long-range interactions. 

\subsubsection{$|g|=1$ limit: classical Ising model}

At the fixed points where $|g|=1$, our construction reduces to a particularly simple statistical mechanics model which allows us to make direct connections to prior work.
When $|g|=1$, the partition function derived in \cref{eq:partition-stat-mech-LR} reduces to 
\begin{equation}
\label{eq:partition-stat-mech-LR-g1}
    \mathcal{Z}_{|g|=1} = \sum_{\{s_i^L, s_i^R\}}  \exp\Bigg[2\sum_{i<j}  J_{ij}^{(\alpha)} s_i^L s_i^R s_j^L s_j^R  \Bigg]~.
\end{equation}
We can define a new spin variable $s_i \equiv s_i^L s_i^R$ and rewrite the partition function as that of a long-range Ising model:
\begin{equation}\label{eq:Ising_ham}
    \mathcal{Z}_{|g|=1} = \sum_{\{s_i\}} \exp \left[ 2 \sum_{i<j} J_{ij}^{(\alpha)} s_i s_j \right]~.
\end{equation}
In terms of the new Ising-spin variables, the \renyi-2 correlators are simply the two-point correlators of a long-range Ising model:
\begin{equation}\label{eq:weak-corr-ising}
    C_{\text{II}, mn} = \langle s_m s_n \rangle_{\mathcal{Z}_{|g|=1}}.
\end{equation}

The special case where $\alpha = \infty$ corresponds to the nearest-neighbor dephasing channel studied in Ref.~\cite{sala_spontaneous_2024}. In this case, we see that $p_{ij}^{(\infty)} = p$ if $|i-j| = 1$ and $0$ otherwise, so we obtain a nearest-neighbor Ising model with coupling strength $J=\mathrm{ArcTanh}\frac{p}{1-p}$. This model can only exhibit a phase transition at zero temperature, which corresponds to $p\to 1/2$. This is precisely the statement in  Ref.~\cite{sala_spontaneous_2024} regarding the absence of SW-SSB in \textit{short-ranged} dephasing channels.

We remark that the partition function presented in \cref{eq:Ising_ham} is not precisely that of a power-law Ising model, whose couplings would decay as $J_{ij}\sim|i-j|^{-\alpha}$; instead, $J_{ij}^{(\alpha)}$ depends on $|i-j|$ through the more complicated expression given in \cref{eq:beta-c-def}. However, when $|i-j|^\alpha \gg p$, we have the following expansion for $\alpha>1$:
\begin{equation}\label{eq:J-expansion}
    J_{ij}^{(\alpha)}=
        \frac{p}{|i-j|^\alpha}+\frac{p^2}{|i-j|^{2\alpha}}+O\left(\frac{1}{|i-j|^{3\alpha}}\right),
\end{equation}
so that the leading contribution is a pure power-law at large site separation. The second term falls off as $2\alpha>2$, making it effectively short-ranged~\cite{thouless1969longrange}. For $0\leq\alpha<1$, we may take $p\mapsto p/N^{1-\alpha}$ in \cref{eq:J-expansion} and observe that the subleading term is suppressed relative to the leading term by a factor of $N^{1-\alpha}$, which diverges in the thermodynamic limit. We therefore expect our model to be well-described at long distances by a power-law Ising model in either regime.

We may build our intuition by recalling results from the (extensive) literature on the classical power-law Ising model. The behavior of this model strongly depends on the power-law exponent $\alpha$. We will be interested in the critical temperature as well as the critical exponents $\beta$ and $\nu$, which are defined via the critical scaling of the correlation length $\xi$ and magnetization $m$ as~\cite{cardy1996scaling}
\begin{equation}
    \xi \sim \frac{1}{(p-p_c)^\nu},\quad m \sim {(p-p_c)^\beta}.
\end{equation}
For the results quoted below, Kac normalization was used for $\alpha<1$. The behavior of the Ising model with power-law interactions for various ranges of $\alpha$ is as follows~\cite{cardy1981onedimensional}:
\begin{itemize}
    \item $0 \leq \alpha < 1$: There is a finite-temperature phase transition from the ferromagnetic to the paramagnetic phase. The critical exponents are expected to take the Curie-Weiss ($\alpha=0$) values $\beta=\nu=\frac{1}{2}$~\cite{cannas1996long}.
    \item $ 1 < \alpha < 1.5$: There is again a finite-temperature  phase transition, with critical exponents given by mean-field theory~\cite{fisher1972critical}: $\beta=\frac{1}{2}$ and $\nu=\frac{1}{\alpha-1}$.
    \item $1.5 < \alpha < 2$: A finite-temperature phase transition exists, however the critical exponents are not mean-field~\cite{fisher1972critical}. They vary continuously with $\alpha$ and have been determined only approximately.
    \item  $\alpha > 2 $: The interactions are effectively short-ranged, and the system behaves like the nearest-neighbor one-dimensional Ising model, which has no finite-temperature phase transition~\cite{thouless1969longrange}.
\end{itemize}
This literature is supportive of the notion that there could be a finite-$p$ transition to SW-SSB in our model so long as the channel $\mathcal{E}_p^{(\alpha)}$ is sufficiently long-ranged, at least at the $g=\pm 1$ fixed points. In \cref{sec:sw-ssb_num}, we will verify this prediction, demonstrating the existence of a finite-$p$ transition which persists for all $g\neq 0$. 

Before proceeding, we discuss a limit in which the finite-$p$ SW-SSB transition can be derived exactly: the $g=1$ line in the $\alpha=0$ phase diagram. In this limit, $J_{ij}^{(\alpha=0)}=\mathrm{ArcTanh}\left[\frac{p/N}{1-p/N}\right]$ for all pairs $(i,j)$, and we are left with an all-to-all Ising model, which is also known as the Curie-Weiss model. In particular, we have a partition function of the form
\begin{equation}
    \mathcal{Z}=\sum_{\{s_i\}}\mathrm{exp}\left[\beta\frac{J_0}{N} \sum_{i<j}s_is_j\right],
\end{equation}
where $J_0=2J_{ij}^{(\alpha=0)}TN$. This model undergoes a magnetic ordering transition when $J_0=T$ (see e.g. Ref.~\cite{kochmanski2013curie}), which amounts to the following condition on $p_c$:
\begin{equation}
    2 \mathrm{ArcTanh}\left[\frac{p/N}{1-p/N}\right]\bigg|_{p=p_c}TN=T.
\end{equation}
This is solved by
\begin{equation}
    p_c = N-\frac{N}{\tanh \left(\frac{1}{2 N}\right)+1}
    = \frac{N}{2} \left (1-e^{-\frac{1}{N}}\right),
\end{equation}
which approaches $\frac{1}{2}$ as $N\to \infty$. We therefore find that there is a transition to SW-SSB at $p_c = 1/2$ along the $g=1$ line in the $\alpha=0$ phase diagram.

\subsection{Numerical results}\label{sec:sw-ssb_num}

The statistical mechanics model introduced in the previous section enables us to efficiently compute the two-point correlator over a large range of finite system sizes. These computations are performed via Markov Chain Monte Carlo (MCMC) simulations utilizing a modified version of the Wolff cluster algorithm~\cite{wolff1989collective}, the details of which are presented in \cref{app:MCMC}. With the method of finite-size scaling, we can extract the critical value $p=p_c(g)$ at which $\rho(\alpha;g,p)$ experiences the onset of SW-SSB and the universal scaling behavior at the transition. For fixed $\alpha$ and $g$, we expect the measured value of the \renyi-2 correlator in an $N$-site chain, denoted $C_\mathrm{II}^{(N)}$, to be related near criticality to a universal scaling function $\tilde{C}_\mathrm{II}$ via~\cite{cardy1996scaling}
\begin{equation}\label{eq:scaling_ansatz}
    C_\mathrm{II}^{(N)}(p)=N^{-2X_\phi} \tilde{C}_\mathrm{II}\left[(p-p_c)N^{y_p}\right],
\end{equation}
where $X_\phi$ and $y_p$ are parameters that capture the critical scaling.

Notice that at criticality, i.e. $p=p_c$, the universal scaling function evaluates to the constant $\tilde C_\mathrm{II}(0)$ and $C_\mathrm{II}^{(N)}(p)$ scales as a pure power-law in system size $N$. We may therefore extract $p_c$ by plotting $C_\mathrm{II}^{(N)}(p)$ vs. $N$ for various $p$ on a log-log scale and numerically identifying the $p\approx p_c$ where the curvature undergoes a sign change. In turn, the slope of this curve estimates $-2X_\phi$. Similarly, we may consider
\begin{equation}
    \partial_p \log C_\mathrm{II}^{(N)}(p)=N^{y_p}\frac{\tilde {C}_\mathrm{II}'[(p-p_c)N^{y_p}]}{\tilde {C}_\mathrm{II}[(p-p_c)N^{y_p}]},
\end{equation}
which also scales as a pure power-law at criticality. From $\partial_p \log C_\mathrm{II}^{(N)}(p\approx p_c)$, we may extract an estimate of $y_p$.

\begin{figure*}
\captionsetup[subfigure]{labelformat=empty,captionskip=-20pt}
 \centering
 \subfloat[\label{fig:phase_1}]{
\includegraphics[width=0.485\textwidth]{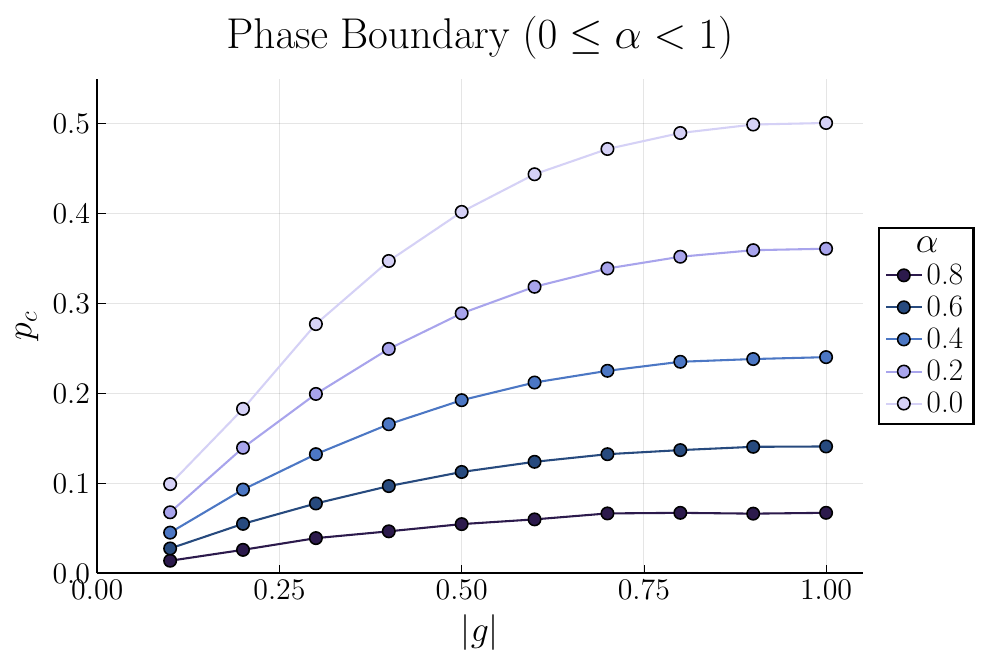}%
\begin{picture}(0,0)
    \put(-247,152){\large (a)}
    \end{picture}
}
     \hfill
     \subfloat[\label{fig:phase_2}]{
     \includegraphics[width=0.485\textwidth]{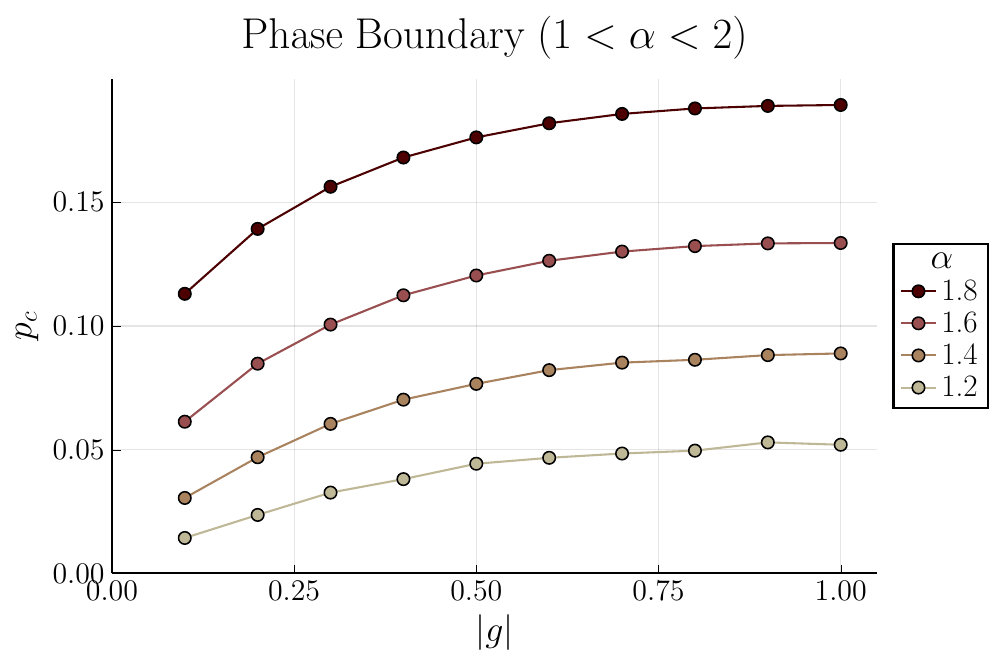}%
     \begin{picture}(0,0)
    \put(-247,152){\large (b)}
    \end{picture}
     }
\caption{The phase boundaries separating the symmetric phase ($p<p_c$) from the SW-SSB phase ($p>p_c$) for various $\alpha$.  (a)  We plot $p_c(g)$ for $\alpha<1$ where $p$ has been Kac normalized. (b) We plot $p_c(g)$ for $\alpha>1$ where $p$ has not been Kac normalized. Notice in particular that $p_c(g)$ decreases with increasing $\alpha$ below $\alpha=1$, while it increases with increasing $\alpha$ above $\alpha=1$. This behavior can be understood from the fact that $\lim_{\alpha\to 1} \frac{p/N^{1-\alpha}}{|i-j|^\alpha}=\lim_{\alpha\to 1} \frac{p}{|i-j|^\alpha}$.}
    \label{fig:phase_diag}
\end{figure*}

We present our numerical results for the phase boundary in \cref{fig:phase_diag}. First, notice that $p_c(g)$ for each $\alpha$ monotonically decreases with decreasing $|g|$. This can be understood from the perspective of the statistical mechanics model \cref{eq:hamiltonian-stat-mech-LR} as follows: the SW-SSB transition corresponds to a finite-temperature spin ordering transition in the composite variable $s^L_is^R_i$. Both terms in \cref{eq:hamiltonian-stat-mech-LR} encourage this spin ordering. Therefore, as $|g|$ decreases from $1$ to $0$, the $\ln|g|$ coefficient of the second term is negative and grows in magnitude. This leads to an earlier onset of spin ordering, and therefore an earlier onset of SW-SSB from the channel perspective. By a similar argument, we also expect $p_c(g)$ to approach zero as $|g|$ approaches zero. This is consistent with the fact that there is no finite-$p$ transition at $g=0$, as the initial pure state $\ket{\psi_{g=0}}$ is the GHZ state, which already spontaneously breaks the strong $\mathbb{Z}_2$ symmetry.

We have also observed numerically (see \cref{app:num_results}) that there is no finite SW-SSB phase for $\alpha>2$. Instead, we find that there is a single point, namely $p=0.5$, at which SW-SSB is observed. This is consistent with the result of Ref.~\cite{sala_spontaneous_2024}, which discussed the case where $\alpha=\infty$ and $g=1$. This is also consistent with the well-known result that there is no phase transition in any short-ranged classical $1D$ Ising model~\cite{thouless1969longrange}; indeed, short-ranged in this case refers to any $\alpha>2$.

We present numerical results for the exponents $X_\phi$ and $y_p$ in \cref{app:num_results}. These exponents are closely related to the usual critical exponents, but the relationship depends sensitively on the choice of $\alpha$. Above $\alpha=1$, the relationship is given by~\cite{flores2015finite}, 
\begin{equation}\label{eq:Xy_coppa}
    X_\phi=\frac{\qq\beta}{\nu},\qquad y_p=\frac{\qq}{\nu},
\end{equation}
where $\qq$ (``koppa'') is an additional critical exponent which captures corrections to finite-size scaling above the upper critical dimension $d_c$~\cite{berche2012hyperscaling,kenna2014new}. In the power-law interacting Ising model, $d_c=2(\alpha-1)$ and $\qq$ takes the values~\cite{flores2015finite}
\begin{equation}
    \qq=\begin{cases}
        \frac{1}{2(\alpha-1)},&1<\alpha<1.5\\
        1,&1.5<\alpha<2\\
    \end{cases},
\end{equation}
so the ratio $\qq/\nu$ takes the value $1/2$ below $\alpha=1.5$. The crossover occurs at $\alpha=1.5$ because $d_c$ is smaller than the spatial dimension when $\alpha<1.5$. 

Below $\alpha=1$, the system is expected to behave like the Curie-Weiss ($\alpha=0$) limit, so that
\begin{equation}\label{eq:Xy_CW}
    X_\phi=\frac{\beta}{\nu_\text{CW}},\qquad y_p=\frac{1}{\nu_\text{CW}},
\end{equation} 
where $\nu_\text{CW}=2$ is the Curie-Weiss finite-size scaling exponent~\cite{botet1982size}, which differs from the bulk correlation-length exponent $\nu=1/2$. Substituting the known values of $\beta$ and $\nu$ for $0\leq \alpha < 1.5$ into \cref{eq:Xy_coppa,eq:Xy_CW}, we see that the power-law Ising model predicts $X_\phi=1/4$ and $y_p=1/2$ throughout this regime. The measured numerical values of $X_\phi$ and $y_p$ are presented in \cref{app:num_results}. Although sampling noise limits our precision, we believe that these results are largely consistent with the theoretical predictions based on the power-law Ising model.

\begin{figure*}
    \centering\includegraphics[width=0.98\textwidth]{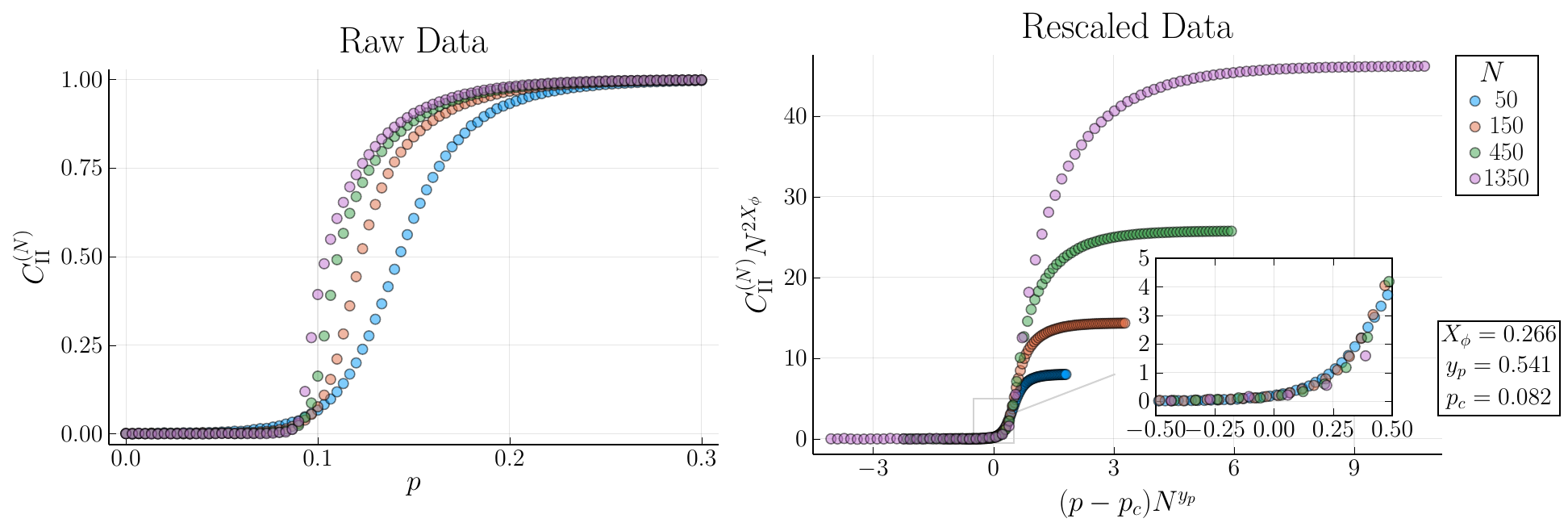}
    \caption{(a) The \renyi-2 correlator $C_\mathrm{II}^{(N)}$ as a function of $p$ at fixed $\alpha=1.4$ and $g=0.6$ before rescaling. (b) The \renyi-2 correlator after rescaling using the values of $p_c$, $X_\phi$, and $y_p$ presented in the table to the lower right. In the inset, we present a magnified view of the same data around $(p-p_c)=0$ to emphasize the scaling collapse,  wherein the finite-size curves $C_\mathrm{II}^{(N)}$ collapse to the universal curve $\tilde{C}_\mathrm{II}$ upon rescaling, as described in the text.}
    \begin{picture}(0,0)
    \put(-247,220){\large (a)}
    \put(-13,220){\large (b)}
    \end{picture}
    \label{fig:scaling_collapse}
\end{figure*}

Finally, we remark that the scaling hypothesis presented in \cref{eq:scaling_ansatz} implies that the finite-size curves $C_\mathrm{II}^{(N)}$ should collapse onto the single universal curve $\tilde{C}_\mathrm{II}$ upon appropriate rescaling, in a phenomenon known as scaling collapse~\cite{cardy1996scaling}. In particular, the scaling hypothesis predicts that there exists a universal scaling function $\tilde C_\mathrm{II}$, which is independent of $N$, that evaluates to $N^{2 X_\phi}C_\mathrm{II}^{(N)}(p)$ at $(p-p_c)N^{y_p}$ in the vicinity of $p=p_c$. We demonstrate scaling collapse for representative data in \cref{fig:scaling_collapse} by plotting $N^{2 X_\phi}C_\mathrm{II}^{(N)}(p)$ vs. $(p-p_c)N^{y_p}$ for various $N$ using the values of $p_c$, $X_\phi$, and $y_p$ extracted via the methods described above. Results for other values of $g$ and $\alpha$ are presented in \cref{app:num_results}. Finally, we note that because we extracted $p_c$, $X_\phi$, and $y_p$ as described above, we use scaling collapse as a consistency check on the measured values rather than a means of directly extracting them.

\section{Mixed-state SPT phase}\label{sec:mixed_SPT}
Having identified a finite-$p$ transition to SW-SSB, we are left with a second puzzle: below $p_c$, where the \renyi-2 correlator detects no symmetry breaking, what is the nature of the phase? 
This question is uninteresting for a trivial initial state $(g>0)$, but becomes quite subtle for a nontrivial initial state $(g<0)$, as we will discuss in this section. 

\subsection{String order parameters}\label{sec:string_order_params}

Though local observables cannot distinguish nontrivial SPT phases from the trivial phase, there often exist non-local string order parameters which can differentiate them~\cite{nijs1989preroughening,kennedy1992hidden,Perez-Garcia2008,pollmann2012detection,duivenvoorden2013from,else2013hidden,bahri2014detecting}. The pure $\mathbb{Z}_2^T$-SPT phase to which $\ket{\psi_g}$ belongs is known to be detectable by the string operator \cite{Perez-Garcia2008,smith2022crossing}
\begin{equation}\label{eq:string_order}
    \mathcal{S}_{mn}=Z_{m-1} Y_m \left( \prod_{i=m+1}^{n-1} X_i \right) Y_n Z_{n+1} .
\end{equation}
Prior work \cite{degroot2022symmetry} has shown that short-ranged finite-depth symmetric channels, applied to pure SPT states, preserve the linear string order parameter $\Tr[\mathcal{S}\rho]$. Although this result was derived for SPTs protected by unitary symmetry, and we are here considering an antiunitary symmetry, it is still a natural starting point from which to begin our characterization of the phase.

Measuring the string order parameter given in \cref{eq:string_order} in $\rho(\alpha;g,p)$ for $p>0$ and $g<0$, we find that it vanishes for any finite $p$. This can be shown analytically, as we do in \cref{ap:string}. The result is
\begin{equation}\label{eq:string_order_result}
\frac{\Tr[\mathcal{S}_{mn}\rho]}{\Tr[\rho]}
=\left(\prod_{i\not\in[m,n]}\prod_{j\in[m,n]}(1-2p_{ij}^{(\alpha)})\right)\frac{\bra{\psi_g}\mathcal{S}\ket{\psi_g}}{\braket{\psi_g|\psi_g}}.
\end{equation}
While the second factor is simply the pure-state value of the string order parameter (which is nonvanishing in the SPT phase), the first factor is exponentially small in the system size, which causes the string order parameter to vanish in the thermodynamic limit. We emphasize that the first factor in \cref{eq:string_order_result} is exponentially small precisely because the channel is power-law correlated; in a model with finite-ranged interactions, this product would contain a finite number of factors as the system size grows and become nonvanishing in the thermodynamic limit. 

In the study of mixed-state SPTs, the \renyi-2 string order parameter has emerged as a useful diagnostic~\cite{xue_tensor_2024,Shah2025instabilityofsteady,kuno2026hierarchymixedsymmetryprotected,shen2026realizationdecoherenceinducedaveragedsymmetryprotected}. In this case, the \renyi-2 string order parameter is given by
\begin{equation}\label{eq:renyi_2_string_order}
    \mathcal{S}_\mathrm{II}:=\frac{\Tr[\mathcal{S}_{mn}\rho^2]}{\Tr[\rho^2]}.
\end{equation}
In the doubled-space picture, we may equivalently write $\mathcal{S}_\mathrm{II}=\langlee \rho|\mathcal{S}_{mn}\otimes \mathbbm{1}|\rho\ranglee/\langlee\rho|\rho\ranglee$. From this perspective, we see that $\mathcal{S}_{\mathrm{II}}$ evaluates the expectation value of $\mathcal{S}_{mn}\otimes \mathbbm{1}$ in $|\rho\ranglee$. Notice that $\mathcal{S}_{mn}\otimes \mathbbm{1}$ consists of a truncated strong symmetry decorated by endpoint operators $Z_{m-1} Y_m\otimes\mathbbm{1}$ and $Y_nZ_{n+1}\otimes\mathbbm{1}$, each of which is neutral under the strong symmetry but charged under complex conjugation. We may therefore interpret this string order parameter as probing non-triviality under a symmetry which acts by $\prod_i X_i$ from one side along with complex conjugation, to be discussed further in \cref{sec:SPT_phase_char}.

In order to compute $\mathcal{S}_\mathrm{II}$, we can once again make use of a mapping to a statistical mechanics model. First, we observe that the string operator $\mathcal{S}_{mn}$ is related to the disorder operator 
\begin{equation}
    \tilde{\mathcal{S}}_{mn}=\prod_{i=m}^{n}X_i
\end{equation}
by $T \mathcal{S}_{mn} T^\dagger = \tilde{\mathcal{S}}_{mn}$, where $T$ is the $CZ$ circuit defined in \cref{eq:T_def}. Because $T\rho(\alpha;g,p)T^\dagger=\rho(\alpha;-g,p)$, we have
\begin{equation}\label{eq:flip_g_S}
    \frac{\Tr[\mathcal{S}_{mn}\rho(\alpha;g,p)^2]}{\Tr[\rho(\alpha;g,p)^2]}=\frac{\Tr[\smash{\tilde{\mathcal{S}}_{mn}}\rho(\alpha;-g,p)^2]}{\Tr[\rho(\alpha;-g,p)^2]}.
\end{equation}
This relation tells us that instead of measuring the \renyi-2 expectation value of $\mathcal{S}_{mn}$ for $g<0$, we can equivalently measure the \renyi-2 expectation value of $\tilde{\mathcal{S}}_{mn}$ for $g>0$.

The latter expectation value maps nicely to an expectation value in the statistical mechanics model of \cref{eq:partition-stat-mech-LR}. In particular, we find
\begin{equation}
\begin{split}
    \mathcal{S}_\mathrm{II}(g)=\bigg\langle \mathrm{exp}\bigg[&\frac{\ln |g|}{2}(s_{m-1}^Ls_{m}^L+s_{n}^Ls_{n+1}^L)\\&-2\smash[b]{\sum_{\substack{i\in[m,n]\\j\notin[m,n]}}}J_{ij}^{(\alpha)} s_i^L s_i^R s_j^L s_j^R\bigg]\bigg\rangle_{\mathcal{Z}},
\end{split}
\end{equation}
\vspace{1em} 

\noindent where we assume $g<0$. (When $g>0$, $\mathcal{S}_\mathrm{II}$ vanishes asymptotically below $p=p_c$.) This quantity can be directly computed in Monte Carlo numerics, and we present representative data in \cref{app:num_results}. We also present an analytical evaluation of the string order parameter at the exactly-solvable point $\alpha=0$, $g=-1$ in \cref{ap:exact_renyi-2}.

\begin{figure}
    \centering
    \includegraphics[width=\linewidth]{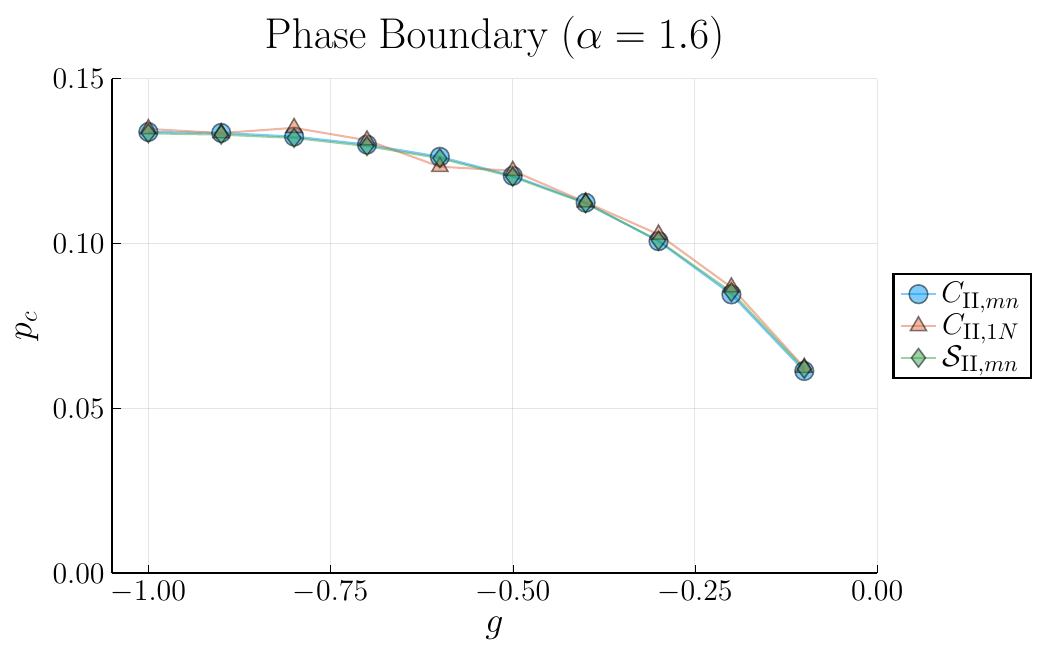}%
    \caption{The critical $p_c$ extracted from finite-size scaling on the observables associated with SW-SSB ($C_{\mathrm{II},mn}$), \renyi-2 coherent information ($C_{\mathrm{II},1N}$), and string order ($\mathcal{S}_{\mathrm{II},mn}$). This figure supports the claim that all three observables probe the same phase transition. We note that the noise in $C_{\mathrm{II},1N}$ is due to the slower convergence of MCMC for this observable. Physically, this effect arises because the edge sites have fewer neighbors than the bulk sites and hence participate in fewer Monte Carlo updates.}
    \label{fig:p_c_comp}
\end{figure}

In both our numerics and the exact solution at the solvable point, we observe a 2nd order phase transition after which the string order parameter becomes vanishingly small in the system size. We expect that this transition coincides with onset of SW-SSB, and we have plotted in \cref{fig:p_c_comp} the critical $p$ with respect to this diagnostic along with that obtained from SW-SSB and the coherent information transition to be discussed in the next section.

\subsection{Coherent information}\label{sec:coh_info}
Since the capability of encoding quantum information at the edges is a salient feature of the pure SPT phase, the \textit{coherent information}~\cite{schumacher1996quantum,lloyd1997capacity} is a valuable probe of the phase continuously connected to the pure state SPT line at $p=0$, $-1\leq g<0$. In this context, the coherent information $I_c$ quantifies the amount of quantum information which can be encoded in the symmetry-protected subspace of the pure-state SPT and recovered after application of the channel $\mathcal{E}_p^{(\alpha)}$. 

Here we briefly recall the definition of coherent information. Coherent information is a function of a quantum channel $\mathcal{E}$ and a state $\rho$ on a Hilbert space, which we call $\mathcal{H}_Q$. Additionally, we introduce a reference Hilbert space $\mathcal{H}_R$ and a state $\ket{\psi}_{QR}\in \mathcal{H}_Q\otimes\mathcal{H}_R$ which purifies $\rho$: $\Tr_R[\ketbra{\psi}{\psi}_{QR}]=\rho$. (We note that the result is independent of the choice of purification.) The coherent information is then defined as the difference of Von Neumann entropies~\cite{schumacher1996quantum,lloyd1997capacity}
\begin{equation}\label{eq:coh_info}
    I_c(\rho,\mathcal{E})=S(\rho_Q)-S(\rho_{QR}),
\end{equation}
where
\begin{equation}
    \rho_{QR}=\left(\mathcal{E}_Q\otimes\mathrm{id}_R\right)(\ketbra{\psi}{\psi}_{QR}), \quad \rho_Q=\Tr_R[\rho_{QR}].
\end{equation}
Intuitively, the first term in \cref{eq:coh_info} captures the quantum correlations preserved by the channel, while the second term captures the inaccessible quantum correlations lost to the environment.

The coherent information provides a natural language for discussing quantum error correction. A noise channel $\mathcal{E}$ is exactly correctable in a given quantum error-correcting code if and only if $I_c(\rho,\mathcal{E})=I_c(\rho,\mathrm{id})$ for every pure $\rho$ in the logical space~\cite{schumacher1996quantum}. Furthermore, it is sufficient to check the case where $\rho$ is the maximally mixed state on the logical space~\cite{knill1997theory}, which can be achieved by taking $\ket{\Psi}_{QR}$ to be a tensor product of logical Bell pairs 
\begin{equation}\label{eq:logical-bell-pairs}
    \ket{\Psi}_{QR}=\bigotimes_{i=1}^k\frac{1}{\sqrt{2}}\left(\ket{\overline{0}_i}\ket{0}_{R_i}+\ket{\overline{1}_i}\ket{1}_{R_i}\right).
\end{equation}
It is instructive to consider the behavior of $I_c$ for $k$ logical qubits in three special cases:
\begin{enumerate}
    \item If we apply the identity channel, then $\rho_{QR}$ is pure and $\rho_Q=\frac{1}{2^k}\mathbbm{1}_{2^k}$, so $I_c=k\log 2$. This is the maximum possible value for $I_c$.
    \item If we apply the totally dephasing channel to one logical qubit, the state on that logical-reference subsystem is given by $\frac{1}{4}(\mathbbm{1}\otimes\mathbbm{1}+Z\otimes Z)$. This subsystem contributes $\log2$ to $S(\rho_{QR})$, which decreases $I_c$ by the same amount.
    \item If we apply the totally depolarizing channel to one logical qubit, the state on that logical-reference subsystem is given by $\frac{1}{4}\mathbbm{1}_4$. This subsystem contributes $2\log2$ to $S(\rho_{QR})$, which decreases $I_c$ by the same amount. Notice that if we were to depolarize every logical qubit, $I_c$ would drop to $-k\log 2$, which is the minimum possible value of $I_c$.
\end{enumerate} 

From the perspective of quantum error correction, an SPT phase is characterized by the presence of a logical subspace robust to local errors which respect the symmetry. If we take $\rho$ to be the maximally mixed state in this logical subspace and $\mathcal{E}$ to be a decoherence channel, the coherent information $I_c(\rho,\mathcal{E})$ sheds light on the fate of the SPT order under decoherence. In particular, constant $I_c$ signals that the SPT order is stable to the noise, as any encoded logical information is fully recoverable. We now apply this principle to characterize the mixed-state phase we obtain for $0<p<p_c(g)$.

To evaluate $I_c$ in our model, we will take $\mathcal{E}$ to be the power law-correlated channel $\mathcal{E}_p^{(\alpha)}$ and $\rho$ to be the maximally mixed state in the codespace spanned by $\{\ket{\psi_g^{\mu\nu}}\mid\mu,\nu\in\{0,1\}\}$, which arises due to the SPT edge modes. In particular, we define
\begin{equation}\label{eq:psi_g_MPS_munu}
	\ket{\psi_g^{\mu\nu}}=\sum_{\{i_j\}=0,1}\bra{\mu_L}M_g^{[i_1]}\cdots M_g^{[i_N]}\ket{\nu_R}\ket{i_1,\ldots,i_N},
\end{equation}
where we have introduced the boundary conditions
\begin{equation}\label{eq:SPT_BCs}
    \begin{split}
        \ket{0_L} &= \begin{pmatrix}
            0\\1
        \end{pmatrix},\qquad \ket{0_R} = \begin{pmatrix}
            g\\-1
        \end{pmatrix},\\
        \ket{1_L} &= \begin{pmatrix}
            -1\\0
        \end{pmatrix},\qquad \ket{1_R} = \begin{pmatrix}
            g\\-g
        \end{pmatrix}.\\
    \end{split}
\end{equation}
The choice of boundary conditions determines the form of the logical operators, and the choice made in \cref{eq:SPT_BCs} brings the logical operators into the following simple form at $g=-1$:
\begin{equation}\label{SPT_logicals}
\def\arraystretch{1.5}
\begin{array}{lclcl}
    \overline{Z}_L:=Z_1, && \overline{Z}_R:=Z_N,\\
     \overline{X}_L:=X_1Z_2, &&  \overline{X}_R:=Z_{N-1}X_N,
\end{array}
\end{equation}
as we prove in \cref{ap:coh_info}. We remark that the symmetric boundary conditions $\ket{L}$ and $\ket{R}$ introduced in \cref{eq:triv_BCs} can be obtained as linear combinations of the vectors in \cref{eq:SPT_BCs}, so that these choices are related by a change of basis in the logical space. Furthermore, neat analogues of \cref{eq:mps-coeff-g-domain,eq:imag_time_symm} hold for $\ket{\psi_g^{\mu\nu}}$. Namely, we find that
\begin{equation}
	\label{eq:mps-coeff-g-domain_munu}
	\bra{\mu_L}M_g^{[i_1]}M_g^{[i_2]}\cdots M_g^{[i_N]}\ket{\nu_R} \propto g^{\text{ DW(}\boldsymbol{i})/2}\delta_{\mu,i_1}\delta_{\nu,i_N},
\end{equation}
as proven in \cref{app:DW_perspective}. In turn, \cref{eq:mps-coeff-g-domain_munu} enables us to define $\ket{\psi^{\mu\nu}_g}$ via imaginary-time evolution
\begin{equation}\label{eq:munu_imag_time}
	\ket{\psi_g^{\mu\nu}}\propto e^{-\frac{1}{4}\ln |g|\sum_{i=1}^{N-1}Z_i Z_{i+1}}\times \begin{cases}
		\ket{\psi_{g=1}^{\mu\nu}},&g>0\\
		\ket{\psi_{g=-1}^{\mu\nu}},&g<0
	\end{cases}.
\end{equation}

\subsubsection{Identification with statistical mechanics}
As we demonstrate in this section, $I_c$ can also be related to an observable in a classical statistical mechanics model. Remarkably, the statistical mechanics model is exactly that which appears in \cref{sec:sw-ssb-stat-mech}, and the observable is nearly identical to the one considered there. This correspondence highlights the close relationship between the onset of SW-SSB and the destruction of SPT order and its associated quantum information in this model.

The observable which maps to our statistical mechanics model is the \renyi-2 coherent information, where the Von Neumann entropies appearing in the definition of $I_c$ are replaced with \renyi-2 entropies $S^{(2)}(\rho)=-\log \Tr[\rho^2]$. For this entropy, the coherent information becomes
\begin{equation}
	I_c^{(2)}=-\log\left(\frac{\Tr[\rho_Q^2]}{\Tr[\rho_{QR}^2]}\right).
\end{equation}
Although this is only a proxy for the von Neumann coherent information, it is widely used (see e.g.~\cite{lyons2024understandingstabilizercodeslocal,hauser2024informationdynamicsdecoheredquantum,weinstein2026intrinsicerrorthresholdsnearly}) because it is computationally tractable and very often captures the essential physics as well as the von Neumann coherent information. 

As we show in \cref{ap:coh_info}, $\Tr[\rho_{QR}^2]$ is precisely equal to the partition function which we wrote in \cref{eq:partition-stat-mech-LR}, and 
\begin{equation}
    \Tr[\rho^2_Q]=\sum_{\{s_i^L,s_i^R\}}e^{-\beta \mathcal{H}}\left(\frac{1- s_1^Ls_1^R }{2} \right)\left(\frac{1- s_N^Ls_N^R }{2}\right),
\end{equation}
with $\beta \mathcal{H}$ defined as in the previous section. Consequently, the \renyi-2 coherent information can be expressed as
\begin{equation}\label{eq:coh_info_result}
    I_c^{(2)} = 2\log 2-\log\left(1+\braket{s_1^Ls_1^Rs_N^Ls_N^R}_{\mathcal{Z}}\right).
\end{equation}
The statistical mechanical observable $\braket{s_1^Ls_1^Rs_N^Ls_N^R}$ which appears in \cref{eq:coh_info_result} is precisely the \renyi-2 correlator appearing in \cref{eq:weak_corr}, with the endpoints configured to be $m=1$ and $n=N$. 

\subsubsection{Implications for SPT order}

We now interpret the result obtained in \cref{eq:coh_info_result} to better understand the mixed SPT phase. From \cref{eq:coh_info_result} and the understanding we have developed for the statistical mechanics model given in \cref{eq:partition-stat-mech-LR}, we identify the following regimes when $1<\alpha<2$:
\begin{equation}\label{eq:coh_info_result_repeated}
    I_c^{(2)} = \begin{cases}
        2\log 2,&p<p_c\\
        2\log 2-\log\left(1+C_{\mathrm{II},1N}\right),&p_c<p<\frac{1}{2}\\
        \log 2,& p=\frac{1}{2}
    \end{cases},
\end{equation}
where $C_{\mathrm{II},1N}=\braket{s_1^Ls_1^Rs_N^Ls_N^R}_{\mathcal{Z}}$ increases continuously from 0 to 1 on the interval $[p_c,\frac{1}{2}]$. 

We posit the following physical mechanisms for the derived values of $I_c$: for $p<p_c$, the logical qubits of the pure-state SPT persist due to symmetry protection. When SW-SSB sets in at $p=p_c$, the symmetry protection is lost (as the protecting symmetry is spontaneously broken), and the coherent information begins to degrade. We substantiate this claim in \cref{sec:SPT_phase_char} by presenting a motivated candidate for this protecting symmetry and demonstrating that it is spontaneously broken at $p=p_c$. The coherent information decreases continuously in the SW-SSB phase until it reaches $\log2$ at $p=1/2$.

The $p=1/2$ point is distinguished because the nearest-neighbor channels $\mathcal{E}^{(\alpha)}_{i,i+1;p}$ contained in $\mathcal{E}_p^{(\alpha)}$ reduce to projectors at this point. Each $\mathcal{E}^{(\alpha)}_{i,i+1;p}$ projects onto the subspace symmetric with respect to $Z_iZ_{i+1}$. It may seem surprising that a qubit of coherent information survives at $p=1/2$, as $\mathcal{E}_{p=1/2}^{(\alpha)}$ implements total dephasing in the original logical subspace
\begin{equation}
    \mathcal{E}_{p=1/2}^{(\alpha)}(\overline{Z}_{L/R})=\overline{Z}_{L/R},\quad \mathcal{E}_{p=1/2}^{(\alpha)}(\overline{X}_{L/R})=0.
\end{equation}
However, there exists a single off-diagonal stabilizer of $\ket{\psi_{g=-1}}$ which is decoherence-free: the product of all bulk stabilizers along with the logical $\overline{X}$ at each end. This operator is given by $Y_1X_2\cdots X_{N-1}Y_N$, which clearly belongs to the symmetric subspace. This delocalized operator, along with either of the logical $\overline{Z}_{L/R}$ operators, generates a single copy of the Pauli algebra. The result is a preserved logical qubit which produces a coherent information of $\log 2$ rather than 0. As one moves away from the $g=-1$ fixed point, the stabilizers become dressed but their global product remains decoherence-free, guaranteeing a preserved logical qubit for all $g<0$.

For $0\leq \alpha<1$, the point $p=1/2$ is not distinguished; instead, we observe that $I_c^{(2)}\to \log 2$ as $p\to\infty$. (Recall that the large-$p$ regime remains physical due to Kac normalization, which divides $p$ by a divergent quantity to obtain the physical probability.) The state retains a coherent information of $\log 2$ due to the same mechanism as in the $\alpha>1$ regime, as the channel always preserves the product of bulk stabilizers. To understand the fact that the coherent information saturates to this quantity only as $p$ grows large, we may once again consider $\mathcal{E}^{(\alpha)}_p\left(\overline {X}_{L/R}\right)$. Because $\overline {X}_{L}$ anti-commutes with $Z_1$, it acquires a factor of
\begin{equation}\label{eq:p-decay-exp}
\begin{split}
        \prod_{\Delta=1}^{N-1}\left(1-\frac{2p/N^{1-\alpha}}{\Delta^\alpha}\right)\overset{N\to\infty}{\longrightarrow}\mathrm{exp}\left[\frac{-2p}{1-\alpha}\right]
\end{split}
\end{equation}
upon application of $\mathcal{E}^{(\alpha)}_p$, as derived in \cref{ap:p-decay-exp} (and similarly for $\overline{X}_R$). We see that this factor is exponentially small in $p$, so that the original logical $\overline{X}$ operators are only asymptotically destroyed by the channel.

We note that the values of $p_c$ presented in \cref{fig:phase_diag} were obtained from measurements of $C_{\mathrm{II},mn}$, with $m$ and $n$ chosen to be well-separated sites in the bulk. In contrast, \cref{eq:coh_info_result_repeated} is defined with respect to the correlator measured at the endpoints of the chain, $C_{\mathrm{II},1N}$. We expect that the $p_c$ obtained through either observable should agree in the thermodynamic limit, and we have verified numerically that this is the case. We demonstrate this agreement for a representative phase boundary in \cref{fig:p_c_comp}.

\subsection{Characterization of the phase}\label{sec:SPT_phase_char}

The results of \cref{sec:string_order_params,sec:coh_info} suggest a novel form of mixed-state SPT order. We find the density matrices $\rho(\alpha;g,p)$ for $g\!<\!0$ and $p\!<\!p_c$ exhibit vanishing string order parameters that are linear in the density matrix, unlike typical decohered SPTs~\cite{degroot2022symmetry}. Nevertheless, from the perspective of coherent information, they retain the same amount of logical quantum information as the initial pure states. Furthermore, the nontriviality of the \renyi-2 string order parameter $\mathcal{S}_\mathrm{II}$ suggests that these density matrices enjoy some form of symmetry protection.

We propose that the mixed-state SPT phase is protected by this candidate symmetry generated by the following transformations:
\begin{equation}\label{eq:T_tilde}
    \tilde{\mathcal{T}}:\rho\mapsto \left(\prod_i X_i \right) \rho^*~.
\end{equation}
The transformation $\tilde{\mathcal{T}}$ is a natural extension of the antiunitary pure-state symmetry $\mathcal{T}:=\prod_i X_i\mathcal{K}$ to density matrices. Notice that $\tilde{\mathcal{T}}$ resembles a strong symmetry insofar as $\prod_i X_i$ acts on the density matrix from a single side. However, it also acts by complex conjugation, which cannot be implemented as a strong symmetry. The transformation $\tilde{\mathcal{T}}$ therefore does not fit neatly into the conventional paradigm of strong and weak symmetries. Note that, since we are considering Hermitian density matrices, symmetry under $\tilde{\mathcal{T}}$ implies symmetry under $\tilde{\mathcal{T}}^\prime:\rho\mapsto \rho^*\left(\prod_i X_i\right)$. We now present evidence which supports the claim that the symmetry $\tilde{\mathcal{T}}$ alone (or, equivalently, $\tilde{\mathcal{T}}^\prime$ alone) is responsible for the novel mixed-state SPT physics we have discussed.

First, we note that $\tilde{\mathcal{T}}$ acts with $\prod_i X_i$ from one side, so it is no longer a symmetry after the onset of SW-SSB, which breaks the strong symmetry $\prod_i X_i$ down to its diagonal subgroup. This observation is consistent with the claim that $\tilde{\mathcal{T}}$ is protecting encoded logical information, as the logical information begins to deteriorate precisely at the onset of SW-SSB. Second, as $\tilde{\mathcal{T}}$ is the natural mixed-state analogue of $\mathcal{T}$, it is also consistent with the string order parameter $\mathcal{S}_{mn}$ given in \cref{eq:string_order}. In particular, the \renyi-2 string order parameter $\mathcal{S}_\mathrm{II}$ is simply the vectorized expectation value of the string order parameter, i.e., $\mathcal{S}_\mathrm{II}=\langle\!\langle \rho|\mathcal{S}\otimes\mathbbm{1}|\rho\rangle\!\rangle$, where the tensor product is between the ket and bra spaces. The nontrivial measured value of $\mathcal{S}_\mathrm{II}$ is therefore also consistent with mixed-state SPT order protected by $\tilde{\mathcal{T}}$ (or $\tilde{\mathcal{T}}'$) alone.

Furthermore, the symmetry fractionalizes to the edges of the system in the mixed SPT phase. To see this, we consider how $\rho(\alpha;g,p)$ transforms under $\tilde{\mathcal{T}}$. We write $\rho(\alpha;g,p)$ diagrammatically as
\begin{equation}
\rho(\alpha;g,p)=
\begin{tikzpicture}
        % ket layer
        \draw (0,0)--(5.2,0);
        \node[gedge] at (0,0) {};
        \node[gedge] at (5.2,0) {};
        \foreach \x in {0.5,1.2,1.9,3.3,4.0,4.7}{
        \draw (\x,0)--(\x,0.8);
        \node[blankg] at (\x,0) {};
        }
        \node[fill=white] at (2.6,0) {$\dots$};
        % bra layer
        \draw (0,-0.5)--(5.2,-0.5);
        \node[gedge] at (0,-0.5) {};
        \node[gedge] at (5.2,-0.5) {};
        \foreach \x in {0.5,1.2,1.9,3.3,4.0,4.7}{
        \draw (\x,-0.5)--(\x,-1.3);
        \node[blankg] at (\x,-0.5) {};
        }
        \node[fill=white] at (2.6,-0.5) {$\dots$};
        % gate
        \draw[BrickRed] (0.5,0.5) -- (-0.3,0.5) -- (-0.3,-1.0) -- (0.5,-1.0);
        \node [style={fill=BrickRed, rounded rectangle, minimum width=50mm, minimum height=1.8mm}] at (2.6,-1.0) {};
        \node [style={fill=BrickRed, rounded rectangle, minimum width=50mm, minimum height=1.8mm}] at (2.6,0.5) {};
        \node[fill=white] at (2.6,-1.0) {$\dots$};
        \node[fill=white] at (2.6,0.5) {$\dots$};
        \node[BrickRed] at (5.35,-1.1) {\fontsize{10}{10}\selectfont$\mathcal{E}^{(\alpha)}_p$};
    \end{tikzpicture}.
\end{equation} 
Because $\mathcal{E}_p^{(\alpha)}$ is real and strongly symmetric under $\prod_i X_i$, we can derive an effective action of $\tilde{\mathcal{T}}$ on the boundary degrees of freedom:
\begin{equation}
\tilde{\mathcal{T}}\rho(\alpha;g,p)=
\begin{tikzpicture}
        %%%%%%%%%%%%%%%%% ket layer
        \draw (-0.4,0)--(5.5,0);
        \node[gedge] at (-0.4,0) {};
        \node[gedge] at (5.5,0) {};
        \foreach \x in {0.8,1.5,2.2,3.6,4.3}{
        \draw (\x,0)--(\x,0.9);
        \node[blankg] at (\x,0) {};
        }
        \node[fill=white] at (2.9,0) {$\dots$};
        %%%%%%%%%%%%%%%%% bra layer
        \draw (0.1,-0.5)--(5.15,-0.5);
        \node[gedge] at (0.1,-0.5) {};
        \node[gedge] at (5.15,-0.5) {};
        \foreach \x in {0.8,1.5,2.2,3.6,4.3}{
        \draw (\x,-0.5)--(\x,-1.4);
        \node[blankg] at (\x,-0.5) {};
        }
        \node[fill=white] at (2.9,-0.5) {$\dots$};
        %%%%%%%%%%%%%%%%% gate
        \draw[BrickRed] (0.5,0.55) -- (-0.7,0.55) -- (-0.7,-1.05) -- (0.5,-1.05);
        \node [style={fill=BrickRed, rounded rectangle, minimum width=50mm, minimum height=1.8mm}] at (2.6,-1.05) {};
        \node [style={fill=BrickRed, rounded rectangle, minimum width=50mm, minimum height=1.8mm}] at (2.6,0.55) {};
        \node[fill=white] at (2.9,-1.05) {$\dots$};
        \node[fill=white] at (2.9,0.55) {$\dots$};
        \node[BrickRed] at (5.35,-1.1) {\fontsize{10}{10}\selectfont$\mathcal{E}^{(\alpha)}_p$};
        %%%%%%%%%%%%%%%%% virtuals
        \node[conj] at (-0.05,0.0) {};
        \node[Vginv] at (0.3,0.0) {};
        \node[conj] at (5.15,0.0) {};
        \node[Vg] at (4.8,0.0) {};
        \node[conj] at (0.45,-0.5) {};
        \node[conj] at (4.8,-0.5) {};
        
    \end{tikzpicture}.
\end{equation}
This is an isomorphic edge algebra to that obtained for the pure state in \cref{eq:edge_alg}. We emphasize that, as in the pure state, both the $\prod_i X_i$ and complex conjugation components of the transformation $\tilde{\mathcal{T}}$ are necessary to produce a nontrivial edge algebra; neither alone is sufficient. In light of the observations we have discussed, we conjecture that the phase for $g<0$ and $0<p<p_c$ is a mixed-state SPT phase protected by the symmetry $\tilde{\mathcal{T}}$ (or $\tilde{\mathcal{T}}^\prime$) alone defined in \cref{eq:T_tilde}.

\section{Proposal for quantum hardware realization}\label{sec:device}

In this section, we will discuss how the SW-SSB phase transition discussed previously can be observed on a quantum device. We will also discuss the practical advantage of our model over other models exhibiting similar phenomena: Namely, its large purity throughout regions of interest in the phase diagram.

\subsection{General protocol}

To observe SW-SSB and its associated phase transition, we ultimately seek to measure the \renyi-2 correlator $C_\mathrm{II}(p)$ for a particular $g$ and $\alpha$ as a function of the decoherence strength $p$. From the form of $C_\mathrm{II}$ given in \cref{eq:correlators}, we see that this requires measuring $\Tr[\rho Z_m Z_n \rho Z_m Z_n]$ and the purity $\Tr[\rho^2]$. We propose the following method of measuring these quantities on a quantum device.

First, we observe that our state of interest can be expressed as
\begin{equation}\label{eq:rho_expansion}
    \rho=\sum_{a}P_a\ketbra{\psi_a}{\psi_a}=\sum_{a}P_a U_a\ketbra{\mathbf{0}}{\mathbf{0}}U_a^\dagger,
\end{equation}
where $\{\ket{\psi_a}\}$ is an ensemble of states, $U_a$ is a quantum circuit which prepares $\ket{\psi_a}$ from the $N$-site product state $\ket{\mathbf{0}} := |0\rangle ^{\otimes N}$, and $P_a\in[0,1]$ is the probability. The \renyi-2 expectation value of a unitary operator $\mathcal{O}$ can then be expressed as
\begin{equation}\label{eq:purity_meas}
    \begin{split}
        \Tr[\rho\mathcal{O}\rho\mathcal{O}]&=\sum_{a,a'}P_{a}P_{a'}\Tr\left[U_{a}\ketbra{\mathbf{0}}{\mathbf{0}}U_{a}^\dagger\mathcal{O} U_{a'}\ketbra{\mathbf{0}}{\mathbf{0}}U_{a'}^\dagger\mathcal{O}^\dagger\right],\\
        &=\sum_{a,a'}P_{a}P_{a'}\left|\bra{\mathbf{0}}U_{a}^\dagger \mathcal{O}U_{a'}\ket{\mathbf{0}}\right|^2.
    \end{split}
\end{equation}
We can see that given $U_a$ and $ U_{a'}$, the term $p(\mathbf{0}|a,a') = \left|\bra{\mathbf{0}}U_{a}^\dagger \mathcal{O}U_{a'}\ket{\mathbf{0}}\right|^2$ is in fact the probability of measuring the state $|\psi_{O;a,a'}\rangle : = U_{a}^\dagger \mathcal{O}U_{a'}\ket{\mathbf{0}}$ in the computational basis and getting the outcome $\ket{\mathbf{0}}$.
We may set $\mathcal{O}=Z_m Z_n$ in \cref{eq:purity_meas} to measure $\Tr[\rho Z_m Z_n \rho Z_m Z_n]$, or set $\mathcal{O}=\mathbbm{1}$ to recover an expression for the purity
\begin{equation}\label{eq:op_meas}
        \Tr[\rho^2]
        =\sum_{a,a'}P_{a}P_{a'}\left|\bra{\mathbf{0}}U_{a}^\dagger U_{a'}\ket{\mathbf{0}}\right|^2.
\end{equation}

With this mapping in mind, we can measure the observable $\Tr[\rho\mathcal{O}\rho\mathcal{O}]$ (with $\mathcal{O}=\mathbbm{1}$ recovering the purity) by applying the following protocol:
\begin{enumerate}
    \item Generate $N_s$ pairs of the form $\{(a,a')\}$, sampled from the distribution $P(a,a')=P_{a}P_{a'}$.
    \item Prepare the state $|\psi_{O;a,a'}\rangle = U_{a}^\dagger \mathcal{O}U_{a'}\ket{\mathbf{0}}$ on the quantum device.
    \item Perform a computational basis measurement on the entire system, noting the number of outcomes $N_{\mathbf{0}}$ which return $\ket{\mathbf{0}}$.
    \item Compute the estimated value $\langlee\rho\mathcal{O}|\mathcal{O}\rho\ranglee\approx \frac{N_{\mathbf{0}}}{N_s}$.
\end{enumerate}

\subsection{Application to $\rho(\alpha;g,p)$}

We will now apply the general protocol presented in the previous section to the model of interest in this work. We first explicitly rewrite the density matrix $\rho(\alpha;g,p)$ in the form of \cref{eq:rho_expansion} as 
\begin{equation}\label{eq:rho_agp_expansion}
    \begin{split}
        \rho(\alpha;g,p)&=\mathcal{E}^{(\alpha)}_p(\ketbra{\psi_g}{\psi_g}),\\
        &=\sum_{\{n_{ij}\}}p_{\{n_{ij}\}}^{(\alpha)}{U}_{\{n_{ij}\}}\ketbra{\mathbf{0}}{\mathbf{0}}{U}_{\{n_{ij}\}}^\dagger,\\
    \end{split}
\end{equation}
where each $n_{ij}\in{0,1}$ is a binary variable assigned to the pair $(i,j)$, and the collection $\{n_{ij}\}$ serves as the label $a$,
\begin{equation}
\begin{gathered}
     U_a = {U}_{\{n_{ij}\}}=\left[\underset{i<j}{\prod}\left(Z_iZ_j\right)^{n_{ij}}\right]W_g,\\P_a = p_{\{n_{ij}\}}^{(\alpha)}=\prod_{i<j}[p_{ij}^{(\alpha)}{n_{ij}} + (1-p_{ij}^{(\alpha)})(1-n_{ij})],
\end{gathered}
\end{equation}
and $W_g$ is a unitary which prepares $\ket{\psi_g}$ from the product state $\ket{\mathbf{0}}$. Note that $a=\{n_{ij}\}$ can be sampled efficiently from the distribution $p_{\{n_{ij}\}}^{(\alpha)}$ by independently sampling each $n_{ij}$ according to the probability distribution $p_{ij}^{(\alpha)} n_{ij} + (1-p_{ij}^{(\alpha)})(1-n_{ij})$.

In Step 2 of the protocol, we prepare the state
\begin{equation}
    W_g^\dagger \left[\underset{i<j}{\prod}\left(Z_iZ_j\right)^{n_{ij}}\right]\mathcal{O}\left[\underset{i<j}{\prod}\left(Z_iZ_j\right)^{n_{ij}'}\right] W_g\ket{\mathbf{0}},
\end{equation}
where $\mathcal{O}$ is set to $\mathbbm{1}$ or $Z_mZ_n$. In either case, the operators between $W_g^\dagger$ and $W_g$ can be simplified to a single layer of $Z$ operations. The circuit which we must implement is therefore dominated by the state preparation unitary $W_g$. We remark that there are multiple methods available to implement this unitary, depending in part on the specifications of the available quantum hardware. On a digital quantum device, one could use sequential generation~\cite{Schon2005}, parallel-sequential circuits~\cite{wei2025state}, a renormalization-based approach~\cite{malz2023preparation}, or constant-depth circuits and fusion measurements~\cite{smith2024constant,Sahay_2025}. One can also use a parametrized quantum circuit trained with classical methods~\cite{Lin2026utilityscalequantum}. 
On an analog quantum device, one could use adiabatic evolution~\cite{zyadi} or counterdiabatic driving~\cite{tang2026weightednestedcommutatorsscalable}. Among these methods, the sequential and RG-based methods have been experimentally realized~\cite{smith2022crossing,scheer2025renormalization}.

\subsection{Resource estimation}

\begin{figure}
\captionsetup[subfigure]{labelformat=empty,captionskip=-20pt}
 \centering
\subfloat[\label{fig:}]{
\includegraphics[width=0.49\textwidth]{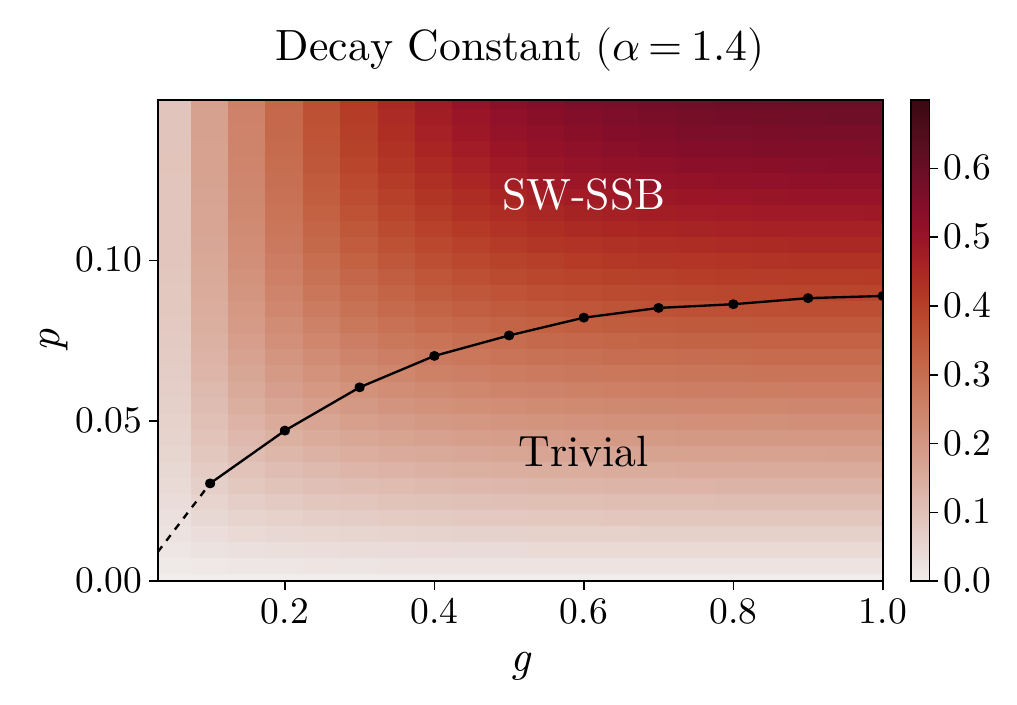}%
     }
\caption{The purity decays in system size as $e^{-\gamma N}$, with decay constant $\gamma$ dependent on $g$, $p$, and $\alpha$. The heatmap shows $\gamma$ at a fixed $\alpha$, plotted as a function of $g$ and $p$.}
    \label{fig:purity}
\end{figure}

With the above protocol, here we discuss the estimation of the number of measurement shots in practice. In particular, we emphasize certain parameter regimes can be probed in far fewer shots than others.

The smallness of the purity, $\Tr[\rho^2]$, presents a fundamental challenge to measuring the \renyi-2 correlator $C_\mathrm{II}$ on a quantum device. Although the ratio $\Tr[\rho Z_m Z_n \rho Z_m Z_n]/\Tr[\rho^2]$ can be $O(1)$, the numerator and denominator are each exponentially small in system size. This scale is set by the purity, which normalizes the correlator (see \cref{eq:correlators}). 
We can make this discussion more precise as follows: 
In our protocol, the purity is estimated as the mean expectation value of outcome all $\ket{0}$. That is to say, given the probability distribution of measurement outcome $p(\mathbf{b}) = \sum_{a,a'} p(\mathbf{b}|a,a')P(a,a')$, where $p(\mathbf{b}|a,a') = |\la \mathbf{b}|\psi_{a,a'}\rangle|^2$ is sampled from the quantum hardware, the purity is estimated as the expectation value $\mu := \mathbb{E}[\delta_{\mathbf{b} = \mathbf{0}}]$ with respect to the probability distribution $p(\mathbf{b})$, or equivalently, $\mathbb{E}[\delta_{\mathbf{b} = \mathbf{0}}] = p(\mathbf{0})$. Note that the numerator of $C_\mathrm{II}(p)$ can be estimated similarly by modifying the circuit $U_a$ accordingly.

Suppose one collects $N_s$ samples of a random variable with variance $\text{Var}=\sigma^2$, so that the standard error is approximately $\sigma /\sqrt{N_s}$. To reach a relative error $\epsilon = \frac{\sigma}{\mu\sqrt{N_s}}$, one would need $N_s \approx \frac{\sigma^2}{\mu^2\epsilon^2}$.
For our protocol, $\text{Var}[\delta_{\mathbf{b} = \mathbf{0}}] = \mathbb{E}[\delta_{\mathbf{b} = \mathbf{0}}^2] - \mathbb{E}[\delta_{\mathbf{b} = \mathbf{0}}]^2 = \mu - \mu^2 \approx \mu$, where we have used $\delta_{\mathbf{b} = \mathbf{0}}^2 = \delta_{\mathbf{b} = \mathbf{0}}$, giving us the estimated shot number to be 
\begin{equation}\label{eq:shot_count}
    N_s \approx \frac{1}{\mu \epsilon^2}
\end{equation}
to reach a relative error $\epsilon$. This result implies that an exponentially large number of measurements is required in order to measure the purity to some fixed precision since the purity is expected to be exponentially small in the system size. 

Nonetheless, the purity may still be measurable for system sizes on the scale of current experiments. Throughout this intermediate regime, the \textit{rate} of exponential decay is especially important for determining whether the purity (and in turn, the \renyi-2 correlator) can be accurately measured. From this perspective, exploring physics beyond the fixed point $g=1$ presents a distinct practical advantage: the rate at which the purity decays is much smaller at small $g$ and $p$. This situation is captured in \cref{fig:purity}, where we plot the decay rate throughout the phase diagram.

To obtain a concrete estimate of the quantum computational resources required to observe the SW-SSB transition on a quantum device, let us focus on the critical line $p=p_c(g)$ for $\alpha=1.4$ and $N=18$. For these parameters, the critical density matrix $\rho(\alpha;g,p_c(g))$ exhibits an approximate purity of $0.38$ at $g=0.1$ and $0.0028$ at $g=1.0$. Based on the estimate in \cref{eq:shot_count}, measuring the purity with a $10\%$ relative error would require $\sim36,000$ shots to probe the phase transition at $g=1$, compared to only $\sim260$ shots at $g=0.1$. This difference can meaningfully affect the practical feasibility of realizing and observing the SW-SSB transitions on quantum hardware.

Finally, as noted above, observing SW-SSB requires measuring $C_{\mathrm{II}}(p)$ at several values of the decoherence strength $p$. The measurement shots collected at these values can also be reused to estimate $C_{\mathrm{II}}(p')$ at a different decoherence strength $p'$ via multiple importance sampling~\cite{VeachGuibas1995}, provided that the $(a,a')$ label associated with each shot is recorded. We describe this classical post-processing procedure in Appendix~\ref{app:reweighting}.

\section{Discussion and outlook}

In summary, we consider power-law noise channels as a new ingredient to explore mixed-state quantum phases and their transitions. This consideration allows us to obtain several novel phenomena that cannot occur in one spatial dimension with finite-depth local channels, and results in a rich phase diagram, with tunable noise strength, power-law exponents, and initial states that interpolate between the trivial and non-trivial SPT phases. Below we will highlight some open questions that are worth further investigation. 

One novel result of our work is the onset of SW-SSB at finite noise strength of power-law channels in one spatial dimension. This conclusion is based on the calculation of the \renyi-2 correlator, a measure that is commonly adopted for diagnosing SW-SSB due to its simple conceptual intuition and its practical computability. As is well-known in literature, one exotic feature of SW-SSB transitions is that they are sensitive to the diagnostic employed; different diagnostics may lead to distinct locations of the critical point and different universality classes governing the transition~\cite{lessa_strong--weak_2024,liu2024diagnosingstrongtoweaksymmetrybreaking,weinstein2025efficient,Zhang2025strong}.

For example, the so-called fidelity correlator is proven to satisfy certain stability theorems~\cite{lessa_strong--weak_2024}, and is argued to serve as a more desirable measure for SW-SSB. Additionally, the divergence of the Markov length, defined via conditional mutual information, has emerged as a powerful probe of mixed-state phases and their transitions~\cite{sang2024stability}. Finally, we note that several recent works have developed a local understanding of SW-SSB~\cite{divi2026localstrongtoweakspontaneoussymmetry,liu2026localdescriptionstrongsymmetries,zhang2026localdiagnosticsstrongtoweakspontaneous}, and it would be worthwhile to consider any applicable local probes when designing an experimental realization of the kind we have described. In a forthcoming work~\cite{fidelity_powerlaw_toappear},
we employ these \renyi-1 quantities to investigate the mixed states induced by a power-law channel. 

% Motivated by analogous mappings in other examples~\cite{lessa_strong--weak_2024}, we expect the fidelity correlator to map, at least at the fixed points, to a disorder-weighted variant of the statistical mechanics model obtained from the \renyi-2 correlator.  Exploring this quantity in the presence of power-law channels presents an interesting direction for future work. 

Another notable feature of our work is the exploration of a pure-state SPT phase subject to a power-law channel. In particular, we investigate the stability of the SPT phase using \renyi-2 coherent information, which captures the stability of quantum memory encoded in the SPT boundary. Notably, \renyi-2 coherent information is a function of the \renyi-2 correlator; the decay in \renyi-2 coherent information corresponds to the onset of SW-SSB of the decohered mixed state. Quantifying SPT order in mixed states based on the capability of storing quantum information, therefore, presents a valuable perspective, which can be employed widely for various SPT states subject to decoherence. We also remark that an open question is to investigate the von Neumann coherent information, which is more intimately connected to the intrinsic threshold of protecting information using error correction protocols. Relatedly, designing a decoder to demonstrate such an error-correction transition is an interesting direction for further exploration.

\begin{acknowledgments}
The authors acknowledge the University of Maryland supercomputing resources (\href{http://hpcc.umd.edu}{http://hpcc.umd.edu}) used in this work. C.F., Z.-Y.W., J.S., and A.V.G.\ acknowledge support from the U.S.~Department of Energy, Office of Science, Accelerated Research in Quantum Computing, Fundamental Algorithmic Research toward Quantum Utility (FAR-Qu).  C.F., Z.-Y.W., J.S., and A.V.G.\ were also supported in part by the DoE ASCR Quantum Testbed Pathfinder program (award No.~DE-SC0024220), NQVL:QSTD:Design:FTL, NSF QLCI (award No.~OMA-2120757), NSF STAQ program, AFOSR MURI,  DARPA SAVaNT ADVENT, ONR MURI, and ARL (W911NF-24-2-0107). C.F., Z.-Y.W., J.S., and A.V.G.\ also acknowledge support from the U.S.~Department of Energy, Office of Science, National Quantum Information Science Research Centers, Quantum Systems Accelerator (award No.~DE-SCL0000121).  Y.-X.W.~acknowledges support from a QuICS Hartree Postdoctoral Fellowship. C.F. acknowledges the Joint Quantum Institute at the University of Maryland for support through a JQI fellowship. T.-C. L. and C.-J. L. acknowledge the support of the RQS postdoctoral fellowship through the National Science Foundation (QLCI grant OMA-2120757).

During the preparation of this manuscript we used OpenAI's GPT-5.5 and GPT-5.6 and Anthropic's Claude Opus 5 and Opus 5.5. Working from author-specified setups and prompts, these models were used to critique drafts, to suggest candidate proof strategies for several of the fixed-point derivations, and to assist in developing the modified Wolff cluster algorithm discussed in \cref{app:MCMC}. All incorporated arguments were re-derived independently by the authors before inclusion, and the authors take full responsibility for the contents of this manuscript.

\end{acknowledgments}
\bibliography{references,library}

\onecolumngrid
\newpage
\appendix

\section{Domain wall perspective}\label{app:DW_perspective}

In this section, we prove the expressions presented in \cref{eq:mps-coeff-g-domain,eq:mps-coeff-g-domain_munu} in the main text, which relate the MPS amplitudes in the computational basis to the number of domain walls in each basis state.

We first note a few special properties of the tensors $M_g^{[0]}$ and $M_g^{[1]}$ defined in \cref{eq:M-tensors-def} that will be useful in this derivation. First, note that these tensors satisfy $\left(M_g^{[i]} \right)^2 = M_g^{[i]}$ for $i=0,1$.
Next, define
\begin{equation}
    D^{[01]} \equiv M_g^{[0]}M_g^{[1]} = \begin{pmatrix}
        0&0\\
        1&g
    \end{pmatrix}.
\end{equation}
We see that $\left(D^{[01]}\right)^2 = g D^{[01]}$ implying $\left(D^{[01]}\right)^n = g^{n-1}D^{[01]}$.
The MPS coefficients involve calculating the contraction of a string of $M_g^{[0]}$ and $M_g^{[1]}$ tensors with the left and right boundary tensors. In a generic string, we can use $\left(M_g^{[i]} \right)^2 = M_g^{[i]}$ to reduce the string to an alternating string composed of the two tensors. Once we have obtained $\left(D^{[01]}\right)^n$ in the bulk of the string, there are four possible cases differing in the tensors appearing near the boundary:
\begin{equation}
    \left(D^{[01]}\right)^n,\quad M_g^{[1]}\left(D^{[01]}\right)^n, \quad\left(D^{[01]}\right)^nM_g^{[0]},\quad\mathrm{and}\quad M_g^{[1]}\left(D^{[01]}\right)^nM_g^{[0]},
\end{equation}
where $n\in[0,\lfloor\frac{N}{2}\rfloor]$.

In \cref{tab:DW_table}, we evaluate each of these expressions in the symmetric and logical-qubit boundary conditions and list the number of domain walls in each case. We find that the MPS coefficient for the basis state $\ket{\boldsymbol{i}}$ evaluates to $g^{\text{\#DW}(\boldsymbol{i})/2}f_g$ for the symmetric boundary conditions and $g^{\text{\#DW}(\boldsymbol{i})/2}\delta_{\mu,i_1}\delta_{\nu,i_N}f_g^{\mu\nu}$ for the logical-qubit boundary conditions. The final factors in each expression are independent of the computational basis vector index $\boldsymbol{i}$ and can be absorbed into the overall normalization.

\begin{table}
    \centering
    \renewcommand{\arraystretch}{1.75}
    \begin{tabular}{lcccccccc}
    \hline\hline
    &$\;\;\;$ & $\prod_{j} M_g^{[i_j]}$ &$\;\;\;$& $\bra{L}\prod_jM_g^{[i_j]}\ket{R}$ &$\;\;\;$& $\bra{\mu_L}\prod_jM_g^{[i_j]}\ket{\nu_R}$ &$\;\;\;$& DW in $\ket{\boldsymbol{i}}$ \\\hline
     Case 1& &$\left(D^{[01]}\right)^n$ & & $g^{n-1/2}f_g$ & & $g^{n-1/2}\delta_{\mu,0}\delta_{\nu,1}f^{\mu\nu}_g$ && $2n-1$ \\
     Case 2& &$M_g^{[1]}\left(D^{[01]}\right)^n$ & & $g^{n}f_g$ & & $g^{n}\delta_{\mu,1}\delta_{\nu,1}f^{\mu\nu}_g$ && $2n$ \\
     Case 3& &$\left(D^{[01]}\right)^nM_g^{[0]}$ & & $g^{n}f_g$ & & $g^{n}\delta_{\mu,0}\delta_{\nu,0}f^{\mu\nu}_g$ && $2n$ \\
     Case 4& &$M_g^{[1]}\left(D^{[01]}\right)^nM_g^{[0]}$ & & $g^{n+1/2}f_g$ & & $g^{n+1/2}\delta_{\mu,1}\delta_{\nu,0}f^{\mu\nu}_g$ && $2n+1$ \\
     Case 5& &$M_g^{[1]}M_g^{[0]}$ &&  $g^{1/2}f_g$ && $g^{1/2}\delta_{\mu,1}\delta_{\nu,0}f^{\mu\nu}_g$ && 1 \\
     Case 6& &$M_g^{[0]}$ &&  $f_g$ && $\delta_{\mu,0}\delta_{\nu,0}f^{\mu\nu}_g$  && 0 \\
     Case 7& &$M_g^{[1]}$ && $f_g$ && $\delta_{\mu,1}\delta_{\nu,1}f^{\mu\nu}_g$ && 0 \\\hline\hline
    \end{tabular}
    \caption{Evaluation of coefficient of the computational basis vector $\ket{\boldsymbol{i}}$ for all possible cases. We also list the number of domain walls (DW) in $\ket{\boldsymbol{i}}$ for comparison with the exponent of $g$. The symbols $f_g:=1+g^{1/2}$ and $f_g^{\mu\nu}:=\left(-g^{1/2}\right)^{\mu+\nu}(g-1)$ are overall constants which are absorbed into normalization. From this table, we read off the results stated in \cref{eq:mps-coeff-g-domain,eq:mps-coeff-g-domain_munu}. }
    \label{tab:DW_table}
\end{table}

\section{Details of the transformation $T$}\label{ap:CZ_trans}

In this section, we will prove several tensor equalities which will enable us to efficiently prove the results quoted in the main text. Namely, we will show that $T$ maps $\ket{\psi_g}$ to $\ket{\psi_{-g}}$ and preserves the $\prod_i X_i$ symmetry.

We will first prove the identity
\begin{equation}
  \begin{tikzpicture}
    \draw (-0.75,0)--(0.75,0);
    \draw (-0.35,0.0) -- (-0.35,0.6);
    \draw (0.35,0.0) -- (0.35,0.6);
    \node[plusg] at (-0.35,0) {};
    \node[plush] at (0.35,0) {};
    \draw (-0.35,0.4) -- (0.35,0.4);
    \node[CZ] at (0.35,0.4) {};
    \node[CZ] at (-0.35,0.4) {};
  \end{tikzpicture}=\begin{tikzpicture}
    \draw (-0.75,0)--(0.75,0);
    \draw (-0.35,0.0) -- (-0.35,0.6);
    \draw (0.35,0.0) -- (0.35,0.6);
    \node[minusg] at (-0.35,0) {};
    \node[plush] at (0.35,0) {};
    \node[Z] at (-0.35,0.4) {};
  \end{tikzpicture},
\end{equation}
where \begin{tikzpicture}
    \draw (-0.35,0.4) -- (0.35,0.4);
    \node[CZ] at (0.35,0.4) {};
    \node[CZ] at (-0.35,0.4) {};
  \end{tikzpicture} denotes the $CZ$ gate. We have
\begin{equation}\label{eq:CZ_identity}
    \begin{split}
        \begin{tikzpicture}
    \draw (-0.75,0)--(0.75,0);
    \draw (-0.35,0.0) -- (-0.35,0.6);
    \draw (0.35,0.0) -- (0.35,0.6);
    \node[plusg] at (-0.35,0) {};
    \node[plush] at (0.35,0) {};
    \draw (-0.35,0.4) -- (0.35,0.4);
    \node[CZ] at (0.35,0.4) {};
    \node[CZ] at (-0.35,0.4) {};
  \end{tikzpicture}&=\begin{pmatrix}0&0\\1&1\end{pmatrix}\otimes\ket{00}+\begin{pmatrix}0&0\\1&h\end{pmatrix}\otimes\ket{01}+\begin{pmatrix}g&g\\0&0\end{pmatrix}\otimes\ket{10}-\begin{pmatrix}1&h\\0&0\end{pmatrix}\otimes\ket{11},\\
  &=\begin{pmatrix}0&0\\1&1\end{pmatrix}\otimes\ket{00}+\begin{pmatrix}0&0\\1&h\end{pmatrix}\otimes\ket{01}-\begin{pmatrix}-g&-g\\0&0\end{pmatrix}\otimes\ket{10}-\begin{pmatrix}1&h\\0&0\end{pmatrix}\otimes\ket{11},\\
  &=\begin{tikzpicture}
    \draw (-0.75,0)--(0.75,0);
    \draw (-0.35,0.0) -- (-0.35,0.6);
    \draw (0.35,0.0) -- (0.35,0.6);
    \node[minusg] at (-0.35,0) {};
    \node[plush] at (0.35,0) {};
    \node[Z] at (-0.35,0.4) {};
  \end{tikzpicture}.
    \end{split}
\end{equation}
This identity shows that we can use $CZ$ gates to convert bulk tensors from $M_g$ to $M_{-g}$.

We now prove the following two identities which will enable us to properly transform the boundary conditions:
\begin{equation}
    \begin{tikzpicture}
    \draw (-0.5,0)--(0.5,0);
    \draw (0,0.0) -- (0,0.5);
    \node[plush] at (0,0) {};
    \node[minusgedge] at (-0.5,0) {};
  \end{tikzpicture}=\begin{tikzpicture}
    \draw (-0.5,0)--(0.5,0);
    \draw (0,0.0) -- (0,0.6);
    \node[plush] at (0,0) {};
    \node[gedge] at (-0.5,0) {};
    \node[Sdag] at (0.0,0.4) {};
  \end{tikzpicture},\qquad\begin{tikzpicture}
    \draw (-0.5,0)--(0.5,0);
    \draw (0,0.0) -- (0,0.5);
    \node[minusg] at (0,0) {};
    \node[minusgedge] at (0.5,0) {};
  \end{tikzpicture}=
\begin{tikzpicture}
    \draw (-0.5,0)--(0.5,0);
    \draw (0,0.0) -- (0,0.6);
    \node[plusg] at (0,0) {};
    \node[gedge] at (0.6,0) {};
    \node[S] at (0.0,0.4) {};
  \end{tikzpicture}.\\
\end{equation}
First, we have
\begin{equation}\label{eq:left_edge_identity}
\begin{split}
\begin{tikzpicture}
    \draw (-0.5,0)--(0.5,0);
    \draw (0,0.0) -- (0,0.5);
    \node[plush] at (0,0) {};
    \node[gedge] at (-0.5,0) {};
  \end{tikzpicture}&=\begin{pmatrix}1&1\end{pmatrix}\otimes \ket{0}+\frac{1}{\sqrt{g}}\begin{pmatrix}1&h\end{pmatrix}\otimes\ket{1},\\
    \Rightarrow \begin{tikzpicture}
    \draw (-0.5,0)--(0.5,0);
    \draw (0,0.0) -- (0,0.5);
    \node[plush] at (0,0) {};
    \node[minusgedge] at (-0.5,0) {};
  \end{tikzpicture}&=\begin{pmatrix}1&1\end{pmatrix}\otimes \ket{0}+\frac{1}{\sqrt{-g}}\begin{pmatrix}1&h\end{pmatrix}\otimes\ket{1},\\
  &=\begin{pmatrix}1&1\end{pmatrix}\otimes \ket{0}-i\frac{1}{\sqrt{g}}\begin{pmatrix}1&h\end{pmatrix}\otimes\ket{1},\\
  &=\begin{tikzpicture}
    \draw (-0.5,0)--(0.5,0);
    \draw (0,0.0) -- (0,0.6);
    \node[plush] at (0,0) {};
    \node[gedge] at (-0.5,0) {};
    \node[Sdag] at (0.0,0.4) {};
  \end{tikzpicture}.
\end{split}
\end{equation}
Next, we have
\begin{equation}
\begin{split}
\begin{tikzpicture}
    \draw (-0.5,0)--(0.5,0);
    \draw (0,0.0) -- (0,0.5);
    \node[plusg] at (0,0) {};
    \node[gedge] at (0.5,0) {};
  \end{tikzpicture}&=\begin{pmatrix}0\\\sqrt{g}+1\end{pmatrix}\otimes \ket{0}+\begin{pmatrix}\sqrt{g}+g\\0\end{pmatrix}\otimes\ket{1},\\
    \Rightarrow \begin{tikzpicture}
    \draw (-0.5,0)--(0.5,0);
    \draw (0,0.0) -- (0,0.5);
    \node[minusg] at (0,0) {};
    \node[minusgedge] at (0.5,0) {};
  \end{tikzpicture}&=\begin{pmatrix}0\\\sqrt{-g}+1\end{pmatrix}\otimes \ket{0}+\begin{pmatrix}\sqrt{-g}-g\\0\end{pmatrix}\otimes\ket{1},\\
  &=\begin{pmatrix}0\\i\sqrt{g}+1\end{pmatrix}\otimes \ket{0}+\begin{pmatrix}i\sqrt{g}-g\\0\end{pmatrix}\otimes\ket{1},\\
  &=\frac{i\sqrt{g}+1}{\sqrt{g}+1}\left[\begin{pmatrix}0\\\sqrt{g}+1\end{pmatrix}\otimes \ket{0}+i\begin{pmatrix}\sqrt{g}+g\\0\end{pmatrix}\otimes\ket{1}\right],\\
  &=\left(\frac{i\sqrt{g}+1}{\sqrt{g}+1}\right)\begin{tikzpicture}
    \draw (-0.5,0)--(0.5,0);
    \draw (0,0.0) -- (0,0.6);
    \node[plusg] at (0,0) {};
    \node[gedge] at (0.6,0) {};
    \node[S] at (0.0,0.4) {};
  \end{tikzpicture}\\
\end{split}
\end{equation}
The quantity in parentheses is an overall multiplicative factor which will be absorbed in the wavefunction normalization.

In summary, we have proven
\begin{equation}
  \begin{tikzpicture}
    \draw (-0.75,0)--(0.75,0);
    \draw (-0.35,0.0) -- (-0.35,0.6);
    \draw (0.35,0.0) -- (0.35,0.6);
    \node[plusg] at (-0.35,0) {};
    \node[plusg] at (0.35,0) {};
    \draw (-0.35,0.4) -- (0.35,0.4);
    \node[CZ] at (0.35,0.4) {};
    \node[CZ] at (-0.35,0.4) {};
  \end{tikzpicture}=\begin{tikzpicture}
    \draw (-0.75,0)--(0.75,0);
    \draw (-0.35,0.0) -- (-0.35,0.6);
    \draw (0.35,0.0) -- (0.35,0.6);
    \node[minusg] at (-0.35,0) {};
    \node[plusg] at (0.35,0) {};
    \node[Z] at (-0.35,0.4) {};
  \end{tikzpicture},\quad\begin{tikzpicture}
    \draw (-0.5,0)--(0.5,0);
    \draw (0,0.0) -- (0,0.6);
    \node[plusg] at (0,0) {};
    \node[gedge] at (-0.5,0) {};
    \node[Sdag] at (0.0,0.4) {};
  \end{tikzpicture}=\begin{tikzpicture}
    \draw (-0.5,0)--(0.5,0);
    \draw (0,0.0) -- (0,0.5);
    \node[plusg] at (0,0) {};
    \node[minusgedge] at (-0.5,0) {};
  \end{tikzpicture},\quad\mathrm{and}\quad 
\begin{tikzpicture}
    \draw (-0.5,0)--(0.5,0);
    \draw (0,0.0) -- (0,0.6);
    \node[plusg] at (0,0) {};
    \node[gedge] at (0.6,0) {};
    \node[S] at (0.0,0.4) {};
  \end{tikzpicture}\propto\begin{tikzpicture}
    \draw (-0.5,0)--(0.5,0);
    \draw (0,0.0) -- (0,0.5);
    \node[minusg] at (0,0) {};
    \node[minusgedge] at (0.5,0) {};
  \end{tikzpicture},
\end{equation}
where we have set $h$ equal to $g$ or $-g$, as will be needed in the following proofs.

We will now prove that $T$ maps $\ket{\psi_g}$ to $\ket{\psi_{-g}}$. We have
\begin{align*}
        T\ket{\psi_g}&=\begin{tikzpicture}
        %%%%% Bottom rail %%%%%
        \draw (0,0)--(6.6,0);
        \node[gedge] at (0,0) {};
        \node[gedge] at (6.6,0) {};
        \foreach \x in {0.5,1.2,1.9,2.6,4.0,4.7,5.4,6.1}{
        \draw (\x,0)--(\x,1.6);
        \node[plusg] at (\x,0) {};
        }
        \node[fill=white] at (3.3,0) {$\dots$};
        %%%%% CZs 1 %%%%%
        \foreach \x in {0.5,1.2,1.9,2.6,4.0,4.7,5.4,6.1}{
        \node[CZ] at (\x,0.4) {};
        }
        \foreach \x in {0.5,1.9,4.0,5.4}{
        \draw (\x,0.4)--(\x+0.7,0.4);
        }
        %%%%% CZs 2 %%%%%
        \foreach \x in {1.2,1.9,2.6,4.0,4.7,5.4}{
        \node[CZ] at (\x,0.8) {};
        }
        \foreach \x in {1.2,2.6,3.3,4.7}{
        \draw (\x,0.8)--(\x+0.7,0.8);
        }
        \node[fill=white] at (3.3,0.8) {$\ldots$};
        %%%%% Zs %%%%%
        \foreach \x in {1.2,1.9,2.6,4.0,4.7,5.4}{
        \node[Z] at (\x,1.2) {};
        }
    %%%%% Ss %%%%%
    \node[S] at (0.5,1.2) {};
    \node[S] at (6.1,1.2) {};
    \end{tikzpicture},\\
    &=\begin{tikzpicture}
        %%%%% Bottom rail %%%%%
        \draw (0,0)--(6.6,0);
        \node[gedge] at (0,0) {};
        \node[gedge] at (6.6,0) {};
        \foreach \x in {0.5,1.9,4.0,5.4}{
        \draw (\x,0)--(\x,2);
        \node[minusg] at (\x,0) {};
        }
        \foreach \x in {1.2,2.6,4.7,6.1}{
        \draw (\x,0)--(\x,1.6);
        \node[plusg] at (\x,0) {};
        }
        \node[fill=white] at (3.3,0) {$\dots$};
        %%%%% CZs 1 %%%%%
        \foreach \x in {0.5,1.9,4.0,5.4}{
        \node[Z] at (\x,0.4) {};
        }
        %%%%% CZs 2 %%%%%
        \foreach \x in {1.2,1.9,2.6,4.0,4.7,5.4}{
        \node[CZ] at (\x,0.8) {};
        }
        \foreach \x in {1.2,2.6,3.3,4.7}{
        \draw (\x,0.8)--(\x+0.7,0.8);
        }
        \node[fill=white] at (3.3,0.8) {$\ldots$};
        %%%%% Zs %%%%%
        \foreach \x in {1.2,1.9,2.6,4.0,4.7,5.4}{
        \node[Z] at (\x,1.2) {};
        }
    %%%%% Ss %%%%%
    \node[S] at (0.5,1.2) {};
    \node[S] at (6.1,1.2) {};
    \end{tikzpicture},\\
    &=\begin{tikzpicture}
        %%%%% Bottom rail %%%%%
        \draw (0,0)--(6.6,0);
        \node[gedge] at (0,0) {};
        \node[gedge] at (6.6,0) {};
        \foreach \x in {0.5,1.9,4.0,5.4}{
        \draw (\x,0)--(\x,1.2);
        \node[minusg] at (\x,0) {};
        }
        \foreach \x in {1.2,2.6,4.7,6.1}{
        \draw (\x,0)--(\x,1.2);
        \node[plusg] at (\x,0) {};
        }
        \node[fill=white] at (3.3,0) {$\dots$};
        %%%%% CZs 2 %%%%%
        \foreach \x in {1.2,1.9,2.6,4.0,4.7,5.4}{
        \node[CZ] at (\x,0.4) {};
        }
        \foreach \x in {1.2,2.6,3.3,4.7}{
        \draw (\x,0.4)--(\x+0.7,0.4);
        }
        \node[fill=white] at (3.3,0.4) {$\ldots$};
        %%%%% Zs %%%%%
        \foreach \x in {1.2,2.6,4.7}{
        \node[Z] at (\x,0.8) {};
        }
    %%%%% Ss %%%%%
    \node[Sdag] at (0.5,0.8) {};
    \node[S] at (6.1,0.8) {};
    \end{tikzpicture},\\
    &=\begin{tikzpicture}
        %%%%% Bottom rail %%%%%
        \draw (0,0)--(6.6,0);
        \node[gedge] at (0,0) {};
        \node[gedge] at (6.6,0) {};
        \foreach \x in {0.5,1.2,1.9,2.6,4.0,4.7,5.4}{
        \draw (\x,0)--(\x,1.2);
        \node[minusg] at (\x,0) {};
        }
        \foreach \x in {6.1}{
        \draw (\x,0)--(\x,1.2);
        \node[plusg] at (\x,0) {};
        }
        \node[fill=white] at (3.3,0) {$\dots$};
        %%%%% CZs 2 %%%%%
        \foreach \x in {1.2,2.6,4.7}{
        \node[Z] at (\x,0.4) {};
        }
        \node[fill=white] at (3.3,0.4) {$\ldots$};
        %%%%% Zs %%%%%
        \foreach \x in {1.2,2.6,4.7}{
        \node[Z] at (\x,0.8) {};
        }
    %%%%% Ss %%%%%
    \node[Sdag] at (0.5,0.8) {};
    \node[S] at (6.1,0.8) {};
    \end{tikzpicture},\\
    &=\begin{tikzpicture}
        %%%%% Bottom rail %%%%%
        \draw (0,0)--(6.6,0);
        \node[gedge] at (0,0) {};
        \node[gedge] at (6.6,0) {};
        \foreach \x in {0.5,1.2,1.9,2.6,4.0,4.7,5.4}{
        \draw (\x,0)--(\x,0.8);
        \node[minusg] at (\x,0) {};
        }
        \foreach \x in {6.1}{
        \node[plusg] at (\x,0) {};
        \draw (\x,0)--(\x,0.8);
        }
        \node[fill=white] at (3.3,0) {$\dots$};
    %%%%% Ss %%%%%
    \node[Sdag] at (0.5,0.4) {};
    \node[S] at (6.1,0.4) {};
    \end{tikzpicture},\\
    &=\begin{tikzpicture}
        %%%%% Bottom rail %%%%%
        \draw (0,0)--(6.6,0);
        \node[minusgedge] at (0,0) {};
        \node[minusgedge] at (6.6,0) {};
        \foreach \x in {0.5,1.2,1.9,2.6,4.0,4.7,5.4,6.1}{
        \draw (\x,0)--(\x,0.5);\node[minusg] at (\x,0) {};
        }
        \node[fill=white] at (3.3,0) {$\dots$};
    \end{tikzpicture}=\ket{\psi_{-g}}.
\end{align*}

Finally, we prove that $T$ commutes with the $\prod_iX_i$ symmetry:
\begin{align*}        T\prod_iX_i&=\begin{tikzpicture}
        %%%%% Xs %%%%%
        \foreach \x in {0.5,1.2,1.9,2.6,4.0,4.7,5.4,6.1}{
        \draw (\x,-0.4)--(\x,1.6);
        \node[X] at (\x,0) {};
        }
        \node[fill=white] at (3.3,0) {$\dots$};
        %%%%% CZs 1 %%%%%
        \foreach \x in {0.5,1.2,1.9,2.6,4.0,4.7,5.4,6.1}{
        \node[CZ] at (\x,0.4) {};
        }
        \foreach \x in {0.5,1.9,4.0,5.4}{
        \draw (\x,0.4)--(\x+0.7,0.4);
        }
        %%%%% CZs 2 %%%%%
        \foreach \x in {1.2,1.9,2.6,4.0,4.7,5.4}{
        \node[CZ] at (\x,0.8) {};
        }
        \foreach \x in {1.2,2.6,3.3,4.7}{
        \draw (\x,0.8)--(\x+0.7,0.8);
        }
        \node[fill=white] at (3.3,0.8) {$\ldots$};
        % \node[fill=white] at (3.3,0.6) {$\ldots$};
        %%%%% Zs %%%%%
        \foreach \x in {1.2,1.9,2.6,4.0,4.7,5.4}{
        \node[Z] at (\x,1.2) {};
        }
    %%%%% Ss %%%%%
    \node[S] at (0.5,1.2) {};
    \node[S] at (6.1,1.2) {};
    \end{tikzpicture},\\
    % \displaybreak
    &=\begin{tikzpicture}
        %%%%% Xs %%%%%
        \foreach \x in {0.5,1.2,1.9,2.6,4.0,4.7,5.4,6.1}{
        \draw (\x,-0.4)--(\x,1.6);
        \node[Y] at (\x,0.4) {};
        }
        \node[fill=white] at (3.3,0) {$\dots$};
        %%%%% CZs 1 %%%%%
        \foreach \x in {0.5,1.2,1.9,2.6,4.0,4.7,5.4,6.1}{
        \node[CZ] at (\x,0.0) {};
        }
        \foreach \x in {0.5,1.9,4.0,5.4}{
        \draw (\x,0.0)--(\x+0.7,0.0);
        }
        %%%%% CZs 2 %%%%%
        \foreach \x in {1.2,1.9,2.6,4.0,4.7,5.4}{
        \node[CZ] at (\x,0.8) {};
        }
        \foreach \x in {1.2,2.6,3.3,4.7}{
        \draw (\x,0.8)--(\x+0.7,0.8);
        }
        \node[fill=white] at (3.3,0.8) {$\ldots$};
        %%%%% Zs %%%%%
        \foreach \x in {1.2,1.9,2.6,4.0,4.7,5.4}{
        \node[Z] at (\x,1.2) {};
        }
    %%%%% Ss %%%%%
    \node[S] at (0.5,1.2) {};
    \node[S] at (6.1,1.2) {};
    \end{tikzpicture},\\
    &=\begin{tikzpicture}
        %%%%% Xs %%%%%
        \foreach \x in {1.2,1.9,2.6,4.0,4.7,5.4}{
        \draw (\x,-0.4)--(\x,1.6);
        \node[X] at (\x,0.8) {};
        }
        \foreach \x in {0.5,6.1}{
        \draw (\x,-0.4)--(\x,1.6);
        \node[Y] at (\x,0.8) {};
        }
        \node[fill=white] at (3.3,0) {$\dots$};
        %%%%% CZs 1 %%%%%
        \foreach \x in {0.5,1.2,1.9,2.6,4.0,4.7,5.4,6.1}{
        \node[CZ] at (\x,0.0) {};
        }
        \foreach \x in {0.5,1.9,4.0,5.4}{
        \draw (\x,0.0)--(\x+0.7,0.0);
        }
        %%%%% CZs 2 %%%%%
        \foreach \x in {1.2,1.9,2.6,4.0,4.7,5.4}{
        \node[CZ] at (\x,0.4) {};
        }
        \foreach \x in {1.2,2.6,3.3,4.7}{
        \draw (\x,0.4)--(\x+0.7,0.4);
        }
        \node[fill=white] at (3.3,0.4) {$\ldots$};
        %%%%% Zs %%%%%
        \foreach \x in {1.2,1.9,2.6,4.0,4.7,5.4}{
        \node[Z] at (\x,1.2) {};
        }
    %%%%% Ss %%%%%
    \node[S] at (0.5,1.2) {};
    \node[S] at (6.1,1.2) {};
    \end{tikzpicture},\\
    &=(-1)^N\;\begin{tikzpicture}
        %%%%% Xs %%%%%
        \foreach \x in {1.2,1.9,2.6,4.0,4.7,5.4}{
        \draw (\x,-0.4)--(\x,1.6);
        \node[X] at (\x,1.2) {};
        }
        \foreach \x in {0.5,6.1}{
        \draw (\x,-0.4)--(\x,1.6);
        \node[X] at (\x,1.2) {};
        }
        \node[fill=white] at (3.3,0) {$\dots$};
        %%%%% CZs 1 %%%%%
        \foreach \x in {0.5,1.2,1.9,2.6,4.0,4.7,5.4,6.1}{
        \node[CZ] at (\x,0.0) {};
        }
        \foreach \x in {0.5,1.9,4.0,5.4}{
        \draw (\x,0.0)--(\x+0.7,0.0);
        }
        %%%%% CZs 2 %%%%%
        \foreach \x in {1.2,1.9,2.6,4.0,4.7,5.4}{
        \node[CZ] at (\x,0.4) {};
        }
        \foreach \x in {1.2,2.6,3.3,4.7}{
        \draw (\x,0.4)--(\x+0.7,0.4);
        }
        \node[fill=white] at (3.3,0.4) {$\ldots$};
        %%%%% Zs %%%%%
        \foreach \x in {1.2,1.9,2.6,4.0,4.7,5.4}{
        \node[Z] at (\x,0.8) {};
        }
    %%%%% Ss %%%%%
    \node[S] at (0.5,0.8) {};
    \node[S] at (6.1,0.8) {};
    \end{tikzpicture}=\prod_iX_iT,
\end{align*}
where $(-1)^N=1$ because we have taken $N$ to be even.

\section{\renyi-2 coherent information mapping}\label{ap:coh_info}

In this section, we present the details of the mapping introduced in the main text between the coherent information and an observable in a statistical mechanics model. Recall from \cref{eq:coh_info} that the coherent information is defined as $I_c=S(\rho_Q)-S(\rho_{QR})$, and we work with the \renyi-2 proxy
\begin{equation}
	I_c^{(2)}=-\log\left(\frac{\Tr[\rho_Q^2]}{\Tr[\rho_{QR}^2]}\right).
\end{equation}

In order to compute $I_c^{(2)}$, we must first express $\rho_{QR}$ in a usable form. To construct the initial pure state, we need to specify a basis of logical states by choosing boundary conditions for the MPS introduced in \cref{eq:psi_g_MPS}. A particularly convenient choice is to configure the boundaries such that the logical $\overline{Z}$ operators are $\overline{Z}_L:=Z_1$ and $\overline{Z}_R:=Z_N$ (for all $g<0$) and the logical $\overline{X}$ operators are $\overline{X}_L:=X_1 Z_2$ and $\overline{X}_R:=Z_{N-1}X_N$ (for $g=-1$). These logical operators are obtained via the choice 
\begin{equation}\label{eq:app-bound-conds}
    \begin{split}
        \ket{0_L} &= \begin{pmatrix}
            0\\1
        \end{pmatrix},\qquad \ket{0_R} = \begin{pmatrix}
            g\\-1
        \end{pmatrix},\\
        \ket{1_L} &= \begin{pmatrix}
            -1\\0
        \end{pmatrix},\qquad \ket{1_R} = \begin{pmatrix}
            g\\-g
        \end{pmatrix}.\\
    \end{split}
\end{equation}
We now verify that this choice indeed yields $\overline{Z}_L:=Z_1$ and $\overline{Z}_R:=Z_N$ by contracting the first or last bulk tensor with each boundary vector:
\begin{equation}
\begin{gathered}
    \begin{tikzpicture}
    \draw (-0.5,0)--(0.5,0);
    \draw (0,0.0) -- (0,0.6);
    \node[plusg] at (0,0) {};
    \node[inner sep = 0.08cm,
    shape = rectangle,
    draw,
    fill,
    label={[label distance=0cm]270:$0_L$}] at (-0.5,0) {};
    \node[Z] at (0.0,0.4) {};
  \end{tikzpicture}=\begin{tikzpicture}
    \draw (-0.8,0)--(0.5,0);
    \draw (0,0.0) -- (0,0.6);
    \node[plusg] at (0,0) {};
    \node[inner sep = 0.08cm,
    shape = rectangle,
    draw,
    fill,
    label={[label distance=0cm]270:$0_L$}] at (-0.8,0) {};
    \node[mZv] at (-0.4,0.0) {};
  \end{tikzpicture}=\begin{tikzpicture}
    \draw (-0.5,0)--(0.5,0);
    \draw (0,0.0) -- (0,0.6);
    \node[plusg] at (0,0) {};
    \node[inner sep = 0.08cm,
    shape = rectangle,
    draw,
    fill,
    label={[label distance=0cm]270:$0_L$}] at (-0.5,0) {};
  \end{tikzpicture},\\
%%%%%%%%%%%%%%%%%%%%%%%%%%%%%%%%%%%%%%%%%%%%%%%%%%%%%%%%%%%%
    \begin{tikzpicture}
    \draw (-0.5,0)--(0.5,0);
    \draw (0,0.0) -- (0,0.6);
    \node[plusg] at (0,0) {};
    \node[inner sep = 0.08cm,
    shape = rectangle,
    draw,
    fill,
    label={[label distance=0cm]270:$1_L$}] at (-0.5,0) {};
    \node[Z] at (0.0,0.4) {};
  \end{tikzpicture}=\begin{tikzpicture}
    \draw (-0.8,0)--(0.5,0);
    \draw (0,0.0) -- (0,0.6);
    \node[plusg] at (0,0) {};
    \node[inner sep = 0.08cm,
    shape = rectangle,
    draw,
    fill,
    label={[label distance=0cm]270:$1_L$}] at (-0.8,0) {};
    \node[mZv] at (-0.4,0.0) {};
  \end{tikzpicture}=-\begin{tikzpicture}
    \draw (-0.5,0)--(0.5,0);
    \draw (0,0.0) -- (0,0.6);
    \node[plusg] at (0,0) {};
    \node[inner sep = 0.08cm,
    shape = rectangle,
    draw,
    fill,
    label={[label distance=0cm]270:$1_L$}] at (-0.5,0) {};
  \end{tikzpicture},\\
%%%%%%%%%%%%%%%%%%%%%%%%%%%%%%%%%%%%%%%%%%%%%%%%%%%%%%%%%%%%
    \begin{tikzpicture}
    \draw (-0.5,0)--(0.5,0);
    \draw (0,0.0) -- (0,0.6);
    \node[plusg] at (0,0) {};
    \node[inner sep = 0.08cm,
    shape = rectangle,
    draw,
    fill,
    label={[label distance=0cm]270:$0_R$}] at (0.5,0) {};
    \node[Z] at (0.0,0.4) {};
  \end{tikzpicture}=M_g^{[0]}\ket{0_R}\otimes \ket{0}-M_g^{[1]}\ket{0_R}\otimes \ket{1}=\begin{pmatrix}
           0 \\ g-1
        \end{pmatrix}\otimes \ket{0}-\begin{pmatrix}
            0\\ 0
        \end{pmatrix}\otimes \ket{1}=\begin{tikzpicture}
    \draw (-0.5,0)--(0.5,0);
    \draw (0,0.0) -- (0,0.6);
    \node[plusg] at (0,0) {};
    \node[inner sep = 0.08cm,
    shape = rectangle,
    draw,
    fill,
    label={[label distance=0cm]270:$0_R$}] at (0.5,0) {};
  \end{tikzpicture},\\
%%%%%%%%%%%%%%%%%%%%%%%%%%%%%%%%%%%%%%%%%%%%%%%%%%%%%%%%%%%%
    \begin{tikzpicture}
    \draw (-0.5,0)--(0.5,0);
    \draw (0,0.0) -- (0,0.6);
    \node[plusg] at (0,0) {};
    \node[inner sep = 0.08cm,
    shape = rectangle,
    draw,
    fill,
    label={[label distance=0cm]270:$1_R$}] at (0.5,0) {};
    \node[Z] at (0.0,0.4) {};
  \end{tikzpicture}=M_g^{[0]}\ket{1_R}\otimes \ket{0}-M_g^{[1]}\ket{1_R}\otimes \ket{1}=\begin{pmatrix}
           0 \\ 0
        \end{pmatrix}\otimes \ket{0}-\begin{pmatrix}
            g(1-g)\\ 0
        \end{pmatrix}\otimes \ket{1}=-\begin{tikzpicture}
    \draw (-0.5,0)--(0.5,0);
    \draw (0,0.0) -- (0,0.6);
    \node[plusg] at (0,0) {};
    \node[inner sep = 0.08cm,
    shape = rectangle,
    draw,
    fill,
    label={[label distance=0cm]270:$1_R$}] at (0.5,0) {};
  \end{tikzpicture}.
\end{gathered}
\end{equation}
We see that acting with $Z_1$ and $Z_N$, respectively, gives an overall plus sign when the $\ket{0}$ boundary conditions are chosen and an overall minus sign when the $\ket{1}$ boundary conditions are chosen.

Next, we will verify the logical action of the specified logical $\overline{X}$ operators at each edge. We specialize to the point $g=-1$ in the following calculation, noting that the operator $V_{g}$ obtained by pulling $X$ through the physical leg of $M_g$ evaluates to $XZ$ in this limit:
\begin{equation}
    \begin{gathered}
        \begin{tikzpicture}
        \draw (-0.5,0)--(1.0,0);
        \draw (0,0.0) -- (0,0.6);
        \draw (0.5,0.0) -- (0.5,0.6);
        \node[minus1] at (0,0) {};
        \node[minus1] at (0.5,0) {};
        \node[inner sep = 0.08cm,
        shape = rectangle,
        draw,
        fill,
        label={[label distance=0cm]270:$0_L$}] at (-0.5,0) {};
        \node[X] at (0.0,0.4) {};
        \node[Z] at (0.5,0.4) {};
      \end{tikzpicture}=-
      \begin{tikzpicture}
        \draw (-0.5,0)--(3.0,0);
        \draw (1.0,0.0) -- (1.0,0.6);
        \draw (2.5,0.0) -- (2.5,0.6);
        \node[minus1] at (1.0,0) {};
        \node[minus1] at (2.5,0) {};
        \node[inner sep = 0.08cm,
        shape = rectangle,
        draw,
        fill,
        label={[label distance=0cm]270:$0_L$}] at (-0.5,0) {};
        \node[Xv] at (0.5,0.0) {};
        \node[Zv] at (0.0,0.0) {};
        \node[Xv] at (1.375,0.0) {};
        \node[Zv] at (1.75,0.0) {};
        \node[Zv] at (2.125,0.0) {};
      \end{tikzpicture}=
      \begin{tikzpicture}
        \draw (-0.5,0)--(1.5,0);
        \draw (0.3,0.0) -- (0.3,0.6);
        \draw (1.1,0.0) -- (1.1,0.6);
        \node[minus1] at (0.3,0) {};
        \node[minus1] at (1.1,0) {};
        \node[inner sep = 0.08cm,
        shape = rectangle,
        draw,
        fill,
        label={[label distance=0cm]270:$0_L$}] at (-0.5,0) {};
        \node[Xv] at (-0.1,0.0) {};
        \node[Xv] at (0.7,0.0) {};
      \end{tikzpicture}=-
      \begin{tikzpicture}
        \draw (-0.5,0)--(1.3,0);
        \draw (0.3,0.0) -- (0.3,0.6);
        \draw (0.8,0.0) -- (0.8,0.6);
        \node[minus1] at (0.3,0) {};
        \node[minus1] at (0.8,0) {};
        \node[inner sep = 0.08cm,
        shape = rectangle,
        draw,
        fill,
        label={[label distance=0cm]270:$1_L$}] at (-0.5,0) {};
        \node[mZv] at (-0.1,0.0) {};
      \end{tikzpicture}=
      \begin{tikzpicture}
        \draw (-0.5,0)--(1.0,0);
        \draw (0,0.0) -- (0,0.6);
        \draw (0.5,0.0) -- (0.5,0.6);
        \node[minus1] at (0,0) {};
        \node[minus1] at (0.5,0) {};
        \node[inner sep = 0.08cm,
        shape = rectangle,
        draw,
        fill,
        label={[label distance=0cm]270:$1_L$}] at (-0.5,0) {};
      \end{tikzpicture},\\
      \begin{tikzpicture}
        \draw (-0.5,0)--(1.0,0);
        \draw (0,0.0) -- (0,0.6);
        \draw (0.5,0.0) -- (0.5,0.6);
        \node[minus1] at (0,0) {};
        \node[minus1] at (0.5,0) {};
        \node[inner sep = 0.08cm,
        shape = rectangle,
        draw,
        fill,
        label={[label distance=0cm]270:$0_R$}] at (1.0,0) {};
        \node[Z] at (0.0,0.4) {};
        \node[X] at (0.5,0.4) {};
      \end{tikzpicture}=
      \begin{tikzpicture}
        \draw (-0.5,0)--(3.0,0);
        \draw (0.0,0.0) -- (0.0,0.6);
        \draw (1.5,0.0) -- (1.5,0.6);
        \node[minus1] at (0.0,0) {};
        \node[minus1] at (1.5,0) {};
        \node[inner sep = 0.08cm,
        shape = rectangle,
        draw,
        fill,
        label={[label distance=0cm]270:$0_R$}] at (3.0,0) {};
        \node[Xv] at (0.375,0.0) {};
        \node[Zv] at (0.75,0.0) {};
        \node[Xv] at (1.125,0.0) {};
        \node[Xv] at (2.0,0.0) {};
        \node[Zv] at (2.5,0.0) {};
      \end{tikzpicture}=-
      \begin{tikzpicture}
        \draw (-0.5,0)--(1.5,0);
        \draw (-0.1,0.0) -- (-0.1,0.6);
        \draw (0.7,0.0) -- (0.7,0.6);
        \node[minus1] at (-0.1,0) {};
        \node[minus1] at (0.7,0) {};
        \node[inner sep = 0.08cm,
        shape = rectangle,
        draw,
        fill,
        label={[label distance=0cm]270:$1_R$}] at (1.5,0) {};
        \node[Zv] at (0.3,0.0) {};
        \node[Xv] at (1.1,0.0) {};
      \end{tikzpicture}=-
      \begin{tikzpicture}
        \draw (-0.5,0)--(1.8,0);
        \draw (-0.1,0.0) -- (-0.1,0.6);
        \draw (1.4,0.0) -- (1.4,0.6);
        \node[minus1] at (-0.1,0) {};
        \node[minus1] at (1.4,0) {};
        \node[inner sep = 0.08cm,
        shape = rectangle,
        draw,
        fill,
        label={[label distance=0cm]270:$1_R$}] at (1.8,0) {};
        \node[Zv] at (0.4,0.0) {};
        \node[mZv] at (0.9,0.0) {};
      \end{tikzpicture}=
      \begin{tikzpicture}
        \draw (-0.5,0)--(1.0,0);
        \draw (0,0.0) -- (0,0.6);
        \draw (0.5,0.0) -- (0.5,0.6);
        \node[minus1] at (0,0) {};
        \node[minus1] at (0.5,0) {};
        \node[inner sep = 0.08cm,
        shape = rectangle,
        draw,
        fill,
        label={[label distance=0cm]270:$1_R$}] at (1.0,0) {};
      \end{tikzpicture}.
    \end{gathered}
\end{equation}
We have just shown that $\overline{X}_L\ket{\psi^{\mu\nu}_{g=-1}}=\ket{\psi^{\bar{\mu}\nu}_{g=-1}}$ and $\overline{X}_R\ket{\psi^{\mu\nu}_{g=-1}}=\ket{\psi^{\mu\bar{\nu}}_{g=-1}}$, where $\bar \mu$ denotes the opposite label from $\mu$. We have therefore verified that the choice of boundary vectors given in \cref{eq:app-bound-conds} gives rises to the desired logical Pauli operators.

Recall that the logical states $\{\ket{\psi_g^{\mu\nu}}:\mu,\nu\in0,1\}$ are defined as
\begin{equation}
    \ket{\psi_g^{\mu\nu}}=\sum_{\{i_j\}=0,1}\bra{\mu_L}M_g^{[i_1]}\cdots M_g^{[i_N]}\ket{\nu_R}\ket{i_1,\ldots,i_N}.
\end{equation}
We demonstrate in \cref{tab:DW_table} that 
\begin{equation}\label{eq:mu_nu_amp}
\begin{split}
    \bra{\mu_L}M_g^{[i_1]}\cdots M_g^{[i_N]}\ket{\nu_R}\propto &g^{\mathrm{DW}(\boldsymbol{i})/2} \delta_{i_1,\mu}\delta_{i_N,\nu},\\
\end{split}
\end{equation}
where $\mathrm{DW}(\boldsymbol{i})$ counts the number of domain walls in the configuration $\boldsymbol{i}$. Namely, 
\begin{equation}
    \mathrm{DW}(\boldsymbol{i})=\sum_{j=1}^{N-1}\left(\frac{1-i_ji_{j+1}}{2}\right).
\end{equation}
Note that it is clear from \cref{eq:mu_nu_amp} and the definition of the logical $\overline{Z}$ operators that $\ket{\psi_g^{\mu\nu}}$ can be decomposed into the form $\ket{\psi_g^{\mu\nu}}=\ket{\mu}\ket{\psi^{\mu\nu}_{g,\mathrm{bulk}}}\ket{\nu}$.

We are now prepared to define the initial state on $QR$ before application of the decoherence channel. Using \cref{eq:logical-bell-pairs}, we have
\begin{equation}
    \begin{gathered}
        \ket{\Psi_g}_{QR}=\frac{1}{2}\left(\sum_{\mu,\nu=0,1}\ket{\psi_g^{\mu\nu}}_Q\ket{\mu,\nu}_R\right)
    \end{gathered}
\end{equation}
so that $\rho_{QR}=\mathcal{E}_p^{(\alpha)}(\ketbra{\Psi_g}{\Psi_g})$. To proceed with the statistical mechanics mapping, we express $\ket{\Psi_g}$ in a more convenient form. When $g<0$, we can write 
\begin{equation}
\begin{split}
    \ket{\psi_g^{\mu\nu}}&=\sum_{\{i_j\}}g^{\frac{1}{4}\sum_{j=1}^{N-1}(1-Z_jZ_{j+1})}\delta_{i_1,\mu}\delta_{i_N,\nu}\ket{i_1,\ldots,i_N},\\
    &=|g|^{\frac{1}{4}\sum_{j=1}^{N-1}(1-Z_jZ_{j+1})}\ket{\psi_{g=-1}^{\mu\nu}},\\&=\mathcal{N}_ge^{-\frac{\ln{|g|}}{4}\sum_{j=1}^{N-1}Z_jZ_{j+1}}\ket{\psi_{g=-1}^{\mu\nu}},
\end{split}
\end{equation}
where we have defined $\mathcal{N}_g:=|g|^{\frac{N-1}{4}}$. This tells us that $\ket{\psi_g^{\mu\nu}}$ can be obtained from the cluster state $\ket{\psi_{g=-1}^{\mu\nu}}$ (up to normalization) via application of the imaginary-time evolution operator $e^{H_g}$, where
\begin{equation}
    H_g = -\frac{\ln{|g|}}{4}\sum_{j=1}^{N-1}Z_jZ_{j+1}.
\end{equation} This in turn allows us to write $\ket{\Psi_g}\propto e^{H_g}\ket{\Psi_{g=-1}}$.

We can then express $\ketbra{\Psi_{g=-1}}{\Psi_{g=-1}}$ as a product of projectors, given by
\begin{equation}\label{eq:cluster_proj}
\begin{split}
    \ketbra{\Psi_{g=-1}}{\Psi_{g=-1}}=&2^{-(N+2)}(1+\overline{Z}_LZ_{A})(1+\overline{Z}_RZ_{B})\prod_{i=1}^{N}(1+S_i),
\end{split}
\end{equation}
where
\begin{equation}
    S_i=\begin{cases}\overline{X}_LX_A,&i=1\\-Z_{i-1}X_iZ_{i+1},& i=2,\ldots,N-1\\\overline{X}_RX_B,&i=N\end{cases}.
\end{equation}
The projectors $\frac{1+S_i}{2}$ for $i=2,\ldots,N-1$ prepare the bulk in the correct state, leaving only degrees of freedom at the edges, which are then prepared into logical Bell pairs with the reference qubits by the remaining projectors. Combining these results and applying commutation relations, we find
\begin{equation}
    \begin{split}
    \ketbra{\Psi_{g}}{\Psi_{g}}&=\mathcal{N}(1+\overline{Z}_LZ_{A})(1+\overline{Z}_RZ_{B})\sum_{\{s_j\}=\pm 1}e^{H_g}\left(\prod_{i=1}^{N}S_i^{\left(\frac{1-s_i}{2}\right)}\right)e^{H_g}\\&=\mathcal{N}(1+\overline{Z}_LZ_{A})(1+\overline{Z}_RZ_{B})\sum_{\{s_j\}=\pm 1}\left(\prod_{i=1}^{N}S_i^{\left(\frac{1-s_i}{2}\right)}\right)e^{-\frac{\ln{|g|}}{2}\sum_{j=1}^{N-1}Z_jZ_{j+1}\left(\frac{1+s_js_{j+1}}{2}\right)},
\end{split}
\end{equation}
where $\mathcal{N}$ is an overall constant introduced to ensure normalization.

Observing that the channel $\mathcal{E}_p^{(\alpha)}$ only acts non-trivially on the $S_i$, we may write
\begin{equation}
    \begin{split}
    \rho_{QR}&=\mathcal{N}(1+\overline{Z}_LZ_{A})(1+\overline{Z}_RZ_{B})\sum_{\{s_j\}=\pm 1}\mathcal{E}_p^{(\alpha)}\left(\prod_{i=1}^{N}S_i^{\left(\frac{1-s_j}{2}\right)}\right)e^{-\frac{\ln{|g|}}{2}\sum_{j=1}^{N-1}Z_jZ_{j+1}\left(\frac{1+s_js_{j+1}}{2}\right)},\\&=\mathcal{N}(1+\overline{Z}_LZ_{A})(1+\overline{Z}_RZ_{B})\sum_{\{s_j\}=\pm 1}\left(\prod_{i=1}^{N}S_i^{\left(\frac{1-s_j}{2}\right)}\right)e^{\sum_{i<j}J_{ij}^{(\alpha)}s_is_j-\frac{\ln{|g|}}{2}\sum_{j=1}^{N-1}Z_jZ_{j+1}\left(\frac{1+s_js_{j+1}}{2}\right)},
\end{split}
\end{equation}
where $J_{ij}^{(\alpha)}$ is defined as in the main text (see \cref{eq:beta-c-def}).

We now evaluate $\rho_Q=\Tr_R[\rho_{QR}]$. The partial trace contributes a factor
\begin{equation}
    \begin{split}
        \Tr_R\left[(1+\overline{Z}_LZ_{A})(1+\overline{Z}_RZ_{B})S_1^{\left(\frac{1-s_1}{2}\right)}S_N^{\left(\frac{1-s_N}{2}\right)}\right]=\delta_{s_1,1}\delta_{s_N,1},
    \end{split}
\end{equation}
so that
\begin{equation}
    \rho_Q=\sum_{\{s_j\}=\pm 1}\left(\prod_{i=2}^{N-1}S_i^{\left(\frac{1-s_j}{2}\right)}\right)e^{\sum_{i<j}J_{ij}^{(\alpha)}s_is_j-\frac{\ln{|g|}}{2}\sum_{j=1}^{N-1}Z_jZ_{j+1}\left(\frac{1+s_js_{j+1}}{2}\right)}\delta_{s_1,1}\delta_{s_N,1}.
\end{equation}

For the present purpose of mapping to a statistical mechanics model, we will consider the \renyi-2 entropy which is defined as $S^{(2)}(\rho)=-\log \Tr[\rho^2]$. For this entropy, the coherent information becomes $I_c=-\log\left(\frac{\Tr[\rho_Q^2]}{\Tr[\rho_{QR}^2]}\right)$.

We now evaluate $\Tr[\rho_{QR}^2]$ by first noting that
\begin{equation}
    \begin{split}
        \rho_{QR}^2&=\mathcal{N}^2(1+\overline{Z}_LZ_{A})(1+\overline{Z}_RZ_{B})\sum_{\{s_j\}=\pm 1}\sum_{\{s_j'\}=\pm 1}\left(\prod_{i=1}^{N}S_i^{\left(\frac{1-s_j}{2}+\frac{1-s_j'}{2}\right)}\right)\\&\qquad\qquad\qquad\qquad\times e^{\sum_{i<j}J_{ij}^{(\alpha)}(s_is_j+s_i's_j')-\frac{\ln{|g|}}{2}\sum_{j=1}^{N-1}Z_jZ_{j+1}\left(\frac{1+s_js_{j+1}}{2}+\frac{1+s_j's_{j+1}'}{2}\right)}.
    \end{split}
\end{equation}
The trace contributes a factor of 
\begin{equation}
\begin{split}
    &\Tr\left[(1+\overline{Z}_LZ_{A})(1+\overline{Z}_RZ_{B})\left(\prod_{i=1}^{N}S_i^{\left(\frac{1-s_j}{2}+\frac{1-s_j'}{2}\right)}\right)e^{-\frac{\ln{|g|}}{2}\sum_{j=1}^{N-1}Z_jZ_{j+1}\left(\frac{1+s_js_{j+1}}{2}+\frac{1+s_j's_{j+1}'}{2}\right)}\right]\\=&\Tr\left[e^{-\frac{\ln{|g|}}{2}\sum_{j=1}^{N-1}Z_jZ_{j+1}\left({1+s_js_{j+1}}\right)}\right]\prod_{i=1}^N\delta_{s_i,s_i'},\\
    =&\sum_{\{s_j''\}=\pm 1}e^{-\frac{\ln{|g|}}{2}\sum_{j=1}^{N-1}s_j''s_{j+1}''\left({1+s_js_{j+1}}\right)}\prod_{i=1}^N\delta_{s_i,s_i'},
\end{split}
\end{equation}
so that 
\begin{equation}
\begin{split}
    \Tr[\rho_{QR}^2]&=\sum_{\{s_j\}=\pm 1}\sum_{\{s_j''\}=\pm 1}\mathrm{exp}\left[2\sum_{i<j}J_{ij}^{(\alpha)}s_is_j-\frac{\ln{|g|}}{2}\sum_{j=1}^{N-1}s_j''s_{j+1}''\left({1+s_js_{j+1}}\right)\right],\\
    &=\sum_{\{s_j\}=\pm 1}\sum_{\{s_j''\}=\pm 1}\mathrm{exp}\left[2\sum_{i<j}J_{ij}^{(\alpha)}s_is_i''s_js_j''-\frac{\ln{|g|}}{2}\sum_{j=1}^{N-1}\left(s_j''s_{j+1}''+s_js_{j+1}\right)\right],\\
\end{split}
\end{equation}
where we have obtained the second equality by redefining $s_i\mapsto s_is_i''$. Repeating this procedure for $\rho_Q$, we find
\begin{equation}
    \begin{split}
        \Tr[\rho_{Q}^2]&=\sum_{\{s_j\}=\pm 1}\sum_{\{s_j''\}=\pm 1}\mathrm{exp}\left[2\sum_{i<j}J_{ij}^{(\alpha)}s_is_i''s_js_j''-\frac{\ln{|g|}}{2}\sum_{j=1}^{N-1}\left(s_j''s_{j+1}''+s_js_{j+1}\right)\right]\delta_{s_1s_1'',1}\delta_{s_Ns_N'',1},\\
        &=\sum_{\{s_j\}=\pm 1}\sum_{\{s_j''\}=\pm 1}\mathrm{exp}\left[2\sum_{i<j}J_{ij}^{(\alpha)}s_is_i''s_js_j''-\frac{\ln{|g|}}{2}\sum_{j=1}^{N-1}\left(s_j''s_{j+1}''+s_js_{j+1}\right)\right]\left(\frac{1-s_1s_1''}{2}\right)\left(\frac{1-s_Ns_N''}{2}\right).\\
    \end{split}
\end{equation}
We can identify $\Tr[\rho^2_{QR}]$ as the partition function $Z_H$ of a statistical mechanics model on $2N$ classical spins with 
\begin{equation}
    H[\{s_j\},\{s_j'\}]=-2\sum_{i<j}J_{ij}^{(\alpha)}s_is_i''s_js_j''+\frac{\ln{|g|}}{2}\sum_{j=1}^{N-1}\left(s_j''s_{j+1}''+s_js_{j+1}\right)
\end{equation}
in units where $\beta =1$. The \renyi-2 coherent information can then be expressed as 
\begin{equation}
\begin{split}
    I_c^{(2)} &= -\log\left(\frac{1}{4}\left(1-\braket{s_1s_1''}-\braket{s_Ns_N''}+\braket{s_1s_1''s_Ns_N''}\right)\right),\\
    &= -\log\left(\frac{1}{4}\left(1+\braket{s_1s_1''s_Ns_N''}\right)\right),\\
    &= 2\log 2-\log\left(1+\braket{s_1s_1''s_Ns_N''}\right),\\
\end{split}
\end{equation}
where expectation values are taken with respect to the Boltzmann weight $\frac{e^{-H}}{Z_H}$ and the second equality is obtained by recognizing that $\braket{s_is_i''}=0$ because $H$ is invariant under the transformations $\{s_i\}\mapsto-\{s_i\}$ and $\{s_i^{''}\}\mapsto-\{s_i^{''}\}$.

\section{Numerical implementation via MCMC}\label{app:MCMC}

In this section, we discuss the numerical algorithm used to produce the data presented in the main text and in \cref{app:num_results}. Having reduced all quantum expectation values of interest to classical correlation functions in the statistical mechanics model given in \cref{eq:hamiltonian-stat-mech-LR}, we performed MCMC numerics to measure each classical correlation function. In particular, we utilized an appropriate variant of the Wolff cluster algorithm~\cite{wolff1989collective} to combat critical slowing-down. We now describe this algorithm:

\begin{enumerate}
    \item Randomly initialize a configuration of the spins $A=\left(\mathbf{s}^L,\mathbf{s}^R\right)$, where $\mathbf{s}^{L/R}\in \{\pm 1\}^N$.
    \item Perform a cluster update, defined as follows:
    \begin{enumerate}
        \item Randomly choose a chain $x\in \{L,R\}$ and a seed site $i_0\in [1,N]$. Initially, the cluster is $C=\{i_0\}$.
        \item Next, we grow the cluster $C$. The following procedure loops until all sites in $C$ have been selected exactly once: begin by selecting a site $i\in C$ which has not yet been selected. For every $j\in [1,N]$ such that $j\notin C$, add $j$ to $C$ with probability
        \begin{equation}\label{eq:pij_cases}
            p_{ij}(A)=\begin{cases}
                1-e^{-\Delta(\beta\mathcal{H}_{ij}[A])},&\Delta(\beta\mathcal{H}_{ij}[A])>0\\
                0,&\text{otherwise}
            \end{cases},
        \end{equation}
        where $\Delta(\beta\mathcal{H}_{ij})$ is the change in $\beta\mathcal{H}_{ij}$ if $s_j^x\mapsto-s_j^x$. In particular, 
        \begin{equation}
            \Delta(\beta\mathcal{H}_{ij}[A])=4J_{ij}^{(\alpha)} s_i^L s_i^R s_j^L s_j^R - (\ln |g|)  s_i^x s_{j}^x\delta_{|i-j|,1}.
        \end{equation}
        This process terminates when the final site currently in the cluster is selected, and all potential new sites are rejected.
    \end{enumerate}
    \item For all $i\in C$, update $A$ by flipping $s_i^x\mapsto -s_i^x$.
    \item Sample each observable in the current spin configuration $A=\left(\mathbf{s}^L,\mathbf{s}^R\right)$.
    \item Repeat steps 2-4 a total of $N_\text{burn-in}+N_\text{samples}$ times.
    \item Discard the first $N_\text{burn-in}$ samples, as the system needs time to thermalize, ``forgetting'' its initial condition and providing unbiased samples.
\end{enumerate}
We emphasize that a move is accepted at every iteration in this algorithm, as opposed to Metropolis MCMC wherein some proposed moves are rejected. This is because all probabilistic features of the algorithm have been transferred to the construction of the cluster.

To verify that this algorithm samples the proper distribution, we demonstrate that it is ergodic and satisfies detailed balance. To see that it is ergodic, notice that there is a finite probability of flipping the single spin $s_i^x$ by rejecting all other spins from the proposed cluster. This fact implies that any two configurations have a finite transition probability.

Next, we prove that the modified Wolff algorithm satisfies detailed balance with respect to the Boltzmann distribution. In particular, let $C\subseteq [1,N]$ be a cluster, let $A$ be an arbitrary spin configuration, and let $B$ be the spin  configuration obtained from $A$ by flipping the spins $\{s_i^x:i\in C\}$. Detailed balance then requires
\begin{equation}
    \frac{p(A\to B)}{p(B\to A)}=e^{-(\beta \mathcal{H}[B]-\beta \mathcal{H}[A])}.
\end{equation}
First, consider the transition probability from $A$ to $B$ given by $p(A\to B)$. This corresponds to building the cluster $C$ and choosing the correct chain $x$ while in $A$, so we have $p(A\to B)=\frac{1}{2}p(\text{build }C|A)$. To compute $p(\text{build }C|A)$, we should integrate probability over measurement histories resulting in $C$, which we denote $\text{hist}\to C$. Each history consists of a seed $i_0$ and a set $\{(i,j,o_{ij})\}$ comprised of a site $i$ currently in $C$, a site $j$ currently outside of $C$, and an outcome $o_{ij}\in \{0,1\}$ which determines whether site $i$ brings site $j$ into $C$. In this notation, we have
\begin{equation}
    p(A\to B)=\frac{1}{2}\frac{|C|}{N}\sum_{\text{hist}\to C}\prod_{(i,j)\in \text{hist}} p(o_{ij}|A),
\end{equation}
where $p(o_{ij}|A):=p_{ij}+o_{ij}(1-2p_{ij})$, where $p_{ij}$ is defined in \cref{eq:pij_cases} and evaluated in the configuration $A$.
Because every history constructs $C$, we know that every $i\in C$ must have tried and failed to bring each $j\notin C$ into the cluster. We may therefore extract a prefactor from the sum
\begin{equation}\label{eq:AtoB}
\begin{split}
    p(A\to B)&=\frac{1}{2}\frac{|C|}{N}\left(\prod_{i\in C}\prod_{j\notin C}(1-p_{ij}(A))\right)\Bigg(\sum_{\text{hist}\to C}\prod_{\substack{(i,j)\in \text{hist}:\\i,j\in C}} p(o_{ij}|A)\Bigg),\\
    &=\frac{1}{2}\frac{|C|}{N}\mathrm{exp}\left[-\sum_{i\in C}\sum_{j\notin C}\Delta(\beta \mathcal{H}_{ij}[A])\delta_{\Delta(\beta \mathcal{H}_{ij}[A])>0}\right]\Bigg(\sum_{\text{hist}\to C}\prod_{\substack{(i,j)\in \text{hist}:\\i,j\in C}} p(o_{ij}|A)\Bigg),
\end{split}
\end{equation}
where we have taken advantage of the fact that the remaining terms in the original sum must be such that $j$ was eventually accepted into $C$. Notice that the reverse transition probability may be written as
\begin{equation}\label{eq:BtoA}
    p(B\to A)=\frac{1}{2}\frac{|C|}{N}\mathrm{exp}\left[\sum_{i\in C}\sum_{j\notin C}\Delta(\beta \mathcal{H}_{ij}[A])\delta_{\Delta(\beta \mathcal{H}_{ij}[A])<0}\right]\left(\sum_{\text{hist}\to C}\prod_{\substack{(i,j)\in \text{hist}:\\i,j\in C}} p(o_{ij}|A)\right),
\end{equation}
where we have used the facts that $\Delta(\beta \mathcal{H}_{ij}[B])=-\Delta(\beta \mathcal{H}_{ij}[A])$ when only one of $i$ or $j$ lies in $C$ and $p(o_{ij}|A)=p(o_{ij}|B)$ when both $i$ and $j$ lie in $C$. We may therefore combine \cref{eq:AtoB,eq:BtoA} to obtain
\begin{equation}
    \frac{p(A\to B)}{p(B\to A)}=\mathrm{exp}\left[-\sum_{i\in C}\sum_{j\notin C}\Delta(\beta \mathcal{H}_{ij}[A])\right]=e^{-(\beta \mathcal{H}[B]-\beta \mathcal{H}[A])}.
\end{equation}
We conclude that the modified Wolff algorithm satisfies detailed balance.

\section{Further numerical results}\label{app:num_results}
In this section, we present additional numerical results referenced in the main text. We also discuss the interpretation of these numerical results.

\begin{figure}
    \centering
    \includegraphics[width=0.48\linewidth]{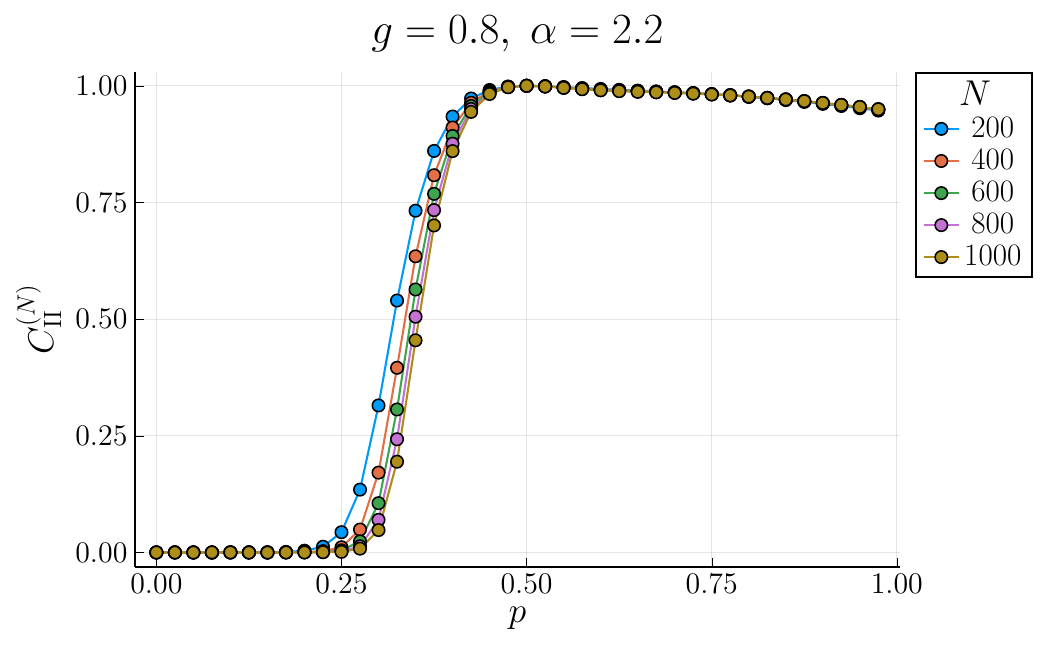}
    \hfill\includegraphics[width=0.48\linewidth]{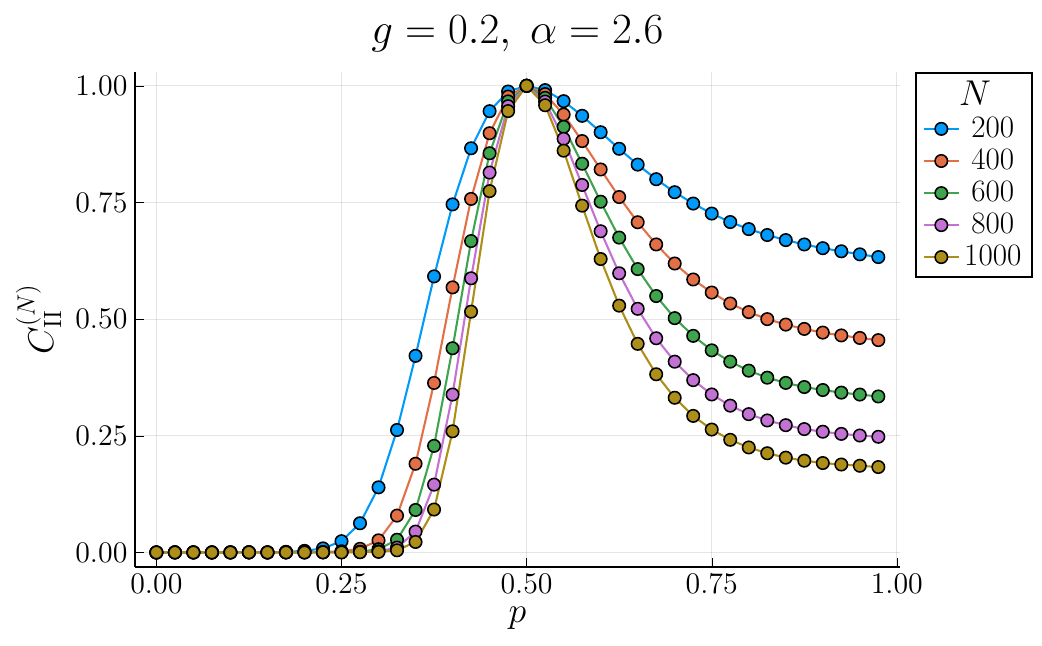}\\
\includegraphics[width=0.48\linewidth]{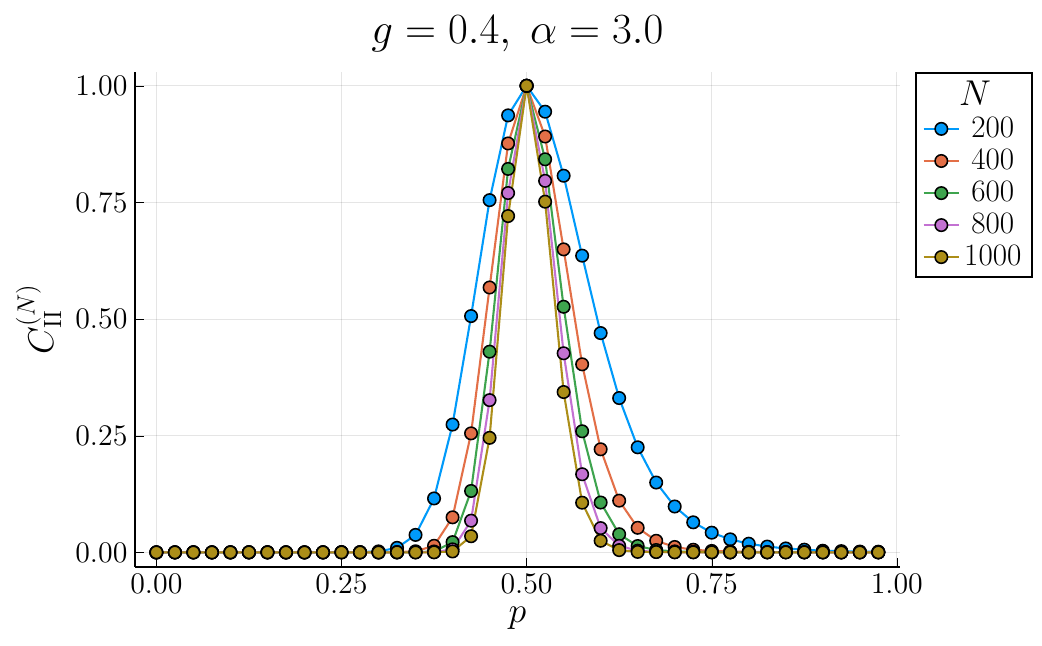}
\hfill\includegraphics[width=0.48\linewidth]{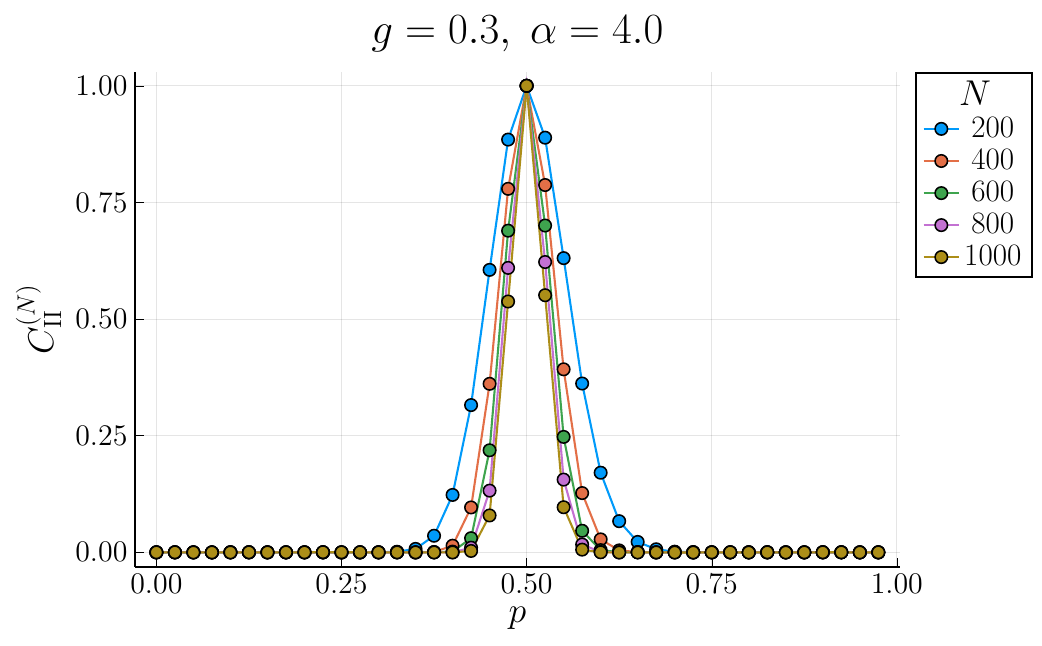}
    \caption{\textbf{Short-Range Regime.} When $\alpha>2$, the \renyi-2 correlator $C_\mathrm{II}^{(N)}$ displays an isolated peak at $p=1/2$ which sharpens with increasing $\alpha$ and increasing $N$. }
    \label{fig:short-range}
\end{figure}

In \cref{fig:short-range}, we plot representative measurements of the \renyi-2 correlator $C_\mathrm{II}$ for several choices of $\alpha>2$. As discussed in the main text, the nearest-neighbor channel corresponds to the $\alpha\to\infty$ limit, and our statistical mechanics mapping suggests that $\mathcal{E}_p^{(\alpha)}$ is effectively short-ranged for $\alpha>2$. The data presented in \cref{fig:short-range} corroborates this expectation, showing that the correlator does not approach a stable finite value as $p$ is increased when tuned to this regime in $\alpha$. Rather, it displays a peak at $p=1/2$ where the nearest-neighbor contributions to $\mathcal{E}_p^{(\alpha)}$ act as projectors onto the operator subspace which commutes with $Z_i Z_{i+1}$.

\begin{figure}
    \centering
    \includegraphics[width=0.9\linewidth]{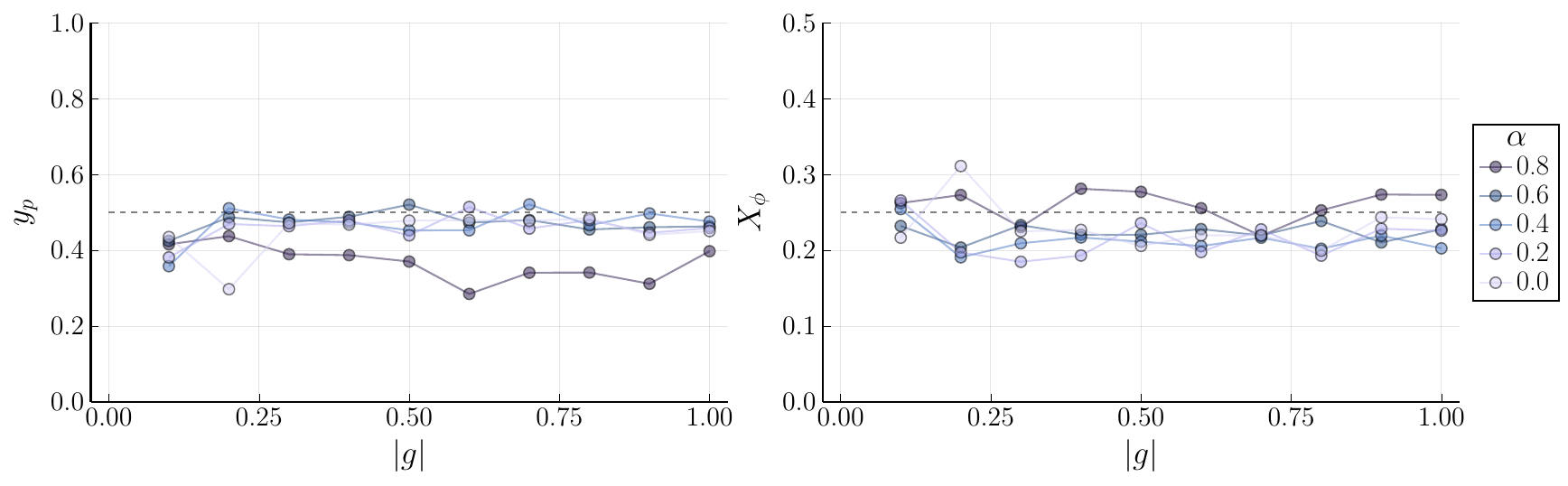}
    \includegraphics[width=0.9\linewidth]{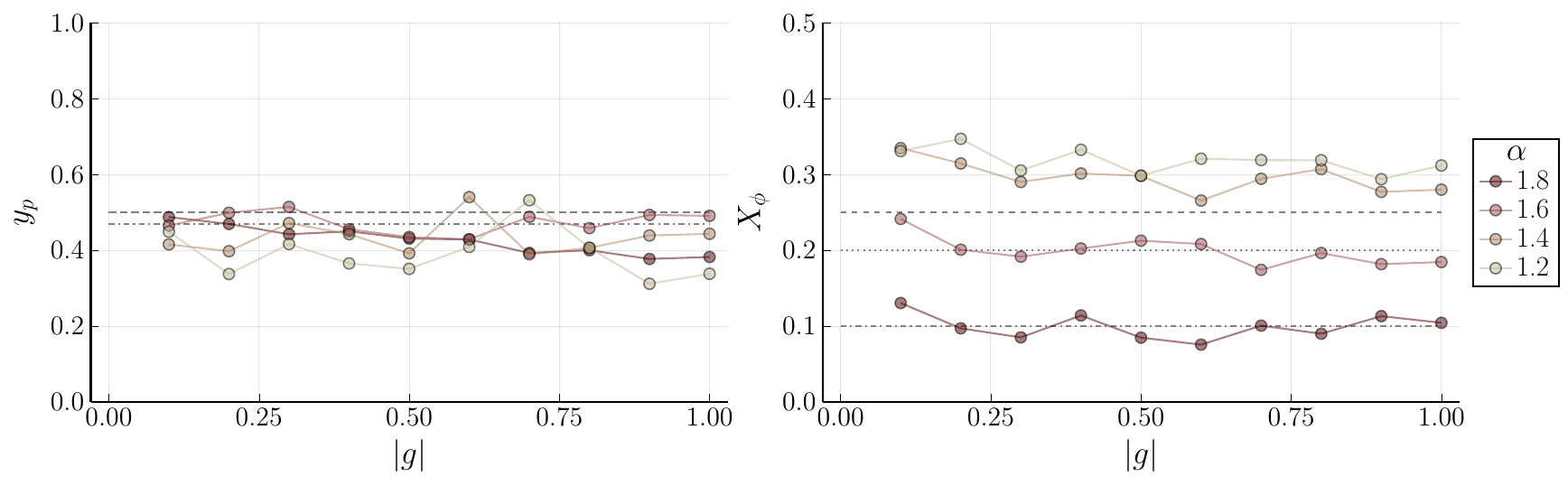}
    \caption{\textbf{Critical Exponents.} The exponents $X_\phi$ and $y_p$ extracted from $C_\mathrm{II}$ near criticality. Fro reference, we include the corresponding values for the power-law Ising model as dashed ($\alpha<1.5$), dotted ($\alpha=1.6$), and dot-dashed ($\alpha=1.8$) lines.}
    \label{fig:crit_exp}
\end{figure}

In \cref{fig:crit_exp}, we present the scaling exponents extracted via MCMC numerics. We compare these values with predictions from the power-law Ising model literature, as denoted by the horizontal lines in each plot. We also note that the data becomes noisier as $\alpha=1$ is approached from either side because finite-size effects are larger in this regime. This can be understood from the fact that the $\alpha=1$ case is marginal: contributions which are subleading away from $\alpha=1$ become comparable in size to the leading contributions (see \cref{eq:scaling-expansion}, for example). 

\begin{figure}
    \centering
    \includegraphics[width=0.44\linewidth]{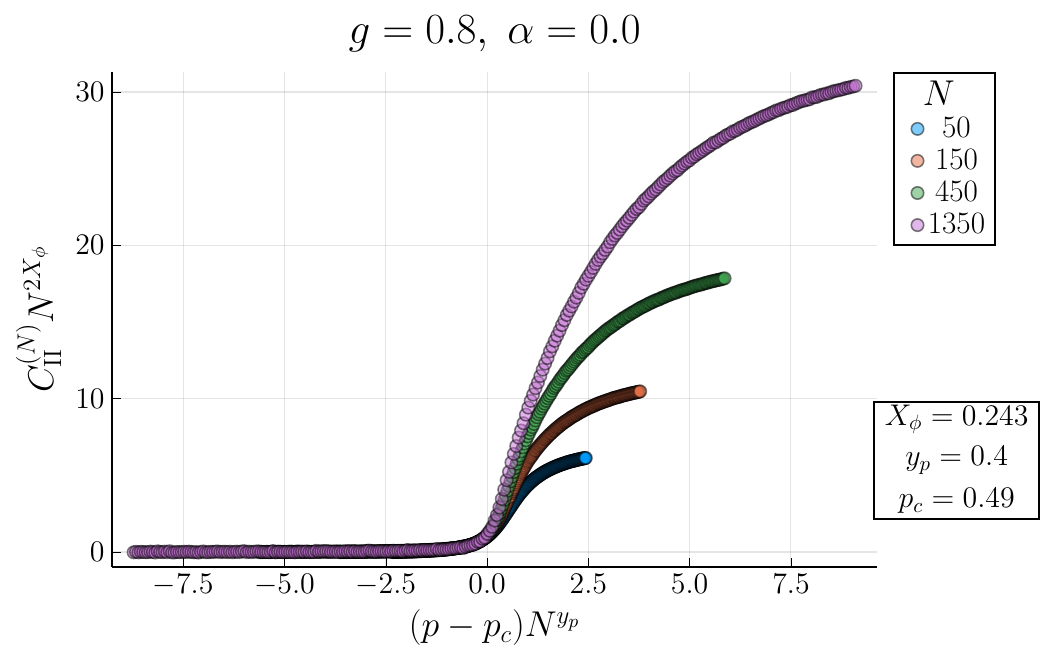}
    \hfill
    \includegraphics[width=0.44\linewidth]{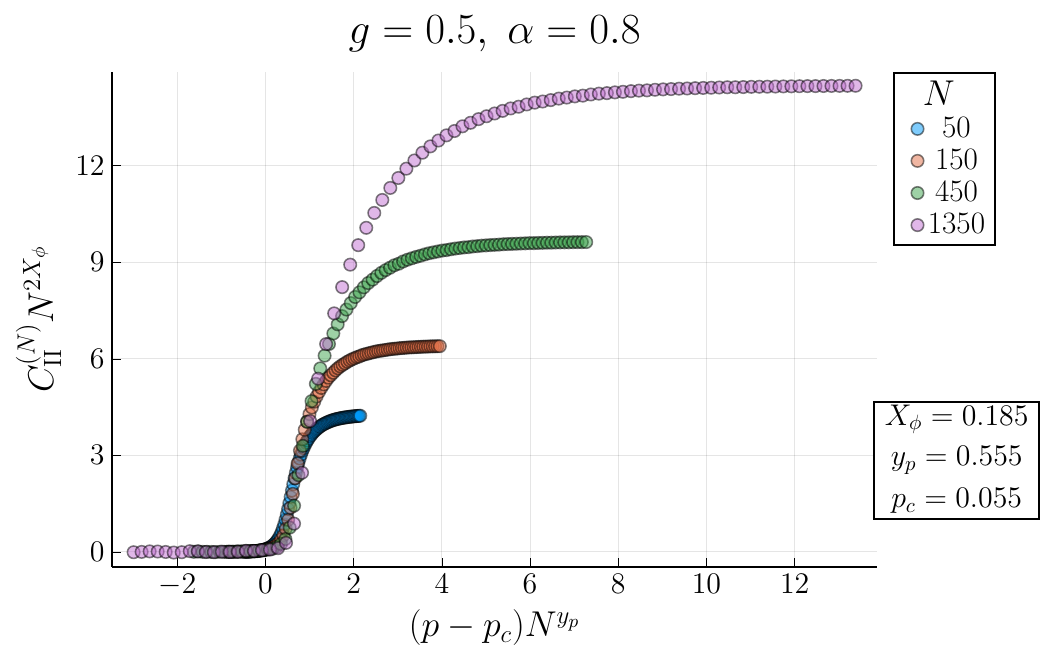}\\
    \includegraphics[width=0.44\linewidth]{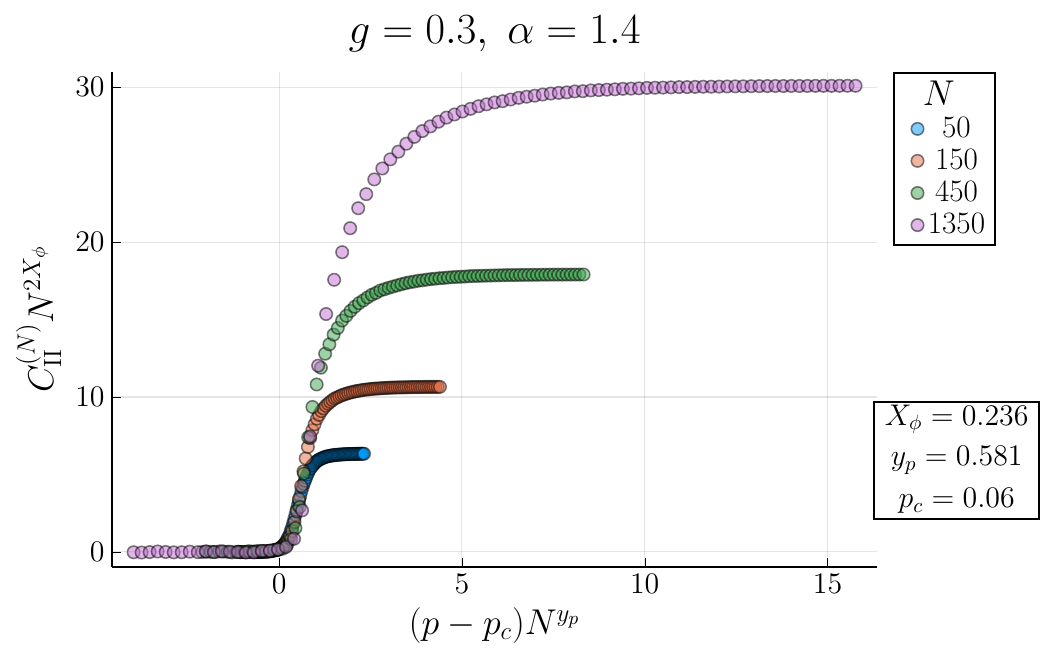}
    \hfill
    \includegraphics[width=0.44\linewidth]{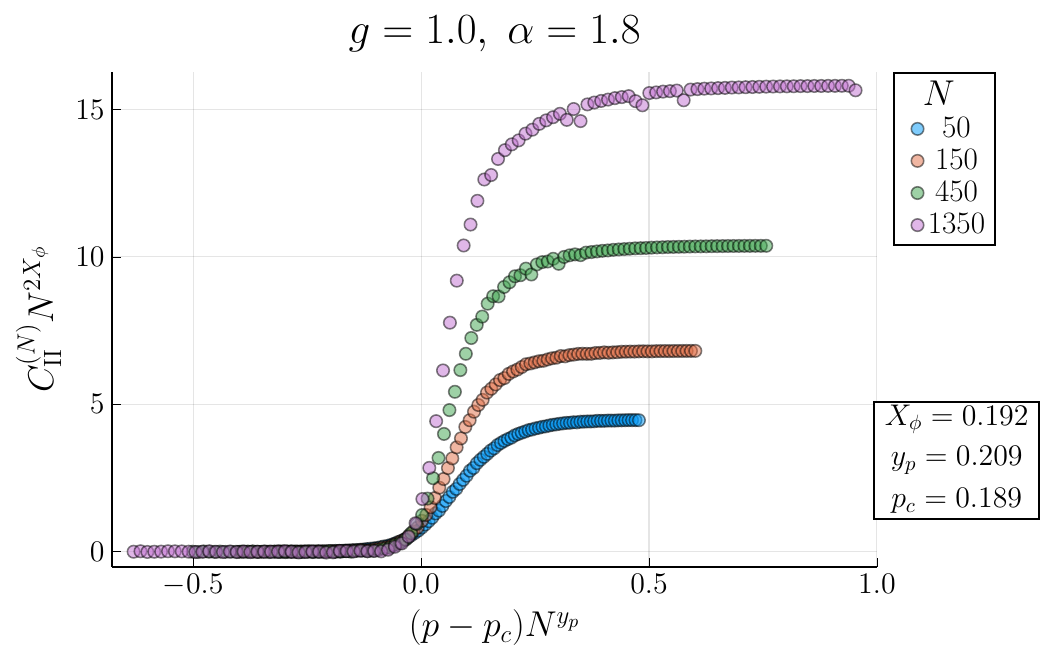}
    \caption{\textbf{Scaling collapse.} Representative scaling collapse data as a function of $p$ for several choices of $\alpha$ and $g$. The values of the scaling parameters used for the plot are presented in the boxes to the lower right of each plot. The rescaled data lies on a single universal curve in a window near $p-p_c=0$.}
    \label{fig:more-scale-coll}
\end{figure}

In \cref{fig:more-scale-coll}, we present additional examples of rescaled data that has undergone scaling collapse. As mentioned in the main text, the scaling exponents were extracted independently and the collapse serves as a consistency check. We also note that it is expected that scaling collapse will improve more slowly with $N$ for a power-law interacting model compared to a short-ranged model.

\begin{figure}
    \centering
    \includegraphics[width=0.44\linewidth]{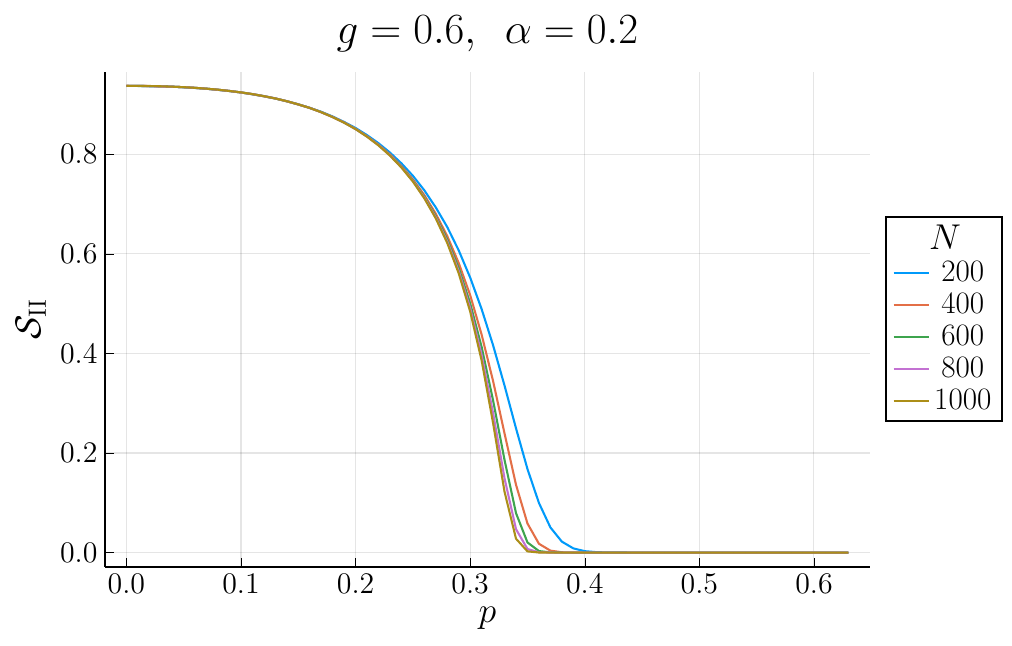}
    \hfill\includegraphics[width=0.44\linewidth]{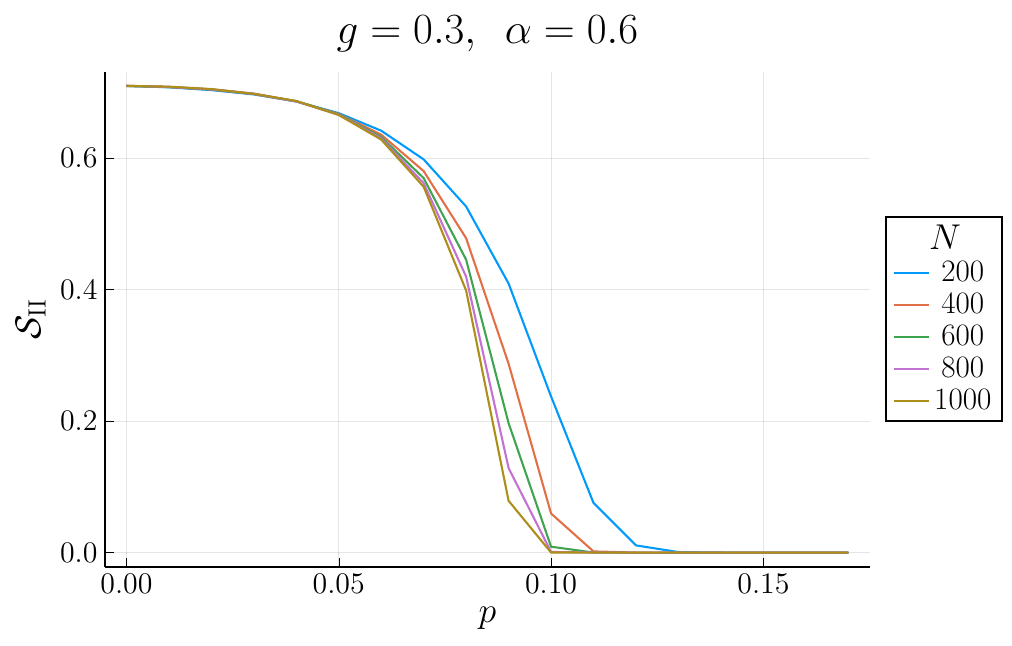}\\
    \includegraphics[width=0.44\linewidth]{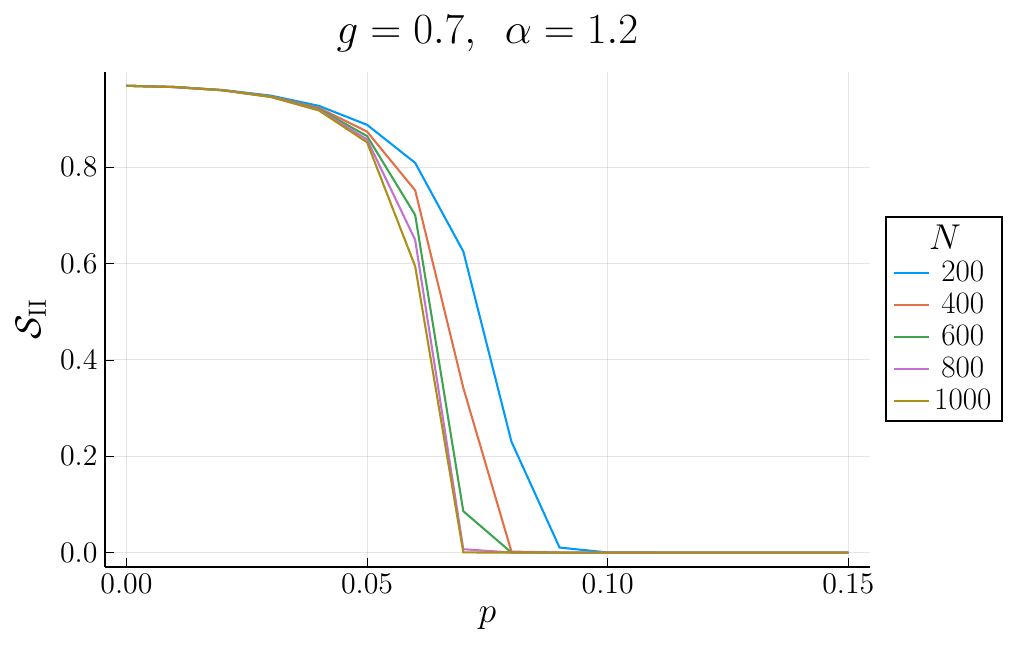}
    \hfill\includegraphics[width=0.44\linewidth]{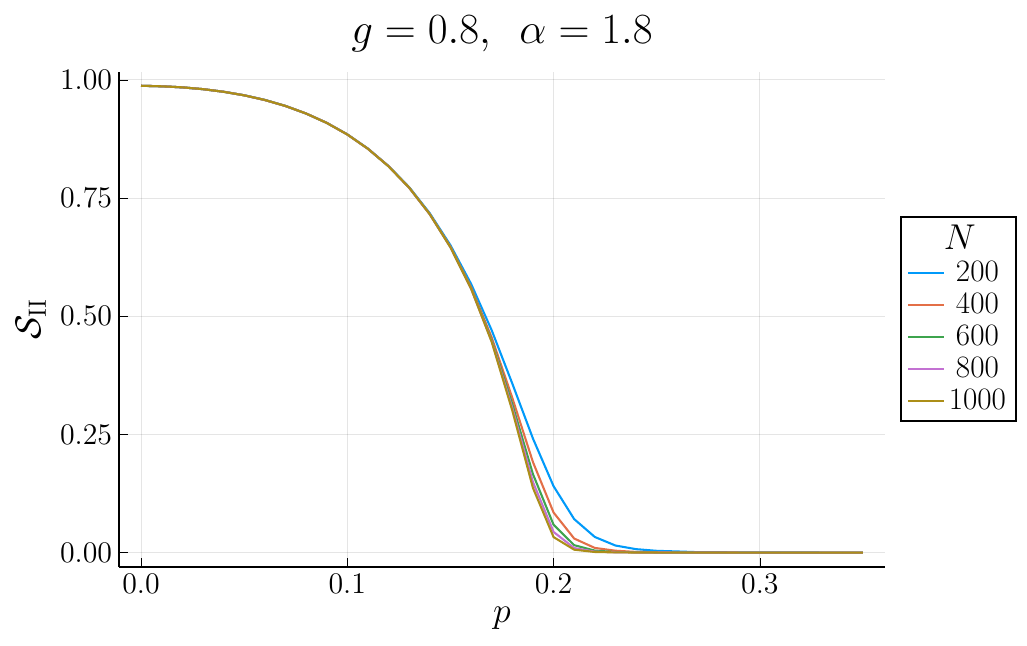}
    \caption{\textbf{String order parameter.} Representative \renyi-2 string order parameter data as a function of $p$ for several choices of $\alpha$ and $g$. Notice that the string order parameter undergoes a second-order phase transition which sharpens as $N$ increases.}
    \label{fig:string-data}
\end{figure}

In \cref{fig:string-data}, we present representative data for the \renyi-2 string order parameter. From the shape of the curves, it is clear that this diagnostic is probing a second-order phase transition which sharpens as $N$ increases.

\clearpage

\section{Mapping linear correlator to nearest-neighbor Ising model}\label{ap:nn-ising}

In this section, we will map the linear two-point correlator in the quantum problem to a two-point correlator in the classical nearest-neighbor Ising model, as presented in \cref{eq:C_I_derivation}. We have
\begin{align}
    C_{\text{I}, mn} &= \frac{\Tr[ Z_m Z_n \rho(\alpha;g,p)]}{\Tr[\rho(\alpha;g,p)]},\\
    &=\frac{\Tr[ Z_m Z_n \mathcal{E}_p^{(\alpha)}(\ketbra{\psi_g}{\psi_g})]}{\Tr[\mathcal{E}_p^{(\alpha)}(\ketbra{\psi_g}{\psi_g})]}.
    \intertext{Using the fact that $\mathcal{E}_p^{(\alpha)}$ is trace-preserving and its Kraus operators $Z_iZ_j$ are Hermitian, we have}
    &=\frac{\Tr[ \mathcal{E}_p^{(\alpha)}(Z_m Z_n) \ketbra{\psi_g}{\psi_g}]}{\Tr[\ketbra{\psi_g}{\psi_g}]},\\
    &=\frac{\bra{\psi_g}Z_m Z_n\ket{\psi_g}}{\braket{\psi_g|\psi_g}},\\
    &=\frac{\sum_{\{i_j\},\{i_j'\}=0,1}\bra{i_1,\ldots,i_N}e^{\beta\sum_{i=1}^{N-1}Z_i Z_{i+1}}Z_m Z_ne^{\beta\sum_{i=1}^{N-1}Z_i Z_{i+1}}\ket{i_1',\ldots,i_N'}}{\sum_{\{i_j\},\{i_j'\}=0,1}\bra{i_1,\ldots,i_N}e^{2\beta\sum_{i=1}^{N-1}Z_i Z_{i+1}}\ket{i_1',\ldots,i_N'}},
\end{align}
where we have used the imaginary-time representation of $\ket{\psi_g}$ introduced in \cref{eq:imag_time_symm} with $\beta=-\frac{1}{4}\ln |g|$ and expressed $\ket{+}^{\otimes N}$ in the computational basis as $\sum_{\{i_j\}=0,1}\ket{s_1,\ldots,s_N}$ (up to normalization, which cancels between the numerator and denominator). Replacing the operator $Z_j$ by its eigenvalue $s_j=(-1)^{i_j}$ and reducing the sum via the product of delta functions $\braket{i_1,\ldots,i_N|i_1',\ldots,i_N'}=\delta_{i_1,i_1'}\cdots\delta_{i_N,i_N'}$, we arrive at
\begin{equation}
    C_{\text{I}, mn}=\frac{\sum_{\{s_i\}=\pm 1} s_m s_n e^{2\beta\sum_{i=1}^{N-1}s_i s_{i+1}}}{\sum_{\{s_i\}=\pm 1}e^{2\beta\sum_{i=1}^{N-1}s_i s_{i+1}}},
\end{equation}
as reported in the main text.

\section{Vanishing of string order parameters linear in the density matrix}\label{ap:string}
Recall that the string order parameter in the $\mathbb{Z}_2^T$ SPT phase is given by~\cite{Perez-Garcia2008}
\begin{equation}
\mathcal{S}_{mn} = Z_{m-1} Y_m \left( \prod_{i=m+1}^{n-1} X_i \right) Y_n Z_{n+1} .
\end{equation}
We will suppress the subscript throughout this appendix for notational simplicity. We want to evaluate the linear version of this order parameter in the mixed state. This is given by
\begin{equation}
\braket{\mathcal{S}}_\rho=\frac{\Tr[\mathcal{S}\rho]}{\Tr[\rho]},
\end{equation}
where $\rho=\rho(\alpha;g,p)$ for $g<0$, which is the decohered SPT. Using the definition of $\rho(\alpha;g,p)$ from \cref{eq:rho_agp}, we have 
\begin{equation}
\begin{split}
\braket{\mathcal{S}}_\rho&=\frac{\Tr[\mathcal{S}\mathcal{E}_p^{(\alpha)}(\ketbra{\psi_g}{\psi_g})]}{\Tr[\mathcal{E}_p^{(\alpha)}(\ketbra{\psi_g}{\psi_g})]},\\
&=\frac{\bra{\psi_g}{\mathcal{E}_p^{(\alpha)}}^\dagger(\mathcal{S})\ket{\psi_g}}{\braket{\psi_g|\psi_g}},\\
\end{split}
\end{equation}
where we have used the fact that $\mathcal{E}_p^{(\alpha)}$ is trace-preserving to simplify the denominator. 

To evaluate $\braket{\mathcal{S}}_\rho$, we must now evaluate $\bra{\psi_g}{\mathcal{E}_p^{(\alpha)}}^\dagger(\mathcal{S})\ket{\psi_g}$. Because the $\mathcal{E}_{ij;p}^{(\alpha)}$ which comprise $\mathcal{E}_{p}^{(\alpha)}$ are Hermitian and commute with one another, we have
\begin{equation}
    \begin{split}
        {\mathcal{E}_p^{(\alpha)}}^\dagger(\mathcal{S})&=\mathcal{E}_p^{(\alpha)}(\mathcal{S}),\\
        &=\prod_{i<j}\mathcal{E}_{ij;p}^{(\alpha)}(\mathcal{S}),\\
        &=\prod_{i<j}p_{ij}^{(\alpha)} Z_iZ_j \mathcal{S} Z_iZ_j+(1-p_{ij}^{(\alpha)})\mathcal{S},\\
        &=\prod_{i<j}\begin{cases}
            (1-2p_{ij}^{(\alpha)})S,&\text{$i$ or $j$, but not both, in $[m,n]$}\\
            S,& \text{otherwise}
        \end{cases},\\
        &=\left(\prod_{i\not\in[m,n]}\prod_{j\in[m,n]}(1-2p_{ij}^{(\alpha)})\right)\mathcal{S}.\\
    \end{split}
\end{equation}
This tells us that 
\begin{equation}
    \braket{\mathcal{S}}_\rho=\left(\prod_{i\not\in[m,n]}\prod_{j\in[m,n]}(1-2p_{ij}^{(\alpha)})\right)\frac{\bra{\psi_g}\mathcal{S}\ket{\psi_g}}{\braket{\psi_g|\psi_g}}.
\end{equation}

We now argue that the prefactor tends to zero in the thermodynamic limit. We may write
\begin{align}
    \ln\left(\prod_{i\not\in[m,n]}\prod_{j\in[m,n]}(1-2p_{ij}^{(\alpha)})\right)&=\sum_{i\not\in[m,n]}\sum_{j\in[m,n]}\ln(1-2p_{ij}^{(\alpha)}).
    \intertext{When $p_{ij}^{(\alpha)}< 1/2$, we may use the inequality $\ln(1-x)\leq -x$ to write}
    &\leq-\sum_{i\not\in[m,n]}\sum_{j\in[m,n]}2p_{ij}^{(\alpha)}.
    \intertext{For simplicity, suppose that the string occupies an extensive region of the chain of size $N/3$, with other fractions giving an equivalent asymptotic result. We may then write (for $1<\alpha<2$)}
    &=-4\sum_{i\in\left[1,\frac{N}{3}\right]}\sum_{j\in\left(\frac{N}{3},\frac{2N}{3}\right]}\frac{p}{|i-j|^\alpha},\\
    &\leq-4p\left(\sum_{\Delta=1}^{N/3}\Delta\frac{1}{\Delta^{\alpha}}\right),
    \intertext{where we have discarded the terms with separation $\Delta$ greater than $N/3$ for simplicity, as the inclusion of these terms does not change the result. Further, note that the additional factor of $\Delta$ appears because there are $\Delta$ pairs of sites crossing each cut separated by a distance $\Delta$. Inserting the asymptotic form of the sum, we find }
    &\approx-4p\left[\frac{1}{(2-\alpha)}\left(\frac{N}{3}\right)^{2-\alpha}+\frac{1}{2}\left(\frac{N}{3}\right)^{1-\alpha}+\zeta(\alpha-1)\right],
\end{align}
Exponentiating both sides, and substituting $p\mapsto p/N^{1-\alpha}$ for the $0\leq\alpha<1$ case, we arrive at  
\begin{align}
    \left(\prod_{i\not\in[m,n]}\prod_{j\in[m,n]}(1-2p_{ij}^{(\alpha)})\right)\lesssim \begin{cases}
        e^{-C_\alpha N^{2-\alpha}},& 1<\alpha<2\\
        e^{-C_\alpha N},&0\leq\alpha<1\\
     \end{cases},
\end{align}
where $C_\alpha:=4p/\left[(2-\alpha)3^{2-\alpha}\right]>0$ when $p>0$ and $\alpha<2$. In either case, therefore, we see that the prefactor vanishes in the thermodynamic limit so long as $p>0$. Notice furthermore that the prefactor does \textit{not} vanish asymptotically when $\alpha>2$. This result is consistent with the fact that the crossover to short-range behavior occurs at $\alpha=2$, and short-range channels preserve linear string order~\cite{degroot2022symmetry}.

\section{\renyi-2 string order parameter at exactly solvable point}\label{ap:exact_renyi-2}

\newcommand{\XN}{\mathcal{P}([N])}
\newcommand{\xbar}{\bar{x}}
\newcommand{\Ssub}{\widetilde{S}}

In this section, we evaluate the \renyi-2 string order parameter at the exactly solvable point $g=-1$ and $\alpha=0$. Via the symmetry transformation introduced in \cref{eq:flip_g_S}, this quantity is equivalent to the \renyi-2 disorder parameter evaluated at $g=1$, and we will present the derivation for this case.

The initial pure state at $g=1$ is $\ket{\psi_{g=1}}=\ket{+}^{\otimes N}$. It will be useful to write the corresponding density matrix $\rho_0=\ketbra{\psi_{g=1}}{\psi_{g=1}}$ as
\begin{equation}
\rho_0
= \left[\frac{1}{2}\left(I+X\right)\right]^{\otimes N}
=\frac{1}{2^N}
\sum_{A \in [N]} X_A,
\qquad
X_A := \prod_{i \in A} X_i,
\end{equation}
where we have introduced
$[N] := \{\text{subsets of } N \text{ sites}\}$. 
In other words, $[N]$ denotes the power set of $\{1,\dots,N\}$.

Consider the action of the channel $\mathcal{E}_p^{(\alpha)}$ for $\alpha=0$. This channel acts on the Pauli string $X_A$ as
\begin{equation}
\mathcal{E}_p^{(\alpha=0)}(X_A)
= X_A
\left(1 - \frac{2p}{N}\right)^{|A|(N-|A|)} .
\end{equation}
By linearity, we have 
\begin{equation}
\mathcal{E}_p^{(\alpha=0)}(\rho_0)
=
\frac{1}{2^N}
\sum_{A \in [N]}
X_A
\left(1-\frac{2p}{N}\right)^{|A|(N-|A|)} .
\end{equation}
In the remainder of this section, we will let $\rho:=\mathcal{E}_p^{(\alpha=0)}(\rho_0)$ for brevity. Because we are measuring a \renyi-2 quantity, we are interested in the square of the density matrix:
\begin{equation}
\rho^2
=
\frac{1}{2^{2N}}
\sum_{A \in [N]}
\sum_{B \in [N]}
X_A X_B
\left(1-\frac{2p}{N}\right)^{|A|(N-|A|)+|B|(N-|B|)} .
\end{equation}
Taking the trace, we find
\begin{align}\label{eq:S_tr_rho_2}
\begin{split}
\Tr[\rho^2]
&=
\frac{1}{2^{2N}}
\sum_{A \in [N]}
\sum_{B \in [N]}
\left(1-\frac{2p}{N}\right)^{|A|(N-|A|)+|B|(N-|B|)}
\Tr[X_A X_B] \\
&=
\frac{1}{2^N}
\sum_{A \in [N]}
\left(1-\frac{2p}{N}\right)^{2|A|(N-|A|)},
\end{split}
\end{align}
where we have used the fact that $\Tr[X_A X_B] = \delta_{A,B}\,2^N$.

Similarly, defining $M$ to be the set of consecutive sites $\{m,\ldots,n\}$ so that $\tilde S=X_M$, we have
\begin{align}\label{eq:S_tr_S_rho_2}
\begin{split}
\Tr[\Ssub \rho^2]
&=
\frac{1}{2^{2N}}
\sum_{A \in [N]}
\sum_{B \in [N]}
\left(1-\frac{2p}{N}\right)^{|A|(N-|A|)+|B|(N-|B|)}
\Tr[X_M X_A X_B], \\
&=
\frac{1}{2^N}
\sum_{A \in [N]}
\left(1-\frac{2p}{N}\right)^{|A|(N-|A|)+|A\oplus M|(N-|A\oplus M|)},
\end{split}
\end{align}
where we have used the property that
$\Tr[\Ssub X_A X_B] = \delta_{A\oplus M,B}\,2^N$,
and $A\oplus M$ denotes the symmetric difference of sets. Equivalently, one could consider $A$ to be a vector in $\mathbb{Z}_2^N$ whose $i$th entry is nonzero if and only if $i\in A$ (and similarly for $B$ and $M$), in which case ``$\oplus$'' denotes addition mod 2.

The \renyi-2 string order parameter at $g=-1$ is equivalent to the ratio $\Tr[\Ssub \rho^2]/\Tr[\rho^2]$. In order to evaluate this quantity, we map it to an expression involving Curie-Weiss partition functions of the form~\cite{kochmanski2013curie}
\begin{equation}\label{eq:CW-part-fun}
Z_n(\beta)
=
\frac{1}{2^n}
\sum_{\{\sigma_i\}}
\exp\!\left(
\frac{\beta}{n}\left(\sum_i \sigma_i\right)^2
\right),
\end{equation}
which has known asymptotics.

To facilitate this mapping, we define a spin configuration $\{\sigma_i^{(A)}\}$ corresponding to the subset $A\subset \{1,\ldots, N\}$ via
\begin{equation}
\sigma_i^{(A)}
=
\begin{cases}
+1, & i \notin A, \\
-1, & i \in A,
\end{cases}
\end{equation}
which allows us to assign a magnetization to the subset $A$:
\begin{equation}
m_N(A) = \sum_{i \in N} \sigma_i^{(A)}.
\end{equation}
We may then express the size of $A$ in terms of the magnetization:
\begin{equation}
|A| = \frac{N - m_N(A)}{2},
\qquad
N-|A| = \frac{N + m_N(A)}{2},
\end{equation}
so that
\begin{equation}
|A|(N-|A|) = \frac{N^2 - m_N(A)^2}{4}.
\end{equation}

The previous equation allows us to change variables, so that $\Tr[\rho^2]$, as given in \cref{eq:S_tr_rho_2}, is written in terms of the magnetization $m_N(A)$:
\begin{align}
\Tr[\rho^2]
&=
\frac{1}{2^N}
\sum_{A \in [N]}
\exp\!\left[
\frac{\beta}{N}\bigl(-N^2 + m_N(A)^2\bigr)
\right]=
e^{-\beta N} Z_N(\beta),
\end{align}
where we have introduced $\beta := -\frac{N}{2}\ln\left(1-\frac{2p}{N}\right)$. 

In order to similarly rewrite $\Tr[\Ssub \rho^2]$, we need a few more expressions. Notice that we have the decomposition
\begin{equation}
m_N(A) = m_{\overline{M}}(A) + m_M(A),
\end{equation}
where $\overline{M}$ denotes the complement of $M$ in $\{1,\ldots,N\}$.
Then
\begin{equation}
|A\oplus M|
=
\frac{N - m_{\overline{M}}(A) + m_M(A)}{2},
\qquad
N-|A\oplus M|
=
\frac{N + m_{\overline{M}}(A) - m_M(A)}{2}.
\end{equation}
Hence
\begin{equation}\label{eq:seg-one}
|A|(N-|A|)
=
\frac{N^2 - \bigl(m_{\overline{M}}(A)+m_M(A)\bigr)^2}{4},
\end{equation}
and
\begin{equation}\label{eq:seg-two}
|A\oplus M|(N-|A\oplus M|)
=
\frac{N^2 - \bigl(m_{\overline{M}}(A)-m_M(A)\bigr)^2}{4}.
\end{equation}
Summing \cref{eq:seg-one,eq:seg-two} gives
\begin{equation}
|A|(N-|A|)+|A\oplus M|(N-|A\oplus M|)
=
\frac{N^2 - m_{\overline{M}}(A)^2 - m_M(A)^2}{2}.
\end{equation}

Finally, we may substitute the previous expression into \cref{eq:S_tr_S_rho_2}, writing $\Tr[\Ssub \rho^2]$ as follows (where $\beta$ is defined as above):
\begin{align}
\begin{split}
\Tr[\Ssub \rho^2]
&=
\frac{1}{2^N}
\sum_{A \in [N]}
\exp\!\left[
\frac{\beta}{N}
\bigl(-N^2 + m_{\overline{M}}(A)^2 + m_M(A)^2\bigr)
\right] \\
&=
e^{-\beta N}
\frac{1}{2^N}
\sum_{A \in [\overline{M}]}
\sum_{B \in [M]}
\exp\!\left[
\frac{\beta}{N}
\bigl(m_{\overline{M}}(A)^2 + m_M(B)^2\bigr)
\right] \\
&=
e^{-\beta N}
\left(
\frac{1}{2^{|\overline{M}|}}
\sum_{A \in [\overline{M}]}
e^{\frac{\beta}{N} m_{\overline{M}}(A)^2}
\right)
\left(
\frac{1}{2^{|M|}}
\sum_{B \in [M]}
e^{\frac{\beta}{N} m_M(B)^2}
\right).
\end{split}
\end{align}
Define $0<x<1$ such that $|M| = xN$. Then
\begin{equation}
\Tr[\Ssub \rho^2]
=
e^{-\beta N}
Z_{(1-x)N}\bigl((1-x)\beta\bigr)\,
Z_{xN}(x\beta).
\end{equation}
Hence
\begin{equation}\label{eq:CW-mapping}
\frac{\Tr[\Ssub \rho^2]}{\Tr[\rho^2]}
=
\frac{
Z_{(1-x)N}\bigl((1-x)\beta\bigr)\,
Z_{xN}(x\beta)
}{
Z_N(\beta)
}.
\end{equation}

The Curie-Weiss partition function may be evaluated via the saddle-point method, and we now quote the standard result for the conventions established in \cref{eq:CW-part-fun}~\cite{kochmanski2013curie}:
\begin{equation}
Z_n(\beta)
=
\frac{1}{2^n}
\sqrt{\frac{2\beta}{f''(m,2\beta)}}\,
e^{-n f(m,2\beta)},
\end{equation}
where $m$ is the average magnetization per site and the function $f$ is defined as
\begin{equation}
f(m,2\beta)
=
\beta m^2 - \ln\bigl[2\cosh(2\beta m)\bigr],
\end{equation}
and
\begin{equation}
f''(m,2\beta)
=
2\beta\left(1-2\beta(1-m^2)\right).
\end{equation}
This model undergoes a phase transition from the disordered to the ordered phase as $\beta$ increases through the critical point $\beta_c=\frac{1}{2}$~\cite{kochmanski2013curie}.

When $T>T_c$ (equivalently $\beta<\beta_c$), we have $m=0$, so
\begin{equation}
f(0,2\beta) = -\ln 2,
\qquad
f''(0,2\beta)=2\beta(1-2\beta),
\end{equation}
and hence
\begin{equation}\label{eq:CW-form-low-beta}
Z_n(\beta)\sim \frac{1}{\sqrt{1-2\beta}},
\end{equation}
which remains nonzero as $n\to\infty$.

When $T<T_c$ (equivalently $\beta>\beta_c$), we have $m\neq 0$, and
\begin{equation}
Z_n(\beta)
=
\frac{1}{2^n}
\sqrt{\frac{1}{1-2\beta(1-m^2)}}
\exp\!\left[-n\Bigl(\beta m^2-\ln\bigl(2\cosh(2\beta m)\bigr)\Bigr)\right].
\end{equation}
Thus, up to prefactors, we have
\begin{equation}\label{eq:CW-form-high-beta}
Z_n(\beta)
\sim
\exp\!\left[-n\Bigl(\beta m^2-\ln\bigl(\cosh(2\beta m)\bigr)\Bigr)\right],
\end{equation}
where $m$ is determined via the self-consistency condition $m=\tanh(2\beta m)$~\cite{kochmanski2013curie}.

We may now apply the behaviors we have just laid out to evaluate $\mathcal{S}_\mathrm{II}$ via \cref{eq:CW-mapping}. First, notice that for $p<\frac{1}{2}$, we have
\begin{equation}
\beta
=
-\frac{N}{2}\ln\left(1-\frac{2p}{N}\right)
<\beta_c=
\frac12,
\end{equation}
and the proper asymptotic behavior to apply is given by \cref{eq:CW-form-low-beta}. We therefore find that for $p<\frac{1}{2}$
\begin{equation}
\frac{\Tr[\Ssub \rho^2]}{\Tr[\rho^2]}
\sim
\sqrt{
\frac{1-2\beta}
{\bigl(1-2(1-x)\beta\bigr)\bigl(1-2x\beta\bigr)}
},
\end{equation}
which remains finite as $N\to\infty$.

Next, we will evaluate the same expression for $p>\frac{1}{2}$. For simplicity, we set $x=\tfrac12$, so that
\begin{equation}\label{eq:Z-ratio-one-half}
    \frac{\Tr[\Ssub \rho^2]}{\Tr[\rho^2]}=\frac{Z_{N/2}(\beta/2)\,Z_{N/2}(\beta/2)}{Z_N(\beta)}.
\end{equation}
We may identify two relevant regimes: $\frac{1}{2}<\beta<1$, where $\beta>\beta_c$ but $\frac{\beta}{2}<\beta_c$, and $\beta>1$, where $\beta,\frac{\beta}{2}>\beta_c$. 

In the $\frac{1}{2}<\beta<1$ regime, we substitute the asymptotic form given in \cref{eq:CW-form-low-beta} for the partition functions present in the numerator and the form given in \cref{eq:CW-form-high-beta} for the partition function appearing in the denominator. In this case, the numerator evaluates to a constant with respect to $N$, and the denominator diverges with $N$, so the ratio is asymptotically vanishing.

In the $\beta>1$ regime, we use the form given in \cref{eq:CW-form-high-beta} for all three partition functions. In particular, we find 
\begin{equation}
\begin{split}
\frac{\Tr[\Ssub \rho^2]}{\Tr[\rho^2]}
&=
\frac{Z_{N/2}(\beta/2)\,Z_{N/2}(\beta/2)}{Z_N(\beta)},\\&\sim
\frac{\exp\!\left[-N\Bigl(\frac{\beta}{2} m_1^2-\ln\bigl(\cosh(\beta m_1)\bigr)\Bigr)\right]}{\exp\!\left[-N\Bigl(\beta m_2^2-\ln\bigl(\cosh(2\beta m_2)\bigr)\Bigr)\right]},\\
&=\exp\!\left[-N\Bigl(\frac{\beta}{2} m_1^2-\beta m_2^2-\ln\bigl(\cosh(\beta m_1)\bigr)+\ln\bigl(\cosh(2\beta m_2)\bigr)\Bigr)\right],
\end{split}
\end{equation}
where $m_1$ solves $m_1 = \tanh(\beta m_1)$ and $m_2$ solves $m_2 = \tanh(2\beta m_2)$.

When $\beta>\beta_c$, the quantity in parentheses is positive, so that 
\begin{equation}
\frac{\Tr[\Ssub \rho^2]}{\Tr[\rho^2]}
\longrightarrow 0.
\end{equation}
We have therefore shown that the \renyi-2 string order parameter $\mathcal{S}_\mathrm{II}$ is finite below $p_c=\frac{1}{2}$, and asymptotically vanishing above.

\section{Large-$p$ decay of $\overline{X}_{L/R}$ for $0\leq\alpha<1$}\label{ap:p-decay-exp}
In this section, we derive the asymptotic scaling of the prefactor appearing in \cref{eq:p-decay-exp}. We begin by taking the logarithm of this prefactor:
\begin{align}
    \ln\left[\prod_{\Delta=1}^{N-1}\left(1-\frac{2p/N^{1-\alpha}}{\Delta^\alpha}\right)\right]&=\sum_{\Delta=1}^{N-1}\ln\left(1-\frac{2p/N^{1-\alpha}}{\Delta^\alpha}\right),
    \intertext{Next, we take $N$ to be large, so we may use the small-$x$ approximation $\log(1-x)\approx -x$:}
    &=-\frac{2p}{N^{1-\alpha}}\sum_{\Delta=1}^{N-1}\frac{1}{\Delta^\alpha},\\
    &=-\frac{2p}{N^{1-\alpha}}\left[\left(\sum_{\Delta=1}^{N}\frac{1}{\Delta^\alpha}\right)-\frac{1}{N^\alpha}\right].
    \intertext{We now replace the sum with its large-$N$ asymptotic form:}
    &=-\frac{2p}{N^{1-\alpha}}\left(-\frac{1}{2 N^\alpha}+\frac{N^{1-\alpha}}{1-\alpha}+\zeta(\alpha)\right),\label{eq:scaling-expansion}\\
    &\overset{N\to\infty}{\longrightarrow}-\frac{2p}{1-\alpha}.
\end{align}
We have just shown that the logarithm of the prefactor evaluates to $-2p/(1-\alpha)$, so by exponentiating we arrive at the result
\begin{equation}
    \prod_{\Delta=1}^{N-1}\left(1-\frac{2p/N^{1-\alpha}}{\Delta^\alpha}\right)\overset{N\to\infty}{\longrightarrow}\mathrm{exp}\left[-\frac{2p}{1-\alpha}\right].
\end{equation}

\section{Classical post-processing via multiple importance sampling}\label{app:reweighting}
In this section, we discuss how experimental measurements of $C_\mathrm{II}(p)$ can also contain information about $C_\mathrm{II}(q)$ for $p\neq q$ which is accessible via classical post-processing.

As mentioned in the main text, observing SW-SSB requires measuring $C_\mathrm{II}(p)$ at several values of the decoherence strength $p$. Suppose that measurement shots are collected at $m$ different decoherence strengths $p_i$, with $i=1,\ldots,m$. The data collected at these values can be reused to estimate the numerator or denominator of $C_\mathrm{II}(q)$ at another decoherence strength $q$ through a simple classical post-processing procedure, thereby increasing the utility of the collected shots over the range of decoherence strengths of interest.

We now describe this procedure, which is an instance of the balance heuristic of multiple importance sampling~\cite{VeachGuibas1995}. Recall that we generate an ensemble of circuit pairs $U_a$ and $U_{a'}$ from the probability distribution $P(a,a';p)$, where we explicitly indicate its dependence on the decoherence strength $p$. We then collect measurement outcomes from the quantum device according to $p(\mathbf{b}|a,a') = \left| \langle \mathbf{b} | \psi_{a,a'} \rangle \right|^{2}$, which is independent of $p$. The numerator or denominator of $C_\mathrm{II}(p)$ is estimated as $p(\mathbf{0};p) =\mathbb{E}_p[\delta_{\mathbf{b}=\mathbf{0}}] =  \sum_{a,a'}p(\mathbf{0}|a,a') P(a,a';p)$, where $\mathbb{E}_p[\bullet] := \sum_{a,a',\mathbf{b}}(\bullet)p(\mathbf{b} |a,a') P(a,a';p)$, using the appropriate choices of $U_a$ and $U_{a'}$.
More concretely, suppose that $N_i$ measurement shots are collected at each decoherence strength $p_i$, for $i=1,\ldots,m$, so that the total number of shots is $N_s=\sum_{i=1}^{m}N_i$. We denote the data from the $\ell$-th shot at $p_i$ by 
\[\mathcal{D}_{i,\ell}=\left(a_{i,\ell},a'_{i,\ell};\mathbf{b}_{i,\ell}\right),\qquad\ell=1,\ldots,N_i,\] 
where $a_{i,\ell}$ and $a'_{i,\ell}$ label the sampled circuit pair and $\mathbf{b}_{i,\ell}$ is the corresponding measurement-outcome bit string.

To construct an estimator for a different decoherence strength $q$ from these data, we first define $\lambda_i:=N_i/N_s$ and the mixture distribution $Q(a,a'):=\sum_{i=1}^m\lambda_iP(a,a';p_i)$. We then have
\begin{align}
p(\mathbf{b};q)
=
\sum_{a,a'}p(\mathbf{b}| a,a')P(a,a';q)
=
\sum_{a,a'}w_q(a,a')p(\mathbf{b}| a,a')Q(a,a'),
\label{eqn:p(q) to weight}
\end{align}
where the weight
\begin{equation}
w_q(a,a')
:=
\frac{P(a,a';q)}{Q(a,a')}
\end{equation}
can be calculated classically and is well defined provided that $Q(a,a')>0$ on the support of $P(a,a';q)$. This condition is automatically satisfied for $p_i\in(0,1)$.
Using the definition of $Q(a,a')$, Eq.~(\ref{eqn:p(q) to weight}) can be rewritten as
\begin{align}
p(\mathbf{b};q)
=
\sum_{a,a'}w_q(a,a')p(\mathbf{b}| a,a')Q(a,a')
=
\sum_{i=1}^m\lambda_i
\sum_{a,a'}w_q(a,a')p(\mathbf{b}| a,a')P(a,a';p_i).
\end{align}
Therefore, the quantity of interest becomes
\begin{align}
p(\mathbf{0};q)
=
\mathbb{E}_q\!\left[\delta_{\mathbf{b}=\mathbf{0}}\right]
=
\sum_{i=1}^m\lambda_i
\mathbb{E}_{p_i}\!\left[
\delta_{\mathbf{b}=\mathbf{0}}w_q(a,a')
\right],
\end{align}
where, as before,
$
\mathbb{E}_p[\bullet]
:=
\sum_{a,a',\mathbf{b}}
(\bullet)\,
p(\mathbf{b}|a,a')P(a,a';p).
$

This relation yields the following estimator. For each $i$, let $I_i$ denote the set of shots for which $\mathbf{b}_{i,\ell}=\mathbf{0}$. An unbiased estimator of $p(\mathbf{0};q)$ is then
\begin{align}
\hat{\mu}
=
\sum_{i=1}^m
\lambda_i\frac{1}{N_i}
\sum_{\ell\in I_i}
w_q(a_{i,\ell},a'_{i,\ell})
=
\frac{1}{N_s}
\sum_{i=1}^m
\sum_{\ell\in I_i}
w_q(a_{i,\ell},a'_{i,\ell}).
\end{align}
It is easy to verify $\mu := \mathbb{E}[\hat{\mu}] = \frac{1}{N_s} \sum_{i=1}^{m} \sum_{\ell=1}^{N_i}\mathbb{E}_{p_i}[X_{i,\ell}] = p(\mathbf{0};q)$, where we have denoted $X_{i,\ell} := w_q(a_{i\ell},a_{i\ell}')\delta_{\mathbf{b}_{i,\ell}=\mathbf{0}}$.

To analyze the variance, we rewrite the estimator as
\begin{align}
    \hat{\mu}
    =
    \frac{1}{N_s}\sum_{i=1}^{m}\sum_{\ell=1}^{N_i}X_{i,\ell}
    =
    \sum_{i=1}^{m}\lambda_i\hat{\mu}_i,
\end{align}
where
\begin{equation}
    \hat{\mu}_i
    :=
    \frac{1}{N_i}\sum_{\ell=1}^{N_i}X_{i,\ell}
    =
    \frac{1}{N_i}\sum_{\ell=1}^{N_i}
    w_q(a_{i,\ell},a'_{i,\ell})
    \delta_{\mathbf{b}_{i,\ell}=\mathbf{0}}.
\end{equation}
The quantities $\hat{\mu}_i$ are mutually independent sample averages with expectations
\begin{align}
    \mu_i
    :=
    \mathbb{E}_{p_i}[X_{i,\ell}]
    =
    \sum_{a,a'}
    w_q(a,a')p(\mathbf{0}| a,a')P(a,a';p_i).
\end{align}
The variance of the estimator is therefore
\begin{align}
    \operatorname{Var}[\hat{\mu}]
    =
    \sum_{i=1}^{m}\lambda_i^2\operatorname{Var}[\hat{\mu}_i]
=
    \sum_{i=1}^{m}\frac{\lambda_i^2}{N_i}
    \operatorname{Var}[X_{i,\ell}]
  =
    \frac{1}{N_s}\sum_{i=1}^{m}\lambda_i v_i,
\end{align}
where
\begin{align}
    v_i
    :=
    \operatorname{Var}[X_{i,\ell}]
   =
    \mathbb{E}_{p_i}\!\left[
    w_q^2
    \delta_{\mathbf{b}=\mathbf{0}}
    \right]
    -
    \mu_i^2.
\end{align}

To obtain a bound on the variance, it is convenient to rewrite
\begin{equation}
    N_s \cdot \text{Var}[\hat{\mu}] = \sum_{i=1}^{m}\lambda_i v_i = S -\sum_{i=1}^m\lambda_i \mu_i^2~,
\end{equation}
where $S =\sum_{a,a'}w_q(a,a')p(\mathbf{0}|a,a')P(a,a';q) = \mathbb{E}_q[w_q \delta_{\mathbf{b}=\mathbf{0}}] = \mu \cdot w_{\text{eff}}$ and the conditional mean weight $w_{\text{eff}} :=\mathbb{E}_q[w_q|\mathbf{b}=\mathbf{0}] =\mathbb{E}_q[w_q \delta_{\mathbf{b}=\mathbf{0}}]/\mathbb{E}_q[ \delta_{\mathbf{b}=\mathbf{0}}]$.
Using Jensen's inequality $\sum_i\lambda_i\mu_i^2 \geq \mu^2$ (recall $\mu = \sum_{i}\lambda_i\mu_i$), we obtain
\begin{equation}
    N_s \cdot \text{Var}[\hat{\mu}] \leq S - \mu^2~.
\end{equation}
Therefore, to estimate $\mu$ to $\epsilon$ relative error, it suffices to take $N_s$ such that $\epsilon^2 \sim \text{Var}[\hat{\mu}]/\mu^2$, or 
\begin{equation}
    N_s \sim \frac{S-\mu^2}{\mu^2\epsilon^2} \leq \frac{S}{\mu^2\epsilon^2} =\frac{w_{\text{eff}}}{\mu\epsilon^2}.
\end{equation}
Finally, using the crude bound $S \leq \mu \cdot [\max_{a,a'}w_q(a,a')]$, we also obtain $N_s \lesssim [\max_{a,a'}w_q(a,a')]/(\mu\epsilon^2)$. 

It is instructive to consider the case where $q$ coincides with one of the $p_i$, say $q=p_1$. This gives us $Q(a,a') = \sum_{i}\lambda_iP(a,a';p_i) \geq \lambda_1 P(a,a';p_1)$, and we have $\max_{a,a'}w_q(a,a') \leq 1/\lambda_1$, or $N_s \lesssim 1/(\lambda_1 \mu \epsilon^2)$.
Since $\lambda_1 N_s = N_1$, this is consistent with $N_1 \sim 1/(\mu\epsilon^2)$, the direct-sampling cost quoted in the main text. The reweighted estimator's guarantee is thus never worse than using the shots at $p_1$ alone, and improves upon it when $w_{\text{eff}} < 1/\lambda_1$.

\end{document}